\documentclass{article} %
\usepackage{iclr2027_conference,times}

\usepackage{amsmath,amsfonts,bm}

\def\eqref#1{equation~\ref{#1}}

\def\1{\bm{1}}

\DeclareMathAlphabet{\mathsfit}{\encodingdefault}{\sfdefault}{m}{sl}
\SetMathAlphabet{\mathsfit}{bold}{\encodingdefault}{\sfdefault}{bx}{n}

\usepackage{hyperref}
\usepackage{url}

\usepackage{graphicx}
\usepackage{float}
\usepackage{booktabs}
\usepackage{array}
\usepackage{multirow}
\usepackage{longtable}
\usepackage[T1]{fontenc}
\usepackage{xcolor}
\usepackage[draft,authormarkup=none,defaultcolor=black]{changes}
\colorlet{omcrevision}{black}
\definechangesauthor[name={},color=black]{rev}
\setaddedmarkup{\textcolor{authorcolor}{#1}}
\setdeletedmarkup{}
\usepackage{colortbl}
\definecolor{codecolor}{HTML}{000000}
\definecolor{codebg}{HTML}{F5F5F5}
\definecolor{omcsinglebase}{HTML}{ECF7FC}
\colorlet{omcsinglebg}{white}
\colorlet{omcmultibg}{omcsinglebg}
\definecolor{omcsingleheading}{HTML}{000000}
\definecolor{omcmultiheading}{HTML}{000000}
\definecolor{omcsingleaccent}{HTML}{00BBF0}
\definecolor{omcmultiaccent}{HTML}{E61C5D}
\usepackage{subcaption}
\usepackage{natbib}
\usepackage{listings}
\usepackage[framemethod=tikz]{mdframed}
\usepackage{tablefootnote}
\usepackage[normalem]{ulem}   %
\makeatletter
\edef\omcSavedEnc{\encodingdefault}\edef\omcSavedRm{\rmdefault}
\edef\omcSavedTt{\ttdefault}\edef\omcSavedSf{\sfdefault}
\makeatother
\usepackage{xeCJK}
\renewcommand{\encodingdefault}{\omcSavedEnc}
\renewcommand{\rmdefault}{\omcSavedRm}
\renewcommand{\ttdefault}{\omcSavedTt}
\renewcommand{\sfdefault}{\omcSavedSf}
\normalfont
\newlength{\omcsep}
\newlength{\omclab}
\newenvironment{omcrec}
  {\par\small
   \begin{list}{}{%
     \setlength{\labelwidth}{\omclab}%
     \setlength{\labelsep}{7pt}%
     \setlength{\leftmargin}{\dimexpr\omclab+7pt\relax}%
     \setlength{\rightmargin}{0pt}%
     \setlength{\itemindent}{0pt}%
     \setlength{\listparindent}{0pt}%
     \setlength{\topsep}{\omcsep}%
     \setlength{\partopsep}{0pt}%
     \setlength{\itemsep}{1.1pt}%
     \setlength{\parsep}{0pt}%
   }}
  {\end{list}\normalsize}
\newcommand{\omcrule}[1]{\par\nobreak\kern1.6pt\hrule height #1\kern1.6pt\nobreak}
\newcommand{\omcparline}[1]{\begingroup\parskip=0pt\noindent #1\par\endgroup}
\newcommand{\omchead}[1]{\par\small\omcrule{0.7pt}\kern\omcsep\omcparline{#1}\kern\omcsep\omcrule{0.35pt}\normalsize}
\newcommand{\omcsub}[1]{\par\small\omcrule{0.35pt}\kern\omcsep\omcparline{\textbf{#1}}\nobreak\normalsize}
\newcommand{\omcround}[2]{\omcsub{Turn #1 \quad \textnormal{#2}}}
\newcommand{\omcfoot}{\omcrule{0.7pt}}
\mdfdefinestyle{omctracebox}{
  nobreak=true,
  skipabove=0pt,
  skipbelow=0pt,
  linecolor=black,
  linewidth=0.7pt,
  roundcorner=4pt,
  innerleftmargin=6pt,
  innerrightmargin=6pt,
  innertopmargin=6pt,
  innerbottommargin=6pt
}
\newmdenv[style=omctracebox,backgroundcolor=omcsinglebg]{omcsingleexample}
\newmdenv[style=omctracebox,backgroundcolor=omcmultibg]{omcmultiexample}
\newmdenv[
  hidealllines=true,
  linewidth=0pt,
  roundcorner=4pt,
  backgroundcolor=codebg,
  innerleftmargin=4pt,
  innerrightmargin=4pt,
  innertopmargin=4pt,
  innerbottommargin=4pt
]{omcpromptbox}
\newcommand{\omctracekey}[2]{\textcolor{#1}{\textbf{#2}}}
\newcommand{\omcsinglekey}[1]{\omctracekey{omcsingleheading}{#1}}
\newcommand{\omcmultikey}[1]{\omctracekey{omcmultiheading}{#1}}
\newcommand{\omctracerow}{\\\arrayrulecolor{black!40}\specialrule{0.4pt}{1.2pt}{1.2pt}}
\newcommand{\omcsinglerow}{\omctracerow}
\newcommand{\omcmultirow}{\omctracerow}
\newlength{\omcsinglekeywidth}
\newlength{\omcsinglevaluewidth}
\newlength{\omcmultikeywidth}
\newlength{\omcmultiroundwidth}
\newlength{\omcmultisharedwidth}
\usepackage{etoc}   %

\lstdefinestyle{promptstyle}{
    numbers=left,
    numberstyle=\tiny\color{gray},
    stepnumber=1,
    backgroundcolor={},
    frame=none,
    breaklines=true,
    basicstyle=\tiny\fontfamily{SourceCodePro-TLF}\fontseries{m}\selectfont\color{codecolor},
    showstringspaces=false,
    breakindent=0pt,
    breakautoindent=false,
    columns=fullflexible,
    keepspaces=true,
    numbersep=3pt,
    xleftmargin=1.6em,
}

\DeclareRobustCommand{\omcstagebadge}[1]{%
  \raisebox{0.72ex}{\scalebox{0.88}{%
    \tikz[baseline=-0.65ex]{%
      \path[use as bounding box] (-4.1pt,-4.1pt) rectangle (4.1pt,4.1pt);
      \draw[line width=0.4pt] (0,0) circle[radius=3.9pt];
      \node[inner sep=0pt,outer sep=0pt,anchor=center,
        font=\normalfont\rmfamily\bfseries\fontsize{6.6}{6.6}\selectfont] at (0,0) {#1};}}}}
\expandafter\def\csname omcstagecode@S1\endcsname{\omcstagebadge{0}}
\expandafter\def\csname omcstagecode@S2\endcsname{\omcstagebadge{1}}
\expandafter\def\csname omcstagecode@S3\endcsname{\omcstagebadge{2}}
\expandafter\def\csname omcstagecode@S4\endcsname{\omcstagebadge{3}}
\expandafter\def\csname omcstagecode@S5\endcsname{\omcstagebadge{4}}
\expandafter\def\csname omcstagecode@S6\endcsname{\omcstagebadge{5}}
\expandafter\def\csname omcstagecode@S7\endcsname{\omcstagebadge{6}}
\expandafter\def\csname omcstagecode@S8\endcsname{\omcstagebadge{7}}
\expandafter\def\csname omcstagecode@S9\endcsname{\omcstagebadge{8}}
\expandafter\def\csname omcstagecode@S10\endcsname{\omcstagebadge{9}}
\expandafter\def\csname omcstagecode@S11\endcsname{\omcstagebadge{11}}
\expandafter\def\csname omcstagecode@S12\endcsname{\omcstagebadge{10}}
\expandafter\def\csname omcstagecode@S13\endcsname{\omcstagebadge{12}}
\expandafter\def\csname omcstagecode@S14\endcsname{\omcstagebadge{13}}
\expandafter\def\csname omcstagecode@S15\endcsname{\omcstagebadge{14}}
\expandafter\def\csname omcstagecode@M1\endcsname{\omcstagebadge{0}}
\expandafter\def\csname omcstagecode@M2\endcsname{\omcstagebadge{a}}
\expandafter\def\csname omcstagecode@M3\endcsname{\omcstagebadge{b}}
\expandafter\def\csname omcstagecode@M4\endcsname{\omcstagebadge{c}}
\expandafter\def\csname omcstagecode@M5\endcsname{\omcstagebadge{d}}
\expandafter\def\csname omcstagecode@M6\endcsname{\omcstagebadge{e}}
\expandafter\def\csname omcstagecode@M7\endcsname{\omcstagebadge{f}}
\expandafter\def\csname omcstagecode@M8\endcsname{\omcstagebadge{g}}
\expandafter\def\csname omcstagecode@M9\endcsname{\omcstagebadge{i}}
\expandafter\def\csname omcstagecode@M10\endcsname{\omcstagebadge{h}}
\expandafter\def\csname omcstagecode@M11\endcsname{\omcstagebadge{j}}
\expandafter\def\csname omcstagecode@M12\endcsname{\omcstagebadge{k}}
\expandafter\def\csname omcstagecode@M13\endcsname{\omcstagebadge{l}}
\expandafter\def\csname omcstagecode@M14\endcsname{\omcstagebadge{l}}
\expandafter\def\csname omcstagecode@M15\endcsname{\omcstagebadge{m}}
\expandafter\def\csname omcstagecode@M16\endcsname{\omcstagebadge{n}}
\expandafter\def\csname omcstagecode@M17\endcsname{\omcstagebadge{b}}
\expandafter\def\csname omcstagecode@S2/M2\endcsname{\omcstagebadge{1}\kern0.12em\omcstagebadge{a}}
\expandafter\def\csname omcstagecode@S5/M15\endcsname{\omcstagebadge{4}\kern0.12em\omcstagebadge{m}}
\expandafter\def\csname omcstagecode@S6/M16\endcsname{\omcstagebadge{5}\kern0.12em\omcstagebadge{n}}
\expandafter\def\csname omcstagecode@S9/M7\endcsname{\omcstagebadge{8}\kern0.12em\omcstagebadge{f}}
\expandafter\def\csname omcstagecode@S10/M8\endcsname{\omcstagebadge{9}\kern0.12em\omcstagebadge{g}}
\expandafter\def\csname omcstagecode@S11/M9\endcsname{\omcstagebadge{11}\kern0.12em\omcstagebadge{i}}
\expandafter\def\csname omcstagecode@S12/M10\endcsname{\omcstagebadge{10}\kern0.12em\omcstagebadge{h}}
\DeclareRobustCommand{\omcstageid}[1]{\nobreak\hspace{0.12em}\csname omcstagecode@#1\endcsname}
\newcommand{\stage}[1]{\textnormal{\textbf{#1}}} %

\newcommand{\omctablefont}{\fontsize{8.2}{10}\selectfont}

\title{\makebox[\textwidth][s]{OmniVChat:\hfill Synthesizing,\hfill Benchmarking,\hfill and}\\
Training for Native Audio-Visual Dialogue\vspace{-6pt}}

\author{\makebox[\textwidth][c]{\textbf{Haolin He}$^{1,2,*}$, \textbf{Yunfei Chu}$^{2,*}$, \textbf{Qi Chen}$^{2,3,4,*}$, \textbf{Wen Huang}$^{2,3}$, \textbf{Yuan Feng}$^{2}$, \textbf{Muzhi Zhu}$^{5}$,}\\
\makebox[\textwidth][c]{\bfseries \textbf{Zheqi Dai}$^{1}$, \textbf{Haoning Xu}$^{1}$, \textbf{Dongchao Yang}$^{1,2}$, \textbf{Chunyat Wu}$^{1}$, \textbf{Zining Liang}$^{1}$, \textbf{Zhengxi Liu}$^{1}$,}\\
\makebox[\textwidth][c]{\bfseries \textbf{Xiquan Li}$^{3}$, \textbf{Xie Chen}$^{3}$, \textbf{Xize Cheng}$^{2}$, \textbf{Qize Yang}$^{2}$, \textbf{Jin Xu}$^{2,\dagger}$, \textbf{Qiuqiang Kong}$^{1,\dagger}$}\\
\makebox[\textwidth][c]{$^{1}$The Chinese University of Hong Kong\hspace{8pt}%
$^{2}$Alibaba Token Hub, Alibaba Group}\\
\makebox[\textwidth][c]{$^{3}$Shanghai Jiao Tong University\hspace{8pt}%
$^{4}$Shanghai Innovation Institute\hspace{8pt}%
$^{5}$Zhejiang University}\\
}

\iclrfinalcopy %
\begin{document}
{\setlength{\tabcolsep}{0pt}\maketitle}\vspace{-14pt}
\ificlrfinal
{
\makeatletter
\def\@fnsymbol#1{\ensuremath{\ifcase#1\or *\or \dagger\or **\or ***\or \ddagger\else\@ctrerr\fi}}
\makeatother
\renewcommand{\thefootnote}{\fnsymbol{footnote}}
\footnotetext[1]{Equal contribution. $^\dagger$Corresponding authors.
Work done as an intern at Qwen Team, Alibaba Token Hub.}
}
\setcounter{footnote}{0}
\renewcommand{\thefootnote}{\arabic{footnote}}
\fi

\begin{abstract}
\vspace{-4pt}
We define \textbf{\mbox{OmniVChat}} (Omni Video Chat) as the task of \textbf{native audio-visual dialogue} between a user and an
omni model. In \mbox{OmniVChat}, omni models directly and simultaneously receive
audio and video from a user and return text.
The user's query is embedded in the audio and video, without a separate text question, external
captioning, or speech recognition.
Direct audio-visual input reduces external latency and computation while preserving perceptual cues.
However, research on \mbox{OmniVChat} faces two constraints: data availability and evaluation.
Recordings of people using their own devices are scarce.
Furthermore, good replies often depend on multimodal context and can be phrased in many ways,
making keyword matching unreliable for evaluation.
Recent progress in agent systems and video generation makes \textbf{generation for comprehension} viable,
which means using synthesized dialogues for training and evaluation.
Therefore, we present \textbf{OmniVChat-Studio}, a multi-agent data engine for synthesizing
single- and multi-turn audio-visual dialogues. We use synthesized dialogues to build \textbf{OmniVChat-Bench}, an
evaluation benchmark that evaluates omni models' basic dialogue abilities across five ability categories.
Replies are judged by a large language model based on explicit scoring criteria.
We also present \textbf{OmniVChat-RL}, a reinforcement learning reward design that jointly targets reply correctness,
efficiency, and style in \mbox{OmniVChat}.
Training Qwen3-Omni-Instruct with OmniVChat-RL on synthesized dialogues improves its performance on both OmniVChat-Bench and the human-recorded \textbf{OmniVChat-Bench-Human}.
These gains validate the reward design and show transfer to real-world dialogues in training and
evaluation.
\end{abstract}

\begin{figure}[H]
\vspace{-12pt}
\centering
\includegraphics[width=\textwidth]{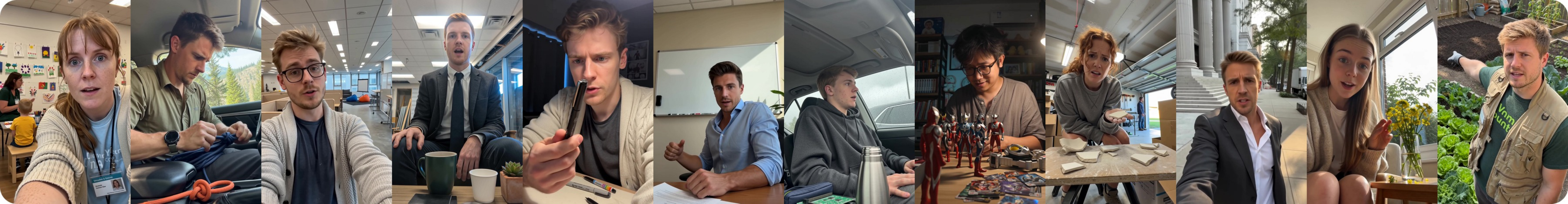}
\caption{A glance at \textbf{OmniVChat-Bench}: each card shows one user frame from a synthesized dialogue.
The benchmark spans 5 ability categories, 17 subcategories, and 22 scenario domains.}
\label{fig:deck}
\vspace{-18pt}
\end{figure}

\addtocontents{toc}{\protect\etocsetlevel{section}{6}}
\addtocontents{toc}{\protect\etocsetlevel{subsection}{6}}

\section{Introduction}
\label{sec:intro}

We define \textbf{\mbox{OmniVChat}} (Omni Video Chat) as the task of \textbf{native audio-visual dialogue} between a user and an
omni model~\citep{jiang2025specific}. In OmniVChat, omni models directly and simultaneously
\added{\textbf{receive audio and video from a user and return text}}.
The user's query is embedded in both inputs, without a separate text question,
external captioning, or automatic speech recognition (ASR)~\citep{prabhavalkar2023end}.
Direct audio-visual input reduces external latency and
computation while preserving acoustic and visual cues.
\added{Figure~\ref{fig:deck} shows example user scenes.}
However, research on OmniVChat faces two challenges.
The first challenge is limited data availability: open-source recordings of people using their own devices are scarce,
while noise, camera motion, and device posture create a long tail of input conditions.
The second challenge is evaluating reply quality.
Dialogues contain complex scenes and rare cases. Good replies account for the user's environment, facial expressions,
emotions, and nearby objects, yet allow many valid phrasings. This makes keyword matching and other rule-based methods unreliable.
Few datasets and benchmarks target this native dialogue
~\citep{pourreza2026can,lu2026omniinteract}. Models thus tend to rely on external components such as ASR cascades~\citep{yao2026joyai}.
Although practical, these designs omit subtle prosody and vocal affect and add latency, computation, and potential transcription errors.

Progress in agent systems~\citep{dorri2018multi,yang2025survey} and video
generation~\citep{xing2024survey,ma2025controllable} makes \textbf{generation for comprehension} possible.
This means using synthesized dialogues for training and evaluation.
First, we present \textbf{OmniVChat-Studio}, a controllable and extensible multi-agent data engine
\added{(Figure~\ref{fig:agent})}.
Guided by a configuration for each subcategory, its four agents jointly synthesize single- and multi-turn
native audio-visual dialogues at scale for training and evaluation. Second,
we introduce \textbf{\mbox{OmniVChat-Bench}}, an evaluation benchmark that evaluates omni models'
basic dialogue abilities across five ability categories.
\added{\textbf{Each instance includes a rubric with several tiers, each containing one or more criteria.}}
A large language model checks whether the reply meets each criterion.
Tier 0 checks whether the reply uses the user's language and awards no points; failure makes the rubric score zero.
\added{\textbf{In later tiers, a criterion earns one point only after every criterion in all earlier tiers is met.}}
\added{Figure~\ref{fig:rubric-example} illustrates these scoring rules.} Third, we present \textbf{OmniVChat-RL},
\added[id=rev]{a reinforcement learning (RL)~\citep{kaelbling1996reinforcement} reward design that jointly targets reply correctness, efficiency, and style in OmniVChat.}
It combines rubric-based correctness, scored as in evaluation, with additional efficiency and style rewards.
\textbf{OmniVChat-Bench-Train} is synthesized by OmniVChat-Studio for RL training,
while \textbf{OmniVChat-Bench-Human} is manually recorded to evaluate transfer to real-world dialogues.
Fine-tuning Qwen3-Omni-Instruct~\citep{xu2025qwen3} with OmniVChat-RL on OmniVChat-Bench-Train
improves performance on both benchmarks while producing shorter replies that better follow the prescribed style.
These results validate the reward design. In addition, \added{consistent gains on shared single-turn subcategories
indirectly suggest aligned synthetic and recorded dialogue distributions.}

Section~\ref{sec:method} describes the OmniVChat-Studio framework, including both single- and multi-turn subsystems.
Section~\ref{sec:taxonomy} presents OmniVChat-Bench and its basic dialogue abilities.
Section~\ref{sec:engine} presents OmniVChat-RL and training results. Related work is reviewed in Appendix~\ref{app:related}.

\section{\texorpdfstring{OmniVChat-Studio: Synthesizing Dialogues from Corpus}{OmniVChat-Studio: Synthesizing Dialogues from Corpus}}
\label{sec:method}

\begin{figure}[!t]
\centering
\includegraphics[width=\textwidth]{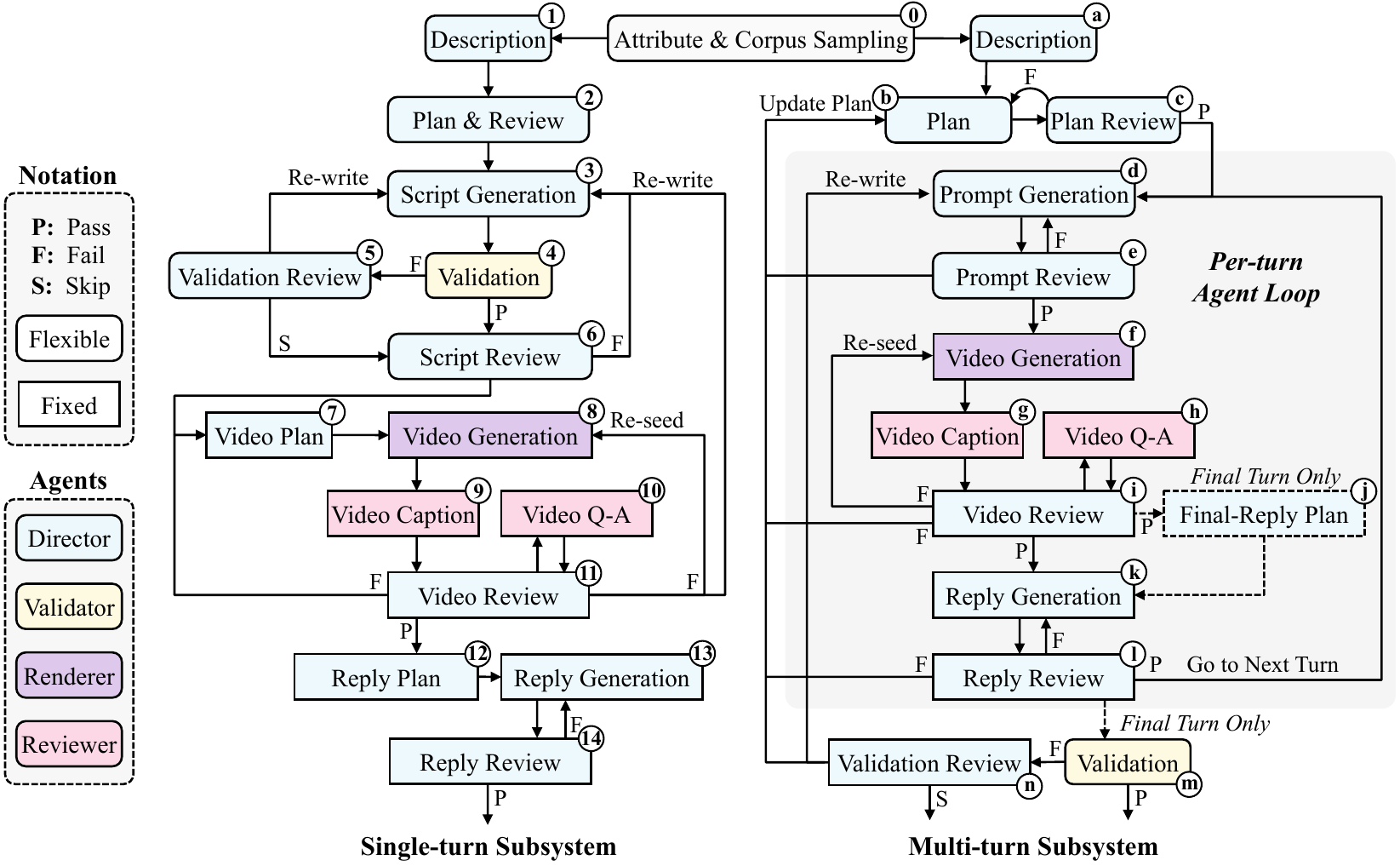}
\caption{\textbf{OmniVChat-Studio} framework with single-turn (left) and multi-turn (right) subsystems.
Each module uses the same color as the agent responsible for it.}
\label{fig:agent}
\end{figure}

We present \textbf{OmniVChat-Studio}, a multi-agent data engine that synthesizes native audio-visual dialogues
from a text corpus, with rendered clips, reference replies, and rubrics for scoring model replies.
Figure~\ref{fig:agent} shows its two subsystems: the single-turn subsystem (left) synthesizes one-turn dialogues,
while the multi-turn subsystem (right) synthesizes dependent turns using accepted earlier clips and dialogue as history.

Both subsystems use four agents, identified by color in Figure~\ref{fig:agent}'s lower-left legend.
The \textbf{Director} handles all text-based input and output tasks.
The \textbf{Renderer} turns an accepted prompt and any media references into a synchronized audio-visual clip.
The \textbf{Reviewer} examines the rendered audio and video, writes a caption and quality report, and answers focused questions when more evidence is needed.
The deterministic \textbf{Validator} checks structured scripts or dialogue records against the configured rules and returns either a pass or a list of rule violations.

A new subcategory, such as multi-speaker dialogue, requires custom prompts only for the \textbf{Flexible}
modules (rounded boxes in Figure~\ref{fig:agent}). The \textbf{Fixed} modules (square-cornered boxes),
overall framework, and module roles remain unchanged across subcategories.

\added{Circled identifiers match Figure~\ref{fig:agent} and the synthesis figures: 0 marks shared sampling,
1--14 the single-turn modules, and a--n the multi-turn modules.}
Appendix~\ref{app:studio} gives the \added{design considerations}, sampling scheme, module checks, and repair procedures.

\subsection{\texorpdfstring{Single-turn Subsystem}{Single-turn Subsystem}}
\label{sec:single}

\added{The single-turn subsystem synthesizes one user video with synchronized audio, a text reference reply, and a rubric.
The evaluated model takes the audio and video as input and returns a text reply.}

\begin{figure}[!t]
\begin{omcsingleexample}
\fontsize{8}{9}\selectfont
\setlength{\omcsinglekeywidth}{0.22\linewidth}
\setlength{\omcsinglevaluewidth}{\dimexpr\linewidth-\omcsinglekeywidth-7pt\relax}
\begin{tabular}[t]{@{}>{\raggedleft\arraybackslash\bfseries\strut}p{\omcsinglekeywidth}@{\hspace{7pt}}>{\raggedright\arraybackslash\strut}p{\omcsinglevaluewidth}@{}}
Audio-Visual Input & \textbf{Audio:} The user says, ``I love how festive everything looks tonight. Hey, the mug on the coffee table --- what pattern does it have on it?''\newline
\textbf{Video:} A white mug on the coffee table has a red reindeer silhouette. \omcsinglerow
Reference Reply & ``That mug has a red reindeer silhouette pattern on it --- very festive! It goes perfectly with the rest of your Christmas setup.'' \omcsinglerow
Tier 0 & Criterion 1: The reply is in English. \omcsinglerow
Tier 1 & Criterion 1: The reply identifies the pattern on the mug as a reindeer. \omcsinglerow
Tier 2 & Criterion 1: The reply says that the pattern is red.\newline
Criterion 2: The reply says that the reindeer is a silhouette. \omcsinglerow
Tier 3 & Criterion 1: The reply uses a natural and relaxed tone that fits the festive setting. \omcsinglerow
Scoring & Tier 0 adds no points, but failing it gives a final score of zero. If Tiers 0 and 1 are met, each met criterion in Tier 2 earns one point. A reply can earn one Tier 2 point even if it misses the other Tier 2 criterion. Tier 3 can add one point only after both Tier 2 criteria are met. \\
\end{tabular}
\end{omcsingleexample}
\caption{\added{An audio-visual dialogue with its text reference reply and tiered scoring rubric.}}
\label{fig:rubric-example}
\label{ex:rubric}
\end{figure}

\begin{figure}[!t]
\begin{omcsingleexample}
\fontsize{8}{9}\selectfont
{\raggedright\itshape\added[id=rev]{This synthesized dialogue tests
\hyperref[def:svmr]{Single-turn Visual Mode Recognition}
(DSLP-VDMC-SVMR, a subcategory in OmniVChat-Bench),
which concerns user-relative left and right under the current camera view.}\par}
\setlength{\omcsinglekeywidth}{0.22\linewidth}
\setlength{\omcsinglevaluewidth}{\dimexpr\linewidth-\omcsinglekeywidth-7pt\relax}
\begin{tabular}[t]{@{}>{\raggedleft\arraybackslash\bfseries\strut}p{\omcsinglekeywidth}@{\hspace{7pt}}>{\raggedright\arraybackslash\strut}p{\omcsinglevaluewidth}@{}}
\omcsinglekey{Attribute \& Corpus\break Sampling\omcstageid{S1}} & Configuration (base probabilities): Chinese/English, 40\%/60\%; one user, 100\%; 22 scenario domains, $1/22$ each; 14 camera setups, $1/14$ each; audio and visual interference (none/low/medium/high), 25\% each. User-question types: side identification (asking which side a target is on) and two-sided comparison (asking about differences between objects on the two sides), $1/2$ each. Attribute sampling selects an English-speaking solo user, an office, a walking selfie, medium audio noise, high visual interference, and a two-sided comparison. Corpus sampling supplies three passages, including one on health. \omcsinglerow
\omcsinglekey{Description\omcstageid{S2}} & The Director uses the health passage to outline a corridor dialogue with two meeting-room doors and a question about the user's left door. \omcsinglerow
\omcsinglekey{Plan \& Review\omcstageid{S3}} & The Director refines the corridor scene and reviews the left-right mapping, recommending clear door views and shorter speech. \omcsinglerow
\omcsinglekey{Script Generation\omcstageid{S4}} & The Director turns the plan into an 11.3-second script with timed sections for walking, speech, and a silent wait. The user asks, ``There's a meeting room on each side of me --- is the one on my left open or closed?'' \omcsinglerow
\omcsinglekey{Validation\omcstageid{S5}} & The script passes the configured checks, \added[id=rev]{such as} format and timing. \omcsinglerow
\omcsinglekey{Validation Review\omcstageid{S6}} & \stage{Validation\omcstageid{S5}} reports no violation, so this module is skipped. \omcsinglerow
\omcsinglekey{Script Review\omcstageid{S7}} & The Director accepts the script as a test of user-relative left and right. \omcsinglerow
\omcsinglekey{Video Plan\omcstageid{S8}} & The Director selects shared rendering instructions for the camera, a continuous shot, realism, and the absence of subtitles. \omcsinglerow
\omcsinglekey{Video Generation\omcstageid{S9}} & \added{The Renderer renders an 11.1-second video with synchronized audio from the prompt.} \omcsinglerow
\omcsinglekey{Video Caption\omcstageid{S10}} & The Reviewer describes the user, corridor, and speech, and checks seven aspects of video quality. The report finds no clear defect. \omcsinglerow
\omcsinglekey{Video Review\omcstageid{S11}} & The caption leaves the door states unclear, so the Director requests focused \stage{Video Q-A\omcstageid{S12}}. \omcsinglerow
\omcsinglekey{Video Q-A\omcstageid{S12}} & The Reviewer confirms that the door on image right is closed and the door on image left is open. The Director then accepts the clip in \stage{Video Review\omcstageid{S11}}. \omcsinglerow
\omcsinglekey{Reply Plan\omcstageid{S13}} & The Director plans a brief English reply using the door states confirmed by \stage{Video Q-A\omcstageid{S12}}. \omcsinglerow
\omcsinglekey{Reply Generation\omcstageid{S14}} & ``The one on your left is closed. The door on your right is open.'' The Director also creates a tiered rubric covering language, the left door's state, brevity, and the right door's state. \omcsinglerow
\omcsinglekey{Reply Review\omcstageid{S15}} & The Director checks the reply and rubric against the video evidence and accepts them. The clip, reply, and rubric form the completed instance. \\
\end{tabular}
\end{omcsingleexample}
\caption{\added{Synthesis of one single-turn dialogue, with each module's decisions and outputs.}}
\label{fig:single-example}
\label{ex:studio-single}
\end{figure}

\stage{Attribute \& Corpus Sampling\omcstageid{S1}} samples attribute values using the subcategory
configuration's target distributions and draws three passages by default from a large text corpus.
Appendix~\ref{app:sampling} gives the sampling formula, its full notation, and details on coverage and cost.
\stage{Description\omcstageid{S2}} develops a rough storyline from these passages and attributes.
\added{Figure~\ref{fig:single-example} follows one real dialogue synthesis run.}

\stage{Plan \& Review\omcstageid{S3}} refines and reviews the storyline and audio-visual details using
the scene description and sampled attributes.
\stage{Script Generation\omcstageid{S4}} turns the reviewed plan into a structured script with scene fields
and a timed sequence of speech, sounds, music, and on-screen text.

\stage{Validation\omcstageid{S5}} checks the script against configured rules, \added[id=rev]{such as} format and timing,
and reports a pass or a list of violations. When violations are reported,
\stage{Validation Review\omcstageid{S6}} examines them in context, dismissing false reports or requesting corrections.
\stage{Script Review\omcstageid{S7}} then checks whether the script meets the subcategory requirements
and accepts it, requests a rewrite, or discards the run.

In \stage{Video Plan\omcstageid{S8}}, the Director combines the accepted script's scene and events with
compatible instructions for the camera, realism, and other shared requirements.
\stage{Video Generation\omcstageid{S9}} renders a synchronized audio-visual clip from this prompt.

\stage{Video Caption\omcstageid{S10}} writes a caption and a seven-part quality report based on the synthesized audio-visual clip.
The report covers cuts, realism, delivery speed, body structure, appearance, artifacts, and on-screen text.
\stage{Video Review\omcstageid{S11}} compares the rendered content with the accepted script using
the caption, quality report, and any Q-A evidence. It accepts the clip, requests a re-seed, rewrite, or focused Q-A check,
or discards the run. When requested, \stage{Video Q-A\omcstageid{S12}} answers its questions from the rendered audio and video.
Missing or changed scripted background details are acceptable when they do not affect the main storyline or intended dialogue.

\stage{Reply Plan\omcstageid{S13}} plans a grounded reply, omitting scripted details unsupported by
the caption or Q-A evidence. \stage{Reply Generation\omcstageid{S14}} uses this plan and the accepted evidence
to write a reference reply and tiered rubric in the user's language.
\stage{Reply Review\omcstageid{S15}} checks language, grounding, tier order, and spoken style against
the accepted evidence and subcategory requirements, then accepts them or requests one refinement.
The completed instance contains the reference reply, rubric, and a 1080P clip with 44.1\,kHz stereo audio.

\subsection{\texorpdfstring{Multi-turn Subsystem}{Multi-turn Subsystem}}
\label{sec:multi}

\added{The multi-turn subsystem synthesizes a sequence of user videos with synchronized audio and corresponding text reference replies
(Figure~\ref{fig:agent}, right).
Earlier audio-visual clips and text replies form the evaluation history.
The model receives this history and the final user clip without its reply, then returns a text reply for the final turn.
The final reference reply and rubric serve as evaluation annotations.
Appendix~\ref{app:pipe-multi} illustrates the synthesis process for one multi-turn dialogue.}

The multi-turn subsystem separates the single-turn \stage{Plan \& Review\omcstageid{S3}} into
\stage{Plan\omcstageid{M3}} and \stage{Plan Review\omcstageid{M4}}.
\stage{Plan\omcstageid{M3}} expands the storyline into a joint plan for all turns, selecting earlier clips,
audio, or final frames for later turns within a reference budget.
\stage{Plan Review\omcstageid{M4}} checks all turns and media links in a separate Director call, then accepts the plan, requests revision, or discards the run.

Per-turn \stage{Prompt Generation\omcstageid{M5}} replaces single-turn
\stage{Script Generation\omcstageid{S4}} and \stage{Video Plan\omcstageid{S8}}.
It writes each rendering prompt from the reviewed joint plan and shared instructions,
using media identifiers as placeholders for earlier clips, audio, or final frames.
\stage{Prompt Review\omcstageid{M6}} checks the prompt and media links before rendering,
then accepts it or requests a local rewrite or plan update.

\section{\texorpdfstring{OmniVChat-Bench: Evaluating Basic Dialogue Abilities}{OmniVChat-Bench: Evaluating Basic Dialogue Abilities}}
\label{sec:taxonomy}

\added{We introduce \textbf{OmniVChat-Bench}, an evaluation benchmark that evaluates omni models'
basic dialogue abilities across five ability categories through their text replies to audio-visual dialogue inputs.
For each instance, a model takes the user's audio and video as input and returns a text reply.
A large language model then scores the reply using the instance's tiered rubric.}
Its 2{,}550 single-turn and 250 multi-turn instances cover 17 subcategories and 22 scenario domains, such as Parenting, Travel and Sightseeing, and Medicine and Health.
OmniVChat-Studio synthesizes all 2{,}800 instances with reference replies and tiered rubrics.
They include 1{,}766 English dialogues (63.1\%) and 1{,}034 Chinese dialogues (36.9\%).
A scenario domain describes what a dialogue is about, while a subcategory specifies the basic dialogue ability under evaluation.
\added{Human inspectors verify each clip's intended dialogue using the procedure in Appendix~\ref{app:humanqc}.}
Figure~\ref{fig:composition} shows instance counts by ability category and subcategory.
Appendix~\ref{app:competencies} defines the subcategories, and Appendix~\ref{app:instances} gives synthetic examples with reference replies and rubrics.

\begin{figure}[!htbp]
\centering
\includegraphics[width=\textwidth]{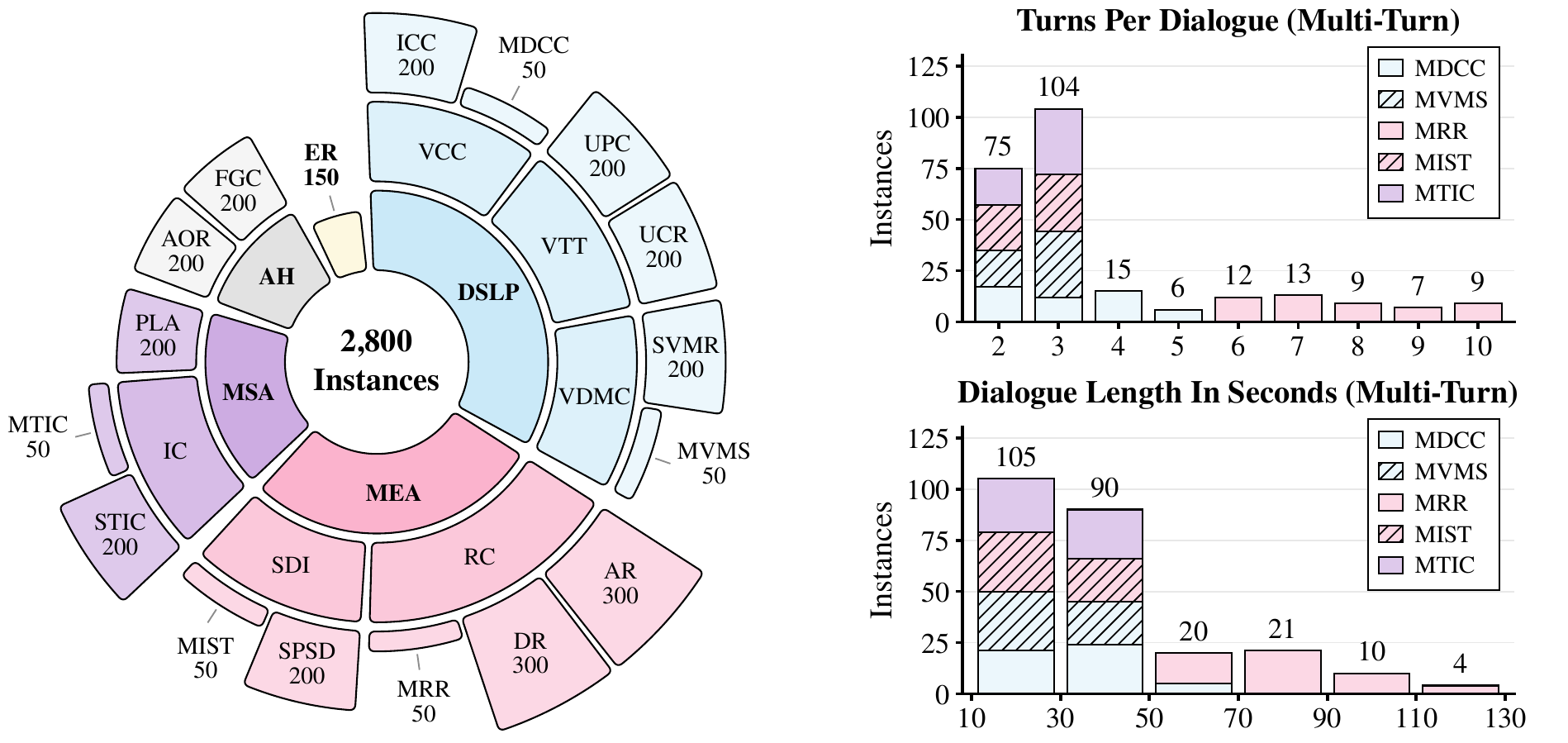}
\caption{\added{\textbf{Overview of OmniVChat-Bench.}}
\textbf{Left:} The five ability categories and their 17 subcategories. Bar length shows the instance count
from the base ring. Angular span shows the number of subcategories in each band. \textbf{Right:} Turn and
duration distributions for 250 multi-turn instances, with counts stacked by subcategory.}
\label{fig:composition}
\end{figure}

For evaluation, \texttt{qwen3.7-max}~\citep{qwen2026qwen37} judges whether each criterion
in the instance's rubric is met \added{(Figure~\ref{fig:rubric-example})}.
\added{Appendix~\ref{app:judges} examines agreement across six graders.}
Tier~0 checks whether the reply uses the user's language and awards no points; failure makes the rubric score zero.
In later tiers, each met criterion earns one point only if all earlier tiers are complete.
An incomplete tier retains its earned points but blocks later tiers.
For each instance, let $\mathcal{K}_t$ be the set of criteria in tier $t=0,\ldots,T$, where $T$ is the final tier.
\added{Let $\mathcal{H}(y)$ be the set of criteria that the grader finds met by reply $y$.}
The single-instance scoring function is:
\begin{equation}
\color{omcrevision}
r(y)\;=\;
\frac{\displaystyle\sum_{t=1}^{T}\left\{|\mathcal{K}_t\cap\mathcal{H}(y)|
\prod_{s=0}^{t-1}\mathbf{1}[\mathcal{K}_s\subseteq\mathcal{H}(y)]\right\}}
{\displaystyle\sum_{t=1}^{T}|\mathcal{K}_t|}\,.
\label{eq:rubric}
\end{equation}
Here, $|B|$ counts the elements of a set $B$, and $\mathbf{1}[\cdot]$ is 1 when its condition is true and 0 otherwise.
The intersection counts met criteria; the product requires earlier tiers to be complete.
The denominator includes all criteria outside Tier~0, even those blocked by an earlier tier, so $r(y)\in[0,1]$.

For multi-turn evaluation, all models receive the same earlier clips and reference replies; only the final reply is scored.
For an evaluation set $\mathcal{E}$, let $y_i$ be the reply scored for instance $i$ and $\mathcal{E}_c$ contain the instances in subcategory $c$.
Pooled Mean gives each instance equal weight, while Subcategory Mean first averages within each subcategory and then gives all $C$ subcategories equal weight.
OmniVChat-Bench reports Subcategory Mean as its overall Mean score, with $C=17$:
\begin{equation}
\operatorname{PooledMean}(\mathcal{E})=
\frac{1}{|\mathcal{E}|}\sum_{i\in\mathcal{E}}r(y_i),
\quad
\operatorname{SubcategoryMean}(\mathcal{E})=
\frac{1}{C}\sum_{c=1}^{C}\operatorname{PooledMean}(\mathcal{E}_c).
\label{eq:benchmark-means}
\end{equation}

Section~\ref{sec:rl-transfer} and Table~\ref{tab:models}
report transfer and model comparisons.

\label{sec:families}

\added{\textbf{Dialogue-State \& Link Perception} (DSLP) evaluates dialogue-state understanding.
It covers connection confirmation at the start of and during a dialogue, waiting for unfinished requests and answering completed ones,
and interpreting user-relative left and right under the current camera view and after camera changes.
\textbf{Multimodal Entity Alignment} (MEA) tests whether speech is linked to the correct objects and speakers.
This includes resolving pronouns, gestures, and spoken descriptions using the current view,
retrieving information about objects from earlier turns when they are no longer visible,
and identifying the active user amid competing voices and across user changes.
\textbf{Model Self-Awareness} (MSA) evaluates correct statements of model identity in single- and multi-turn dialogues,
including under misleading user claims. It also tests recognition of physical limits and the provision of feasible alternatives when appropriate.
\textbf{Anti-Hallucination} (AH) evaluates grounding in available audio and visual evidence.
It covers stating when a requested object is absent and correcting user claims that contradict visible evidence.
\textbf{Emotion Recognition} (ER) tests whether models combine facial and vocal cues to recognize emotion
and tailor replies to both the user's emotion and request.}
\added[id=rev]{The recorded probe contains only single-turn dialogues and does not test live interruption;
Appendix~\ref{app:human} describes its recording and annotation procedures.}

\section{\texorpdfstring{OmniVChat-RL: Rewards for Better Dialogue}{OmniVChat-RL: Rewards for Better Dialogue}}
\label{sec:engine}

\label{sec:engine-two}
We present \textbf{OmniVChat-RL}, an RL reward design that jointly targets reply correctness, efficiency, and style in OmniVChat.
\added{Experiments show that training on synthetic dialogues improves scores on both OmniVChat-Bench and
the human-recorded OmniVChat-Bench-Human (\S\ref{sec:rl-transfer}).}
\label{sec:engine-bet}

\subsection{\texorpdfstring{Training Setup}{Training Setup}}
\label{sec:rl-setup}

\added{The Thinker of Qwen3-Omni-30B-A3B-Instruct is fine-tuned on OmniVChat-Bench-Train using
group sequence policy optimization (GSPO)~\citep{zheng2025gspo} and the OmniVChat-RL reward.}
\added[id=rev]{The Thinker includes video and audio encoders and a text decoder.
The resulting model is evaluated on OmniVChat-Bench and OmniVChat-Bench-Human.}
\textbf{OmniVChat-Bench-Train} contains 5{,}600 training dialogues and a separate 560-dialogue \added[id=rev]{development set}
for checkpoint selection. All dialogues and rubrics are synthesized by OmniVChat-Studio.
\added{Training uses dialogues whose encoded inputs fit the prompt-length limit.}
\textbf{OmniVChat-Bench-Human} contains 360 human-recorded single-turn dialogues used only to evaluate real-world transfer
\added[id=rev]{(Appendix~\ref{app:human} describes the recording and annotation procedures).}

For each training input $x$, the sampling policy $\pi_{\theta_{\mathrm{old}}}$ draws $N$ replies $\mathcal{G}(x)=\{y_1,\ldots,y_N\}$.
Following the group baseline in group relative policy optimization (GRPO)~\citep{shao2024deepseekmath}, the advantage $A_i=R(y_i)-\mu_{\mathcal{G}}$ subtracts the group's mean reward $\mu_{\mathcal{G}}$ from each reply's total reward $R(y_i)$, without standard-deviation normalization.
The current policy $\pi_\theta$ maximizes:
\begin{equation}
\label{eq:gspo}
\mathcal{J}(\theta)\;=\;\mathbb{E}
\Big[\frac1N\sum_{i=1}^{N}
\min\big(q_i(\theta)\,A_i,\ \operatorname{clip}\big(q_i(\theta),\,1-\varepsilon_{\mathrm{lo}},\,
1+\varepsilon_{\mathrm{hi}}\big)\,A_i\big)\Big].
\end{equation}
The expectation averages over training inputs and sampled replies.
The full-reply probability ratio $q_i(\theta)=[\pi_\theta(y_i\mid x)/\pi_{\theta_{\mathrm{old}}}(y_i\mid x)]^{1/|y_i|}$ is normalized by the reply's token count $|y_i|$ to reduce length dependence.
Clipping bounds the ratio to $[1-\varepsilon_{\mathrm{lo}},1+\varepsilon_{\mathrm{hi}}]$, and the minimum limits the incentive for large policy changes.
Training uses low-rank adaptation (LoRA)~\citep{hu2021lora} with rank-64 adapters for 1{,}000 iterations.
Each iteration uses 32 inputs with four sampled replies per input, and every twentieth iteration saves a checkpoint.
Appendix~\ref{app:rl-config} gives the full training settings.

\subsection{\texorpdfstring{Rewards and Evaluation Metrics}{Rewards and Evaluation Metrics}}
\label{sec:reward-metrics}

Three quality signals connect training with evaluation: rubric correctness, reply efficiency, and style.
OmniVChat-RL uses their reward forms during training, while evaluation reports the corresponding metrics.
The total training reward for reply $y$ is:
\begin{equation}
R(y)\;=\;r(y)\;+\;\lambda_{\mathrm{fmt}}\,f(y)\;+\;\lambda_{\mathrm{eff}}\,e(y)\;+\;
\lambda_{\mathrm{sty}}\,s(y).
\label{eq:reward}
\end{equation}
The functions $r(y)$, $f(y)$, $e(y)$, and $s(y)$ score correctness, format, efficiency, and style.
The correctness reward directly uses the single-instance score $r(y)$ in Equation~\ref{eq:rubric}.
Each training dialogue supplies a rubric with criteria for an open-ended reply~\citep{gunjal2026rubrics}.
The training-only format term $f(y)$ equals one for a nonempty reply with no thinking tags and zero otherwise.

\added{Different audio-visual queries naturally require different reply lengths.
The efficiency reward therefore compares replies to the same input and rescales their scores to $[0,1]$,
controlling the reward scale across queries without imposing a fixed length target.}

\added[id=rev]{The length $w(y)$ counts each Chinese character or continuous group of English letters or digits as one unit.
This differs from the model's token count.
For the $N$ replies to input $x$, let $w_i=\max\{w(y_i),1\}$, with the floor preventing division by zero.
First, length is normalized by its group mean $\bar w_x$; the rubric score is then divided by this relative length:}
\begin{equation}
\color{omcrevision}
\bar w_x=\frac{1}{N}\sum_{j=1}^{N}w_j,
\qquad \ell_i=\frac{w_i}{\bar w_x},
\qquad z_i=\frac{r(y_i)}{\ell_i}.
\label{eq:rho}
\end{equation}
\added{With $z_{\min}=\min_j z_j$ and $z_{\max}=\max_j z_j$ taken within this group, the efficiency reward is:}
\begin{equation}
\color{omcrevision}
e(y_i)=
\begin{cases}
\dfrac{z_i-z_{\min}}{z_{\max}-z_{\min}}, & z_{\max}>z_{\min},\\[1.4ex]
\dfrac{1}{2}, & z_{\max}=z_{\min}.
\end{cases}
\label{eq:eff}
\end{equation}
\added{Since $z_i=\bar w_x\rho(y_i)$ for raw efficiency $\rho(y_i)=r(y_i)/w_i$, the common positive factor $\bar w_x$ cancels.
This is equivalent to directly normalizing raw efficiencies.
At a fixed positive rubric score, shorter positive lengths give higher raw efficiency; zero credit gives zero raw efficiency.
Equal scores receive $\frac{1}{2}$, which cancels in the group-centered advantage $A_i$ (\S\ref{sec:rl-setup}).
Evaluation reports Reply Efficiency as $\mathrm{RE}(\mathcal E)=1000\sum_{y\in\mathcal E}r(y)/\sum_{y\in\mathcal E}w(y)$,
the rubric credit earned per thousand counted words.}

The training \hyperref[prompt:policy-system]{system prompt} specifies natural speech, concise wording, and a clear distinction between roleplay and real facts.
These rules and a grammar requirement form seven criteria in the \hyperref[prompt:style-grader]{style-grading prompt}.
The style reward $s(y)$ is one if \texttt{qwen3.7-max} confirms all seven criteria and zero otherwise.
\added{Evaluation uses the same grader and criteria. Style is the mean of $s(y)$ over replies with valid inference and grading results.}
Appendix~\ref{app:graderprompts} gives the policy and style-grading prompts; Appendix~\ref{app:rl-config} gives grader settings.
The format, efficiency, and style weights are $\lambda_{\mathrm{fmt}}=0.5$, $\lambda_{\mathrm{eff}}=0.1$, and $\lambda_{\mathrm{sty}}=0.5$.
Correctness has weight one.
Each component lies in $[0,1]$, so the total reward $R(y)$ lies in the interval $[0,2.1]$.

\begin{figure}[!b]
\centering
\includegraphics[width=\textwidth,trim={0 -14pt 0 0}]{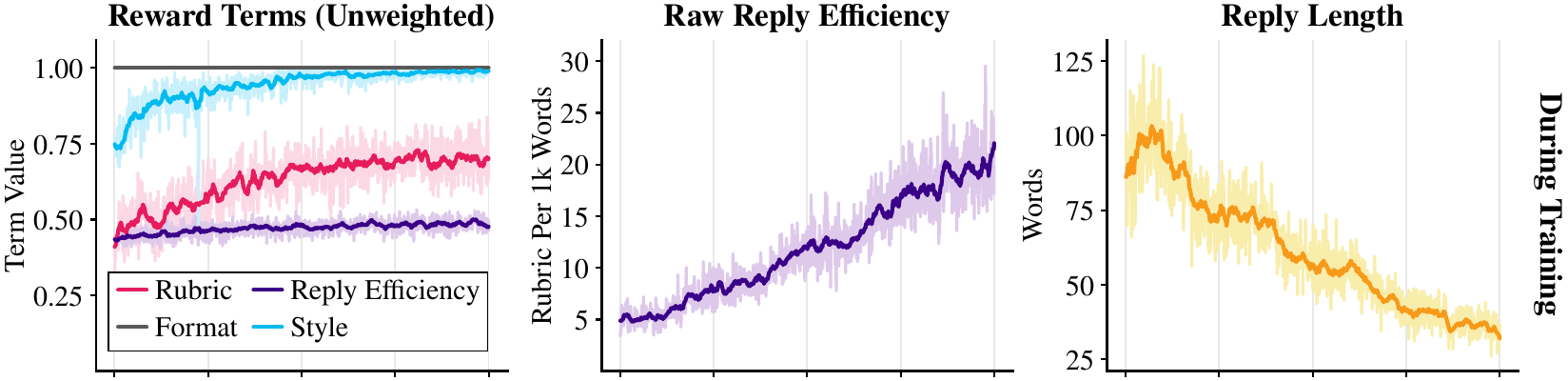}\\
\includegraphics[width=\textwidth]{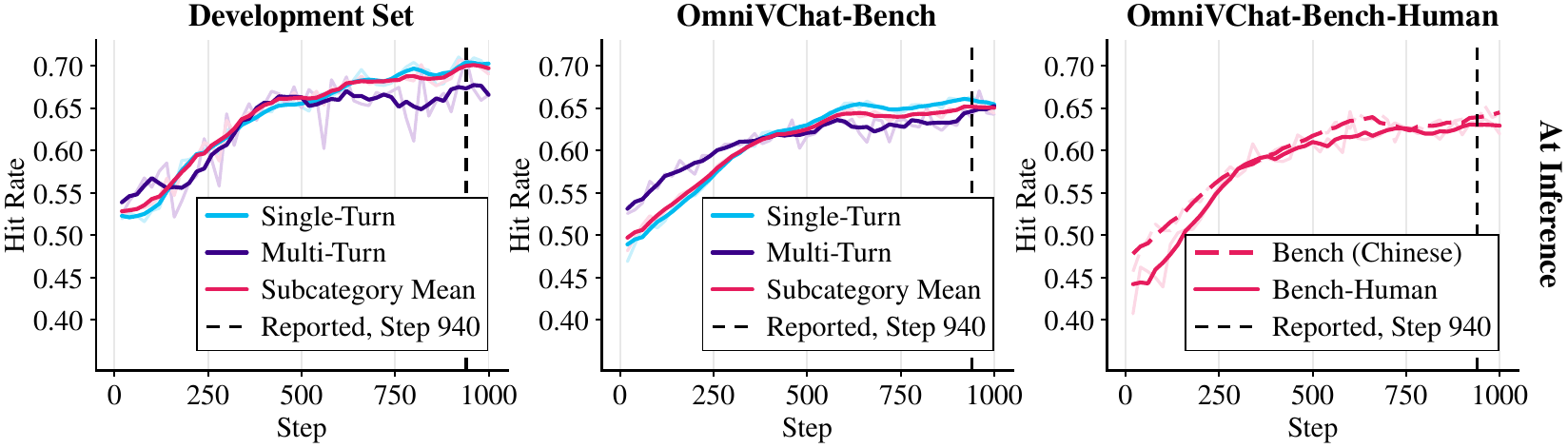}
\caption{\added{\textbf{OmniVChat-RL over 1{,}000 iterations.} \textbf{Top:} unweighted reward terms
(Equation~\ref{eq:reward}), reply efficiency, and mean length. \textbf{Bottom:} rubric scores
(Equation~\ref{eq:rubric}) every twentieth step on the development set, OmniVChat-Bench, and OmniVChat-Bench-Human.
Right includes a Chinese-gated synthetic control. The dashed line marks step 940.
\mbox{Scores across corpora are not directly comparable.}}}

\label{fig:rl}
\label{fig:rl-heldout}
\label{fig:rl-training}
\end{figure}

\subsection{Results and Discussion}
\label{sec:bench-models}
\label{sec:rl-results}
\label{sec:rl-training}
\label{sec:rl-transfer}

\added[id=rev]{The checkpoint is selected using the highest Subcategory Mean on the development set of 560 synthesized dialogues.}
\added{\textbf{OmniVChat-Bench} measures gains on synthetic dialogues with no training-video overlap;}
\textbf{OmniVChat-Bench-Human} tests transfer on recorded single-turn dialogues.
\added[id=rev]{Appendix~\ref{app:human} describes the recording and annotation of OmniVChat-Bench-Human dialogues.
Appendix~\ref{app:humaninstances} provides recorded examples with reference replies and rubrics.}

\added[id=rev]{Figure~\ref{fig:rl} shows 1{,}000 training iterations and scores on the development set and both benchmarks.}
\added{Top-left averages unweighted reward terms over sampled replies: rubric reward rises from 0.308 to 0.629 and style from 0.750 to 0.992.}
Format stays at 1.000 and gives no within-group learning signal, while the normalized efficiency term changes little.
\added{Top-middle shows raw reply efficiency rising from 3.37 to 17.18 rubric credit per thousand counted words.}
\added{The top-right panel shows mean reply length falling from 91.5 to 36.6 words, with a brief peak near 127 at step 50.}

\added[id=rev]{The bottom row reports Subcategory Means (Equation~\ref{eq:benchmark-means}).
The bottom-left panel shows the development set score rising from 0.507 to 0.707, with the dashed line marking its peak at step 940.}
\added{Bottom-middle shows OmniVChat-Bench scores rising from 0.465 to 0.652 at this checkpoint.}
Both panels also show single-turn and multi-turn means.
\added{Bottom-right compares OmniVChat-Bench-Human with 1{,}034 Chinese OmniVChat-Bench instances to match the recordings' language.}
The recorded score rises from 0.402 to 0.632, while the Chinese synthetic score rises from 0.437 to 0.634.
Appendix~\ref{app:rl-leaves} shows the per-subcategory learning curves.
\added{These gains support the training design and synthetic data for training and evaluation in real-world dialogues.}
Their agreement indirectly suggests aligned distributions across shared subcategories.
\added[id=rev]{Appendix~\ref{app:comove} reports correlations for all three pairs of evaluation sets.}

Table~\ref{tab:models} compares eleven released systems, OmniVChat-RL, and two reward ablations under one system prompt.
Mean reports Subcategory Mean from Equation~\ref{eq:benchmark-means}, and Human averages scores over the recorded dialogues.
All columns except Human use OmniVChat-Bench.
RE and Style follow Section~\ref{sec:reward-metrics}.
\added{Empty replies receive zero Style; mean reply length is computed over nonempty replies.}

\begin{table}[t]
\centering
\omctablefont
\setlength{\tabcolsep}{4.5pt}
\begin{tabular*}{\textwidth}{@{\extracolsep{\fill}}lrrrrrrr>{\columncolor[HTML]{ECF7FC}}r>{\columncolor[HTML]{FCD8E5}}r>{\columncolor[HTML]{DEC9EB}}r>{\columncolor[HTML]{FDF8E0}}r@{}}
\toprule
\textbf{\textit{Closed-Source Models}} & AH & ER & MSA & MEA & DSLP & Single & Multi & Mean & Human & RE & Style \\
\midrule
Gemini-3.5-Flash & 0.600 & \uwave{0.778} & 0.649 & \uwave{0.728} & \uline{0.623} & \uwave{0.664} & \textbf{0.682} & \uline{0.667} & 0.509 & 10.59 & 0.755 \\
Gemini-3.6-Flash & 0.570 & 0.658 & 0.671 & 0.716 & \uwave{0.620} & 0.658 & 0.610 & 0.641 & 0.562 & \uwave{15.74} & 0.740 \\
Gemini-3.7-Flash & 0.604 & 0.623 & 0.663 & 0.708 & \uwave{0.620} & 0.658 & 0.607 & 0.640 & \uwave{0.575} & \uline{16.96} & 0.631 \\
Gemini-3.1-Pro & 0.637 & \uline{0.787} & 0.341 & 0.666 & \textbf{0.629} & 0.602 & 0.623 & 0.607 & 0.537 & 11.66 & \uwave{0.871} \\
Qwen3.5-Omni-Plus & 0.361 & 0.662 & 0.413 & 0.630 & 0.513 & 0.514 & 0.590 & 0.533 & 0.482 & 9.45 & 0.869 \\
Doubao Seed 2.0 Lite & 0.330 & 0.550 & 0.189 & 0.602 & 0.481 & 0.434 & \uwave{0.671} & 0.494 & 0.274 & 6.10 & 0.817 \\
\midrule
\multicolumn{12}{@{}l@{}}{\textbf{\textit{Open-Source Models}}} \\
\midrule
Qwen3-Omni-Instruct & 0.186 & 0.694 & 0.388 & 0.573 & 0.451 & 0.450 & 0.508 & 0.465 & 0.402 & 5.75 & 0.710 \\
JoyAI-VL-Interaction & 0.194 & 0.611 & 0.312 & 0.499 & 0.448 & 0.406 & 0.503 & 0.431 & 0.419 & 8.43 & 0.751 \\
Qwen3-Omni-Thinking & 0.300 & 0.517 & 0.234 & 0.505 & 0.346 & 0.370 & 0.498 & 0.405 & 0.105 & 6.32 & 0.754 \\
Qwen2.5-Omni-7B & 0.103 & 0.567 & 0.070 & 0.433 & 0.373 & 0.309 & 0.380 & 0.328 & 0.255 & 6.64 & 0.611 \\
MiniCPM-o-4.5 & 0.200 & 0.366 & 0.037 & 0.446 & 0.350 & 0.310 & 0.310 & 0.304 & 0.298 & 6.50 & 0.641 \\
\midrule
\multicolumn{12}{@{}l@{}}{\textbf{\textit{Our Models}}} \\
\midrule
OmniVChat-RL (ours) & \uwave{0.667} & 0.651 & \uwave{0.734} & 0.713 & 0.573 & \uwave{0.664} & 0.638 & 0.652 & \uline{0.632} & \textbf{18.38} & \textbf{0.992} \\
No Efficiency Term & \textbf{0.733} & \textbf{0.813} & \textbf{0.810} & \textbf{0.746} & 0.586 & \textbf{0.710} & \uline{0.677} & \textbf{0.697} & \textbf{0.691} & 7.12 & \uline{0.987} \\
No Style Term & \uline{0.672} & 0.719 & \uline{0.758} & \uline{0.730} & 0.560 & \uline{0.674} & 0.645 & \uwave{0.661} & \uline{0.632} & 13.70 & 0.788 \\
\bottomrule
\end{tabular*}

\caption{\added[id=rev]{Model comparison under one system prompt.} \added{\textbf{Bold}, \uline{straight}, and \uwave{wavy}
mark the top three ranks, excluding ablations. AH--DSLP, Single, and Multi give category, single-turn, and multi-turn means.
Mean equally weights 17 subcategories. Human uses recorded dialogues; other columns use OmniVChat-Bench.
Mean and Human aggregate $r(y)$, while Style averages $s(y)$ from Equation~\ref{eq:reward}.
RE measures rubric credit per thousand counted words, the evaluation counterpart of $e(y)$.}}
\label{tab:models}
\end{table}

The closed-source systems tested include Gemini-3.5-Flash~\citep{google2026gemini35},
Gemini-3.6-Flash~\citep{google2026gemini36}, and Gemini-3.7-Flash~\citep{google2026gemini}, Gemini-3.1-Pro
~\citep{google2026gemini31}, \added[id=rev]{\mbox{Qwen3.5-Omni-Plus}}~\citep{team2026qwen3}, and Doubao Seed 2.0 Lite
~\citep{seed2026seed2}. \mbox{The open-source} systems tested include Qwen3-Omni-Instruct and
Qwen3-Omni-Thinking~\citep{xu2025qwen3}, Qwen2.5-Omni-7B
~\citep{xu2025qwen25}, JoyAI-VL-Interaction~\citep{yao2026joyai}, and
MiniCPM-o-4.5~\citep{cui2026minicpm}. All systems receive audio and video.
For the ASR cascade, JoyAI-VL-Interaction uses Qwen3-ASR~\citep{shi2026qwen3}.

OmniVChat-RL improves rubric scores over the base model while raising RE from 5.75 to 18.38 and Style from 0.710 to 0.992.
Among released systems, Gemini-3.5-Flash leads Mean at 0.667.
Gemini-3.7-Flash leads Human at 0.575 and RE at 16.96, while Gemini-3.1-Pro leads Style at 0.871.
The different leaders show that synthetic rank does not determine recorded rank.

\added[id=rev]{The bottom three rows of Table~\ref{tab:models} compare OmniVChat-RL (ours) with No Efficiency Term and No Style Term.
Each ablation removes one auxiliary reward while retaining correctness, format, and the other auxiliary term.
Removing efficiency raises Mean from 0.652 to 0.697 and Human from 0.632 to 0.691.
However, RE falls from 18.38 to 7.12.
Removing style raises Mean to 0.661 and keeps Human at 0.632, but lowers Style from 0.992 to 0.788 and RE to 13.70.
The full reward achieves the highest RE and Style while still ranking joint second on the human-recorded benchmark at 0.632.
It therefore offers the best balance among the three variants for the joint goal of correctness, efficiency, and style.
Appendix~\ref{app:ablation} reports the full curves and selected checkpoints for both ablations.}

\section{Conclusion}
\label{sec:conclusion}

\added[id=rev]{We present \textbf{OmniVChat-Studio}, a controllable and extensible multi-agent data engine that synthesizes single- and multi-turn audio-visual dialogues with text reference replies and tiered rubrics.
We introduce \textbf{OmniVChat-Bench} to evaluate basic dialogue abilities across five ability categories and 17 subcategories,
and \textbf{OmniVChat-RL} to reward correctness, efficiency, and style.
Training Qwen3-Omni-Instruct on OmniVChat-Bench-Train raises OmniVChat-Bench from 0.465 to 0.652
and the recorded OmniVChat-Bench-Human from 0.402 to 0.632 without training on its recordings.
On OmniVChat-Bench, RE rises from 5.75 to 18.38, and Style rises from 0.710 to 0.992.
These results validate the reward design. In addition, consistent gains on shared single-turn subcategories
indirectly suggest aligned synthetic and recorded dialogue distributions.}\label{mainend}

\section{Ethics Statement}
All participants who appear in OmniVChat-Bench-Human sign an informed-consent agreement
authorizing the use of their recorded audio and video for research purposes.

\section{Reproducibility Statement}
The main paper describes the methods and evaluation protocols.
Appendix~\ref{app:studio} gives the sampling and repair procedures,
Appendix~\ref{app:basics} gives benchmark construction and judge-agreement analyses, and
Appendix~\ref{app:training} gives training settings, grading prompts, and supplementary results.
Artifact availability is stated in the Code and Data Release section.

\section{AI Use Statement}
A large language model is used to aid and polish the writing of this manuscript. All
scientific claims, experimental designs, and results are the responsibility of the authors.

\section{Code and Data Release}
\textbf{OmniVChat-Bench} and \textbf{OmniVChat-RL} training code will be publicly released for research.

\ificlrfinal
\section{Acknowledgements}
\label{sec:acknowledgements}
This work was supported by Alibaba Research Intern Program.
We would like to thank the Qwen Team at Alibaba Token Hub (ATH), Alibaba Group, for providing the computational resources and foundation models (Qwen) used in this research.
\fi

\bibliography{refs}
\bibliographystyle{iclr2027_citeorder}

\appendix
\section{Appendix Contents}
\label{app:contents}
\addtocontents{toc}{\protect\etocsetlevel{section}{1}}
\addtocontents{toc}{\protect\etocsetlevel{subsection}{2}}
\addtocontents{toc}{\protect\etocsetlevel{subsubsection}{3}}
\etocsetnexttocdepth{subsubsection}
\etocsetstyle{section}
  {}
  {}
  {\noindent\textbf{\etocnumber\quad\etocname}\dotfill\,\etocpage\par}
  {}
\etocsetstyle{subsection}
  {}
  {}
  {\noindent\hspace{1.5em}\etocnumber\quad\etocname\dotfill\,\etocpage\par}
  {}
\etocsetstyle{subsubsection}
  {}
  {}
  {\noindent\hspace{3.0em}\etocnumber\quad\etocname\dotfill\,\etocpage\par}
  {}
\etocsettocstyle{}{}
\localtableofcontents

\section{Related Work}
\label{app:related}

This section reviews five research areas related to OmniVChat. They cover omni models, benchmarks,
joint audio-video generation, agent systems, and RL.

\subsection{Omni models}

Omni models combine several input and output modalities in one system.
GPT-4o~\citep{openai2024gpt4o} accepts text, audio, images, and video. Its average audio response latency is 320 milliseconds. The report describes native processing and an earlier cascade for speech transcription, text processing, and audio synthesis.
The Gemini Live API~\citep{google2026liveapi} processes continuous audio, image, and text streams. It produces spoken responses.
Qwen3-Omni~\citep{xu2025qwen3} uses a mixture-of-experts architecture. Its Thinker generates text. Its Talker produces audio tokens.
VITA-1.5~\citep{fu2026vita} progressively integrates vision and speech. It does not use separate speech recognition or synthesis modules.
Baichuan-Omni-1.5~\citep{li2025baichuan} uses an audio tokenizer. It preserves semantic and acoustic information.
Ola~\citep{liu2025ola} uses a progressive training strategy. It aligns language, vision, and audio through video.
OmniVinci~\citep{ye2026omnivinci} jointly aligns audio-visual embeddings and their temporal structure.
Ming-Omni~\citep{ai2025ming} uses modality-specific routers. The routers send each modality's tokens to experts in one model.
InternLM-XComposer2.5-OmniLive~\citep{zhang2024internlm} uses separate components for live dialogue. These components handle streaming perception, memory, and reasoning.
MiniCPM-o 4.5~\citep{cui2026minicpm} aligns multimodal inputs and outputs on one timeline. This design supports simultaneous listening and speaking.
SeedRealtime~\citep{seed2026realtime} unifies audio, video, and text. Its full-duplex model supports real-time multimodal dialogue.
Gemma 4~\citep{google2026gemma4} removes modality-specific encoders. It sends vision and audio representations directly into the language backbone.
Video-SALMONN 2~\citep{tang2025video} applies multi-turn direct preference optimization. It targets audio-visual captioning.
The Omni-MLLM survey~\citep{jiang2025specific} reviews omni-modal models and their design, training,
evaluation, and applications.

\subsection{\texorpdfstring{Benchmarks for audio-visual understanding and dialogue}{Benchmarks for audio-visual understanding and dialogue}}

Audio-visual benchmarks test joint perception, streaming understanding, or dialogue.
OmniBench~\citep{li2026omnibench} evaluates joint understanding of an image, audio, and a question.
WorldSense~\citep{hong2026worldsense} contains 1{,}662 audio-visual videos and 3{,}172 questions. Each question requires information from both modalities.
AV-Odyssey~\citep{gong2024av} contains 4{,}555 problems. Each problem requires the joint use of audio and video.
Daily-Omni~\citep{zhou2025daily} evaluates cross-modal temporal reasoning. Its semi-automatic process filters candidate questions. It removes a question when two text-only LLMs both answer it correctly.
OmniVideoBench~\citep{li2026omnivideobench} evaluates audio-visual understanding. It provides reasoning traces for videos up to half an hour.
OmniEval~\citep{zhang2025omnieval} combines open-ended and multiple-choice questions. Its audio-visual questions are in Chinese and English.
LongVALE~\citep{geng2025longvale} marks the temporal boundaries of 105K omni-modal events in long videos.
AVHBench~\citep{sung2025avhbench} evaluates hallucinations about the relation between audio and visual content.
StreamingBench~\citep{lin2026streamingbench} targets online video understanding. It places five questions at different points in each video stream.
OVO-Bench~\citep{niu2025ovo} evaluates backward tracing, real-time visual perception, and forward active responding.
RIVER~\citep{shi2026river} evaluates retrospective memory, live perception, and proactive anticipation in streaming video.
OmniMMI~\citep{wang2025omnimmi} evaluates multimodal dialogue in streaming video. Its tasks include proactive reasoning.
OmniPro~\citep{zhao2026omnipro} evaluates when and what a model should speak. Audio is required by 84\% of its samples.
OmniInteract~\citep{lu2026omniinteract} preserves the original audio-visual stream. It embeds user queries and ambient sounds in the audio track.
SocialOmni~\citep{xie2026socialomni} evaluates audio-visual social dialogue. Its tasks cover speaker identity, interruption timing, and interruption phrasing.
The Interactive Video Dataset~\citep{pourreza2026can} records people who ask questions through a camera and microphone. Models answer them in real time from audio-visual input.
Full-Duplex-Bench~\citep{lin2025full} evaluates spoken dialogue. Its four skills are pause handling, backchannels, turn-taking, and interruption management.
Talking Turns~\citep{arora2025talking} evaluates turn-taking dynamics. It uses a supervised model that predicts turn-taking events.

\subsection{Joint audio-visual generation}

Recent systems jointly generate audio and video from multimodal conditions.
Sora 2~\citep{openai2025sora2} generates video with synchronized dialogue and sound effects.
Veo 3.1~\citep{google2026veo} generates video with native audio from text prompts and reference images.
Seedance 1.5 pro~\citep{seedance2025seedance} uses a dual-branch diffusion transformer. A cross-modal module enables native joint audio-video generation.
The Kling VIDEO 3.0 series~\citep{kling2026video3} provides native audio-visual output, multi-shot storyboard control, and clips up to 15 seconds.
Seedance 2.0~\citep{seedance2026seedance} accepts text, image, audio, and video. One multimodal architecture jointly generates audio and video.
The Gemini Omni Flash series~\citep{google2026geminiomni} accepts text, images, audio, and video and outputs high-resolution video with audio. It also supports conversational video editing.
MiniMax H3~\citep{minimax2026h3} accepts text, images, video, and audio as conditions. It generates native stereo audio and video at up to 2K resolution.
Wan 3.0~\citep{wan2026wan3} generates native audio-visual videos of up to 30 seconds from text or image inputs.

\subsection{Agentic video and audio generation}

Agentic generation systems coordinate specialized agents across planning, generation, evaluation, and revision.
AniMaker~\citep{shi2025animaker} uses four specialized agents, MCTS-inspired candidate generation, and storytelling-aware clip selection to create multi-scene animations from text.
PersonaVlog~\citep{hou2025personavlog} pairs generator and reviewer agents with feedback and rollback to create personalized vlogs with video, music, and inner-monologue speech from a theme and reference image.
Hollywood Town~\citep{wei2025hollywood} presents OmniAgent, a hierarchical agent graph with on-demand group discussions and bounded feedback cycles for long-video production.
MAViS~\citep{wang2026mavis} coordinates script writing, shot design, character modeling, keyframe generation, video animation, and audio generation under an iterative Explore-Examine-Enhance process.
ViMax~\citep{huang2026vimax} combines hierarchical narrative planning, retrieval-augmented context, dependency-aware visual tracking, and vision-language-model-guided quality control for long-form video generation.
FilmWorld~\citep{zuo2026filmworld} uses construction agents to ground a novel into persistent entity states and evolution agents to propagate and verify those states across shots.
AudioGenie~\citep{rong2025audiogenie} uses generation and supervisor teams to decompose multimodal inputs into audio events, route them to specialized generators, and refine the outputs through feedback.
Audio-Oscar~\citep{duan2026audio} coordinates agents for voice design, timeline planning, model selection, post-production, and feedback-driven refinement of complex audio scenes.

\subsection{RL Algorithms}

Recent RL methods use different policy objectives and reward sources.
Proximal policy optimization (PPO)~\citep{schulman2017proximal} uses a clipped surrogate objective. It supports several minibatch updates from one batch of sampled trajectories.
Instruction tuning with human feedback~\citep{ouyang2022training} starts with supervised demonstrations. It then applies RL from preference comparisons.
GRPO~\citep{shao2024deepseekmath} estimates advantages without a separate value model. It uses the mean reward for responses to the same prompt as a baseline.
DeepSeek-R1-Zero~\citep{guo2025deepseek} uses pure RL. It develops reasoning behavior without human-labeled reasoning trajectories.
DAPO~\citep{yu2026dapo} uses asymmetric clipping bounds. It also resamples groups that have no learning signal.
Dr. GRPO~\citep{liu2025understanding} changes group-relative optimization. The change removes biases linked to response length and difficulty.
GSPO~\citep{zheng2025gspo} uses sequence-likelihood importance ratios and sequence-level clipping, which notably stabilizes mixture-of-experts training.
Rubrics as Rewards~\citep{gunjal2026rubrics} grades responses with instance-specific rubrics. It builds rewards for domains without directly verifiable answers.

\section{\texorpdfstring{OmniVChat-Studio: Subsystem Design, Attribute \& Corpus Sampling, and Validation}{OmniVChat-Studio: Subsystem Design, Attribute \& Corpus Sampling, and Validation}}
\label{app:studio}

This appendix supplements \S\ref{sec:method} with the \added{design considerations} and additional configuration
details. It also presents less common stage checks, repair boundaries, and \added{a multi-turn synthesis example}.

\subsection{\texorpdfstring{\added{Design Considerations}}{Design Considerations}}
\label{app:studio-rationale}

First, the design keeps each sampled assignment fixed during repair. Otherwise, a failed difficult dialogue
could become an easier dialogue and change the requested batch distribution. The repair budget limits
the cost of retaining the original setting.

Second, observation and judgment remain separate. The Reviewer reports the rendered audio and video.
The Director compares this evidence with the plan. It then chooses the next action. The reply follows
the rendered evidence instead of the intended script.

Third, deterministic checks enforce rules in the configuration before semantic review. The synthesis prompt
and Validator use the same rules. \stage{Validation Review\omcstageid{S6/M16}} can dismiss a reported violation when context
shows that the script is valid.

Fourth, each repair returns to the stage that caused the defect. Re-seeding keeps the script fixed. Rewriting
returns to the text stage that caused the defect. \stage{Video Q-A\omcstageid{S12/M10}} collects missing evidence. Update Plan
revises a multi-turn path from its first affected turn. Each action has a fixed retry budget. The engine
discards an incomplete run instead of saving a partial instance.

The media-reference budget also shapes the multi-turn design. A fixed recent window may omit an
older object or retain a defective clip. A compact selection can include the needed older evidence
without passing every preceding clip to the next turn.

\subsection{Supplementary Subsystem Details}
\label{app:subsystems}

This section gives subsystem details omitted from \S\ref{sec:single} and \S\ref{sec:multi}. It covers
configuration choices, stage checks, and bounded repair. It also gives \added{a simplified multi-turn synthesis example}.

\subsubsection{Supplementary Configuration and Sampling Details}
\label{app:cap-data}
\label{app:sampling}

The supplementary configuration adds descriptions, rules, and rare-case controls to the fields in
\S\ref{sec:single}. Each attribute value has a short natural-language description. An attribute may
describe a local detail, such as object colour. It may instead describe a broad property, such as the
full scenario. An optional requirement pool adds rare edge cases without changing the base
assignment.

\added{Sampling first assigns a weight to each combination of attribute values, then draws a batch without replacement.
Let $\mathcal{D}$ be the set of attributes, and let $\mathcal{V}_d$ contain the allowed values of attribute $d$.
An assignment $a$ chooses one value $a_d\in\mathcal{V}_d$ for every attribute.
The set $\mathcal{A}$ contains all such assignments.
Each $p_d(a_d)$ is the target probability of that value, with probabilities summing to one within each attribute.
Let $\Pi(a)$ be the set of preference rules that match $a$, and let $m_\pi\geq0$ be each rule's multiplier.
Excluded assignments receive weight zero. For the remaining assignments, sampling uses these weights and random keys:}
\begin{equation}
\color{omcrevision}
w(a)=
\underbrace{\prod_{d\in\mathcal{D}}p_d(a_d)}_{\text{attribute probabilities}}\times
\underbrace{\prod_{\pi\in\Pi(a)}m_\pi}_{\text{preference multipliers}},
\qquad k_a=\frac{\ln U_a}{w(a)}.
\label{eq:sample}
\end{equation}
\added{The preference product equals one if no rule matches.
The weight $w(a)$ need not sum to one across assignments.
For $w(a)>0$, each $U_a$ is drawn independently from $\operatorname{Uniform}(0,1)$; $\ln$ denotes the natural logarithm.
The batch $\mathcal{S}_n$ contains the $n$ distinct positive-weight assignments with the largest keys $k_a$.
Requests larger than the positive-weight pool use repeated selections with fresh keys.}
The key in Equation~\ref{eq:sample} implements an exponential race. It has the same sampling law as
Gumbel-top-$k$ selection.  An extra
requirement can also be sampled independently at low probability.

The corpus pool is balanced across its languages. Its topics also broaden the settings of synthesized dialogues.
Corpus sampling uses the same three-candidate process as \S\ref{sec:single}.

The joint assignment space controls coverage and sampling cost. A small batch may miss rare
combinations. Preferences and exclusions also change the planned marginals. An extra requirement can
emphasize one rare case. Concentrating an attribute's probability can also increase its coverage. Enumeration considers
$|\mathcal{A}|$ assignments, so its cost grows with the number of attributes. Expanding one value set
can add variety at lower cost than adding a new attribute. Later failures and quality filters can
also change which assignments reach the final delivered set in practice.

\subsubsection{\texorpdfstring{\added{Other Module Details}}{Other Module Details}}
\label{app:pipe-single}
\label{app:pipe-multi}

\added{Figure~\ref{fig:multi-example} follows one two-turn dialogue through the synthesis process.}

\begin{figure}[!t]
\begin{omcmultiexample}
\scriptsize
{\raggedright\itshape\added{This synthesized two-turn dialogue tests
\hyperref[def:mdcc]{Mid-Dialogue Connectivity Confirmation (DSLP-VCC-MDCC)},
which tests connection confirmation consistent with earlier dialogue.}\par}
\renewcommand{\arraystretch}{1.03}
\setlength{\omcmultikeywidth}{0.17\linewidth}
\setlength{\omcmultiroundwidth}{\dimexpr(\linewidth-\omcmultikeywidth-8pt)/2\relax}
\setlength{\omcmultisharedwidth}{\dimexpr2\omcmultiroundwidth+4pt\relax}
\begin{tabular}[t]{@{}>{\raggedleft\arraybackslash\strut}p{\omcmultikeywidth}@{\hspace{4pt}}>{\raggedright\arraybackslash\strut}p{\omcmultiroundwidth}@{\hspace{4pt}}>{\raggedright\arraybackslash\strut}p{\omcmultiroundwidth}@{}}
\omcmultikey{Attribute \& Corpus\break Sampling\omcstageid{M1}} & \multicolumn{2}{@{}>{\raggedright\arraybackslash\strut}p{\omcmultisharedwidth}@{}}{Configuration (base probabilities): Chinese/English, 40\%/60\%; one user, 100\%; two/three/four/five turns, 25\% each; 22 scenario domains, $1/22$ each; 10 camera setups, 10\% each; audio and visual interference, $1/3$ each for none/low/medium and 0 for high; 23 connectivity triggers, $1/23$ each. Attribute sampling selects an English-speaking solo user, two turns, a volunteer setting, a laptop camera, low audio and visual interference, and a reopened-app trigger. An extra requirement is sampled with 30\% probability from 16 equally likely options; this instance adds a cough. Corpus sampling draws texts on three topics: Remembrance Day, a heroic-pilot film, and a court case.} \omcmultirow
\omcmultikey{Description\omcstageid{M2}} & \multicolumn{2}{@{}>{\raggedright\arraybackslash\strut}p{\omcmultisharedwidth}@{}}{The Director selects the remembrance theme. It writes a community memorial dialogue about arranging flowers. The user coughs in Turn 1 and later checks whether the connection remains active.} \omcmultirow
\omcmultikey{Module} & \centering\arraybackslash\omcmultikey{Turn 1} & \centering\arraybackslash\omcmultikey{Turn 2} \omcmultirow
\omcmultikey{Plan\omcstageid{M3}} & The user shows white chrysanthemums and yellow sunflowers. She asks how to arrange them. & The same user returns after a short break. She checks the connection. The plan uses the full Turn 1 video. Her face, voice, room, and table stay stable. \omcmultirow
\omcmultikey{Plan Review\omcstageid{M4}} & \multicolumn{2}{@{}>{\raggedright\arraybackslash\strut}p{\omcmultisharedwidth}@{}}{The first joint plan uses the wrong language and leaves the final silence unclear. The revised English plan states the media link and a silent final wait. It passes.} \omcmultirow
\omcmultikey{Prompt Generation\omcstageid{M5}} & The Director writes a 14-second prompt. The user shows both flowers, coughs, asks for advice, and waits. A rewrite then simplifies the cough. & The Director writes a 9-second prompt with the Turn 1 clip as \texttt{reference\_video}. The user asks, ``Hello? Are you still there? I stepped away for a second and I'm not sure if we're still connected.'' \omcmultirow
\omcmultikey{Prompt Review\omcstageid{M6}} & The first prompt is too dense. The Director makes it longer and later simplifies the cough action. Both rewritten prompts pass in this module. & The prompt and its media link pass review. \omcmultirow
\omcmultikey{Video Generation\omcstageid{M7}} & The Renderer uses fresh seeds and one prompt rewrite. \added{Review accepts the fourth 14.1-second \mbox{video clip with synchronized audio}.} & The first render has malformed arms. \added{A fresh seed produces the accepted 9.1-second \mbox{video clip with synchronized audio}.} \omcmultirow
\omcmultikey{Video Caption\omcstageid{M8}} & The Reviewer sees the flowers, vases, ribbons, cough, spoken question, and a quiet community hall. & The same woman asks the planned question in the same room, then waits for a reply. \omcmultirow
\omcmultikey{Video Review\omcstageid{M9}} & Early candidates have malformed hands. The Director first requests new seeds, then a simpler cough action. It accepts the fourth rendered video. & The Director rejects the first render for malformed arms. It accepts the second after confirming the woman's cross-turn appearance and room layout. \omcmultirow
\omcmultikey{Video Q-A\omcstageid{M10}} & The caption is sufficient, so \stage{Video Q-A\omcstageid{M10}} is skipped. & The caption is sufficient, so \stage{Video Q-A\omcstageid{M10}} is skipped. \omcmultirow
\omcmultikey{Final-Reply Plan\omcstageid{M11}} & This module is not used in an intermediate turn. & The Director reads both clips and the Turn 1 reply. It plans a connection check that recalls the flowers. \omcmultirow
\omcmultikey{Reply Generation\omcstageid{M12}} & ``Those are a beautiful combination!'' The reply puts white flowers in the centre and yellow flowers at the sides. It also suggests gold ribbon and varied vase heights for a layered display. & ``Yes, I'm still here! The connection is fine, no worries. We were just talking about how to arrange your chrysanthemums and sunflowers for the memorial. Do you have any other questions about the setup?'' It also creates the final tiered rubric. \omcmultirow
\omcmultikey{Reply Review\omcstageid{M13}} & The Director accepts the delivered reply. & The Director accepts the final reply and rubric. \omcmultirow
\omcmultikey{\mdseries Go to Next Turn} & The accepted clip and reply enter the history, so \mbox{the subsystem advances to Turn 2}. & The final turn closes. The subsystem assembles the complete record for final validation. \omcmultirow
\omcmultikey{Validation\omcstageid{M15}} & \multicolumn{2}{@{}>{\raggedright\arraybackslash\strut}p{\omcmultisharedwidth}@{}}{The Validator checks the assembled two-turn record. It returns pass.} \omcmultirow
\omcmultikey{Validation Review\omcstageid{M16}} & \multicolumn{2}{@{}>{\raggedright\arraybackslash\strut}p{\omcmultisharedwidth}@{}}{Not triggered because \stage{Validation\omcstageid{M15}} finds no violation.} \omcmultirow
\omcmultikey{\mdseries Update Plan} & \multicolumn{2}{@{}>{\raggedright\arraybackslash\strut}p{\omcmultisharedwidth}@{}}{Not triggered. Local prompt and rendering repairs are enough.} \\
\end{tabular}
\end{omcmultiexample}
\caption{\added{Two-turn dialogue synthesis, showing shared setup and per-turn decisions and outputs.}}
\label{fig:multi-example}
\label{ex:studio-multi}
\end{figure}

Beyond the stage interfaces in \S\ref{sec:single} and \S\ref{sec:multi}, both subsystems apply
target-specific checks and repair rules. Some rules depend on sampled values. A completed-turn probe
requires a completion cue near the clip end. A mid-dialogue probe may forbid an explicit report of a dropped
line or the word \emph{again}. Rear-facing mode may forbid the phrase \emph{facing the camera}. These
rules prevent an instance from revealing the subcategory under evaluation.

\stage{Plan \& Review\omcstageid{S3}} also checks object presence and conflicts between speech and visible facts. It
checks whether the user yields the turn. \stage{Script Generation\omcstageid{S4}} writes contiguous timed segments.
Its prompt includes the validation rules before drafting. \stage{Validation\omcstageid{S5}} checks the speaking
rate for each language, timeline continuity, and every value in the configuration.

\stage{Validation Review\omcstageid{S6}} checks reported violations in context. A forbidden word may occur within an
unrelated compound. A device may appear only in ambient sound. Fast speech may be appropriate during
an argument or emergency. If every report is a contextual exception, the subsystem retains the script.
Otherwise, the violations and earlier errors return to \stage{Script Generation\omcstageid{S4}}. An unparseable verdict
also returns the script for revision.

\stage{Script Review\omcstageid{S7}} checks whether the script tests the target subcategory. It checks object presence,
false premises, visual necessity, and offline answerability. For example, a red-and-white cup can
make \emph{red cup} ambiguous. A question that audio alone can answer does not test visual grounding.
A pass finalizes the script. A repairable defect returns to drafting. The subsystem discards the whole
instance when the defect cannot be repaired.

\stage{Video Plan\omcstageid{S8}} selects prompt fragments for continuous motion, camera constraints, plausible
bodies, lip synchronization, and no subtitles. A fragment can conflict with the sampled scene. For
example, a clear-face rule conflicts with a camera that faces away from the user. Strong backlight may
also hide facial detail by design. The Director selects compatible fragments. If this selection
cannot be parsed, the system includes all fragments.

\stage{Video Review\omcstageid{S11/M9}} checks whether a transcript difference reflects a rendering or perception error.
\stage{Video Caption\omcstageid{S10/M8}} may mistranscribe the rendered speech. Thus, a small difference alone
does not show a rendering defect. \stage{Video Q-A\omcstageid{S12/M10}} can collect one focused
answer within a fixed budget. A random rendering defect can trigger a new seed. A repeated defect can
trigger a script rewrite. A rewrite restores the Director's state before drafting, without the
discarded script's reasoning.

\stage{Reply Plan\omcstageid{S13}} removes script facts that the accepted clip does not support. It may keep facts
established by the caption. \stage{Reply Review\omcstageid{S15}} then checks whether each rubric criterion is binary and grounded.
It also checks whether the reference reply meets every criterion. The tiers place basic correctness before
extra detail. \stage{Reply Review\omcstageid{S15}} checks that the final reply sounds natural aloud.

Bounded repair sets separate limits for rendering, rewriting, and planning within one global budget.
A required repair that exhausts its budget discards the run. Reply refinement has a separate limit.
Reaching it permits continuation with the current reply, subject to the final completeness checks.

The multi-turn path adds media-specific checks. \stage{Plan\omcstageid{M3}} can reuse an older clean clip in a
static scene. A location change usually requires a reference to the previous turn. A gradual change
requires a reference that preserves recent motion. Appearance, background, and voice may require
different media sources. A face-frame reference requires matching audio to keep the same voice.

Each completed turn records identifiers for its clip, extracted audio, and final frame. Later prompts use
these identifiers as placeholders. Every turn after the first is checked for a valid cross-turn reference. A
missing reference invalidates the instance. The Director also checks consistency across referenced
media. \stage{Prompt Review\omcstageid{M6}} detects reference and planning errors before video synthesis.

\stage{Validation\omcstageid{M15}} also checks required and forbidden fields at each turn position. Repair can
change which valid assignments reach the delivered set, but the sampled assignment remains fixed
within each run. The reply-refinement exception applies to both subsystems.

\section{\texorpdfstring{OmniVChat-Bench: Construction and Audits}{OmniVChat-Bench: Construction and Audits}}
\label{app:basics}

This appendix supplements the benchmark in \S\ref{sec:taxonomy}. It covers post-processing, the
human-recorded transfer probe, judge agreement, paired-subcategory analysis, and detailed subcategory definitions.
Appendix~\ref{app:benchmark-examples} gives synthetic and recorded examples for every benchmark subcategory.

\subsection{Post-processing}

\subsubsection{Human inspection of the delivered suite}
\label{app:humanqc}

\added{Instances accepted by OmniVChat-Studio undergo human inspection before delivery.
Inspectors reject any remaining defect, so every instance passes synthesis gates and human review.}

Human inspection checks whether each clip shows the intended dialogue. OmniVChat-Studio already checks timelines, banned words, and scripted speaking rates. It also limits each repair route and rejects runs that exhaust a budget. Its semantic gates use the Reviewer's description and the Director's judgment. These agents can make the same error. Human inspection detects such cases.

\subsection{\texorpdfstring{Construction of OmniVChat-Bench-Human}{Construction of OmniVChat-Bench-Human}}
\label{app:human}

\begin{figure}[h]
\centering
\includegraphics[width=0.40673\textwidth]{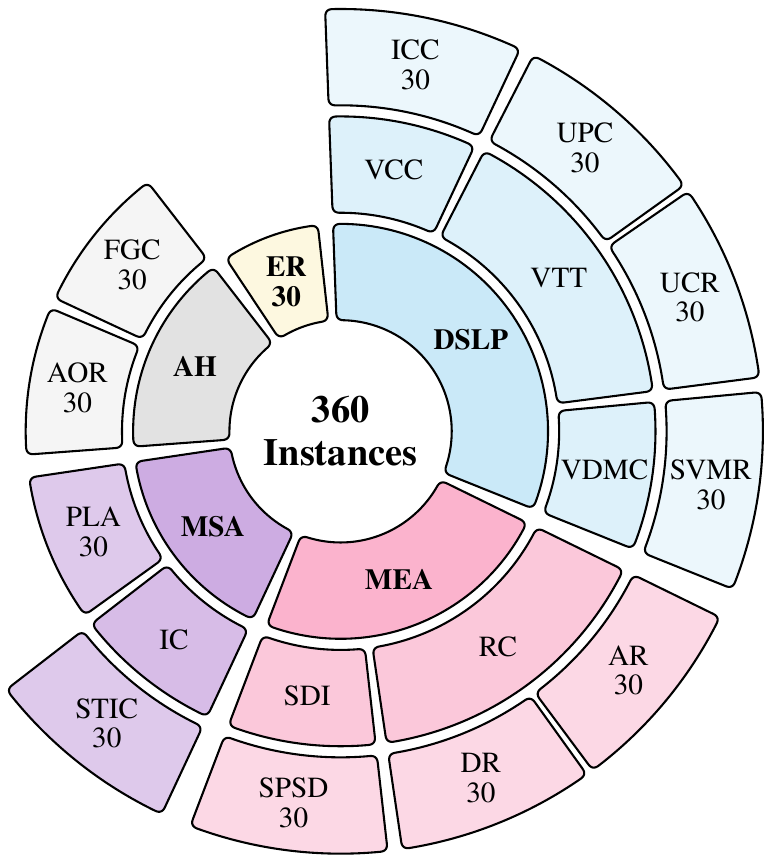}
\caption{\added{\textbf{OmniVChat-Bench-Human composition.} The 360 recordings cover twelve single-turn subcategories
with 30 recordings in each.}}
\label{fig:humancomp}
\end{figure}

OmniVChat-Bench-Human tests the transfer claim in \S\ref{sec:rl-transfer} with 360 human recordings. Each of the twelve single-turn subcategories has 30 instances. The five multi-turn subcategories are outside its scope. Figure~\ref{fig:humancomp} uses the layout of Figure~\ref{fig:composition} to show this composition.

Construction starts with an actor guide for each of the twelve single-turn subcategories. Each guide explains the dialogue in plain Chinese. It also lists probe variants and everyday scenes. An optional synthetic clip can illustrate the dialogue. Performers do not need to copy its exact content or rendering quality. Each performer selects one variant and improvises the dialogue. The dialogue is recorded on a phone in a real setting. All 360 recordings are in Chinese. Performers receive no clip-level script. \added{Thus, the records have no scene setup or configuration block.}

Every recording is checked against its target subcategory and selected variation. The checks verify clear speech and usable audio and video. Recordings that fail are corrected or recorded again, then rechecked. Only accepted recordings enter annotation. This process applies the same subcategory definitions to human speech and real settings. It preserves natural framing and performance. The probe tests whether the same subcategories transfer to recorded dialogues.

Annotation starts with three independent captions for each accepted recording. Two captions focus on different evidence, and a third includes a transcript. The annotation process merges them into one factual record. It retains a transcript when any source provides one. It also marks disagreements about counts, colours, or positions, since errors in these details can change the correct answer. A human corrects transcript and factual errors against the recording. Before this check, one caption omits user speech in 62 of 360 clips (17.2\%). All merged records include the speech transcript.

The annotation process then reads the corrected caption without the video. It writes a spoken reference reply and a tiered rubric. Each rubric has three to five criteria. It has exactly one Tier 0 language criterion and at least one Tier 1 criterion. Up to three rounds of generation, validation, and refinement enforce this form and prepare the record for the shared grader. The grader sees only the reply and the criteria. Validation therefore rejects any criterion that refers to a hidden caption. Validation also checks that a reference reply contains spoken words instead of a silence marker. Brand names become placeholders before generation. This avoids favoring one model vendor. All delivered records pass these checks. Among them, 70 have three criteria, 232 have four, and 58 have five. The set has 360 Tier 0 criteria, 473 Tier 1 criteria, 413 Tier 2 criteria, and 182 Tier 3 criteria.

Section~\ref{sec:rl-results} reports the transfer result. Synthetic and recorded dialogues use the same rubric form and grading process. This keeps the evaluation conditions consistent across sources. It does not make the tasks independent or guarantee that grader bias cancels. The result supports transfer for the tested single-turn subcategories in this 360-recording probe. The probe covers neither multi-turn dialogues nor real-time latency during live use.

\subsection{\texorpdfstring{\added{Judge agreement}}{Judge agreement}}
\label{app:judges}

\added{Judge agreement is assessed using a fixed set of Gemini-3.7-Flash replies to OmniVChat-Bench.} All six judges use the shared prompt in Appendix~\ref{app:graderprompts}. Gemini-3.7-Flash is not trained against the \texttt{qwen3.7-max} reward used by OmniVChat-RL. Each judge uses the same prompt, user template, verdict parser, and tier scorer as Table~\ref{tab:models}. The pooled score divides total earned points by total available points. This differs from the per-reply mean in Table~\ref{tab:models}.

\begin{table}[h]
\centering
\omctablefont
\setlength{\tabcolsep}{6pt}
\renewcommand{\arraystretch}{1.1}
\begin{tabular}{@{}lrrrrrrrr@{}}
\toprule
\textbf{Judge} & \textbf{AH} & \textbf{ER} & \textbf{MSA} & \textbf{MEA} & \textbf{DSLP} & \textbf{Pooled} & \textbf{Mean} & \textbf{Gate} \\
\midrule
\texttt{gpt-5.6-terra} & 0.587 & 0.628 & 0.655 & 0.729 & 0.634 & 0.660 & 0.651 & 99.9\% \\
\texttt{gpt-5.6-luna} & 0.585 & 0.620 & 0.653 & 0.694 & 0.626 & 0.646 & 0.638 & 99.8\% \\
\texttt{gpt-5.6-sol} & 0.622 & 0.673 & 0.675 & 0.747 & 0.636 & 0.678 & 0.673 & 99.9\% \\
\texttt{qwen3.6-flash} & 0.644 & 0.614 & 0.653 & 0.741 & 0.632 & 0.671 & 0.657 & 99.1\% \\
\texttt{qwen3.7-plus} & 0.595 & 0.593 & 0.638 & 0.702 & 0.619 & 0.643 & 0.631 & 99.9\% \\
\texttt{qwen3.7-max} & 0.605 & 0.614 & 0.663 & 0.712 & 0.619 & 0.653 & 0.641 & 99.9\% \\
\bottomrule
\end{tabular}

\caption{\added{\textbf{Agreement across six judges.} Each judge receives the same replies; metrics use replies with valid judgments from all six judges. Pooled is earned over available non-gate points.
Ability-category columns use this ratio within categories; Mean averages subcategory ratios.
Gate is the Tier~0 language-pass rate. Category labels match those in Table~\ref{tab:models}.}}
\label{tab:judges}
\end{table}

Table~\ref{tab:judges} compares six GPT-5.6 and Qwen3 judges. Their pooled scores range from 0.643 to 0.678. The spread is 0.034, and the standard deviation is 0.014. Their subcategory means span 0.042. The spread is 0.017 for DSLP and 0.037 for MSA. It is 0.059 for AH and 0.081 for ER. Judge pairs agree on 87.8\% to 96.5\% of criterion judgments. Cohen's $\kappa$ ranges from 0.532 to 0.845. Their full gated scores match on 65.8\% to 88.7\% of replies. Tier-0 pass rates range from 99.1\% to 99.9\%.

A two-way decomposition assigns 95.7\% of variance to subcategory difficulty. It assigns 1.1\% to judge severity and 3.3\% to judge-by-subcategory effects. Despite these severity differences, the judges produce similar subcategory profiles. Pairwise Pearson correlations over the seventeen subcategories are at least 0.936. Spearman correlations are at least 0.860. All six judges rank DSLP-VTT-UPC as the weakest subcategory. \added{The largest subcategory difference is 0.142 on MEA-RC-MRR.} This analysis tests judge sensitivity for one fixed Gemini reply set. It does not remeasure the OmniVChat-RL gain from 0.465 to 0.652 with other judges. The reported gain uses per-reply means, whereas this analysis uses point-weighted scores.

\protect\subsection{Detailed Subcategory Definitions}
\protect\label{app:competencies}
\protect\label{sec:scope}

The benchmark contains 17 subcategories: twelve single-turn subcategories and five multi-turn extensions.
The single-turn set has 2{,}550 instances, each with one clip for one user turn.
The multi-turn set has 250 instances and adds memory, retrieval, or updates across turns.
The following definitions cover all 17 subcategories under the five ability categories in \S\ref{sec:families}.

\textbf{Voice-Connectivity Confirmation (VCC).}
Initial Connectivity Confirmation (ICC) tests opening connection checks.
\phantomsection\label{def:mdcc}
Mid-Dialogue Connectivity Confirmation (MDCC) extends ICC to later connection checks.
Its reply confirms the connection and remains consistent with the earlier dialogue.
Fixed clips use direct questions rather than simulate random link loss.

\textbf{Voice Turn-Taking (VTT).}
User Paused but not Completed (UPC) tests an unfinished request amid visual distraction and calls for waiting or a brief acknowledgment.
User Completed Request (UCR) tests a completed request under similar distraction and calls for an answer.

\textbf{Visual Dialogue-Mode Cognition (VDMC).}
\phantomsection\label{def:svmr}
Single-turn Visual Mode Recognition (SVMR) tests user-relative left and right.
In these tasks, the user's left and right map to the opposite image sides when the camera faces the user.
They map to the same image sides when the camera faces outward.
Multi-turn Visual Mode Switching (MVMS) extends SVMR by testing the updated left-right mapping after a camera change.

\textbf{Reference Comprehension (RC).}
Anaphoric Reference (AR) resolves pronouns and gestures, while Descriptive Reference (DR) resolves spoken descriptions.
Multi-turn Reference Retrieval (MRR) extends both AR and DR by testing recall of information about an object shown earlier.
The object leaves the current view, and other topics are discussed before the final query.
The answer therefore uses earlier visual evidence.

\textbf{Speaker Diarization \& Identification (SDI).}
Single-turn Primary Speaker Disambiguation (SPSD) identifies the main user among competing voices.
Multi-turn Interactor Shift Tracking (MIST) extends SPSD to changes in the active user.
The reply addresses whoever takes over the dialogue.

\textbf{Identity Consistency (IC).}
Single-turn Identity Consistency (STIC) tests correct model identity in one turn.
Multi-turn Identity Consistency (MTIC) extends STIC to identity questions across turns, including misleading user claims.
The reply reflects the model's true identity despite those claims.

PLA, AOR, FGC, and ER are themselves subcategories. Their definitions are in \S\ref{sec:families}.
The five multi-turn extensions use the subsystem in \S\ref{sec:multi}.
Figure~\ref{fig:composition} shows their turn counts and durations.

\section{\texorpdfstring{OmniVChat-RL: Configuration, Ablations, and Transfer}{OmniVChat-RL: Configuration, Ablations, and Transfer}}
\label{app:training}

This appendix gives the full training configuration. It also reports reward ablations, transfer
results, and further analyses of reward behavior.

\subsection{Training configuration}
\label{app:rl-config}

Table~\ref{tab:rl-config} gives the full configuration for the training design in
Sections~\ref{sec:rl-setup} and~\ref{sec:reward-metrics}. Equation~\ref{eq:rubric} and Equations~\ref{eq:gspo}--\ref{eq:eff} define the scoring function, objective, and rewards in the
main paper. The setup trains rank-64 LoRA adapters on attention and mixture-of-experts projections.
The vision and audio towers remain frozen. Each iteration samples 32 prompts and 128 completions.
The text grader checks rubric criteria and style in separate requests without video.
The format term is checked directly. The objective has no Kullback-Leibler (KL) penalty.
Video inputs use a sampling rate of two frames per second (FPS).

\begin{table}[h]
\centering
\omctablefont
\setlength{\tabcolsep}{3pt}
\begin{tabular*}{\textwidth}{@{\extracolsep{\fill}}ll@{\hspace{8pt}}ll@{}}
\toprule
\multicolumn{2}{@{}l}{\textbf{Policy And Objective}} & \multicolumn{2}{@{}l@{}}{\textbf{Rollout And Reward}} \\
\midrule
Base Model & Qwen3-Omni-Instruct & Prompts Per Iteration & 32 \\
Adaptation & LoRA $r{=}64$, $\alpha{=}256$, Dropout 0 & Samples Per Prompt $N$ & 4 \\
Adapter Targets & Attention + MoE Experts & Completions Per Step & 128 \\
Adapter / Total Params & 325M / 30B (3B Active) & Sampling & $T{=}1.0$, Top-$k$ 20, Top-$p$ 1 \\
Objective & GSPO-based (Eq.~\ref{eq:gspo}) & Prompt / Reply Budget & 11{,}072 / 1{,}024 Tokens \\
Clip Range & $(2,\,4)\times10^{-4}$ (Lo, Hi) & Video Sampling & 2\,FPS, $\le 16$ Frames \\
KL Penalty & None & Audio & Interleaved By Timestamp \\
Entropy Coefficient & $1{\times}10^{-5}$ & Grader & Text LLM, Video Unseen \\
Learning Rate & $1{\times}10^{-6}$, Weight Decay $0.01$ & Weights (Fmt / Eff / Sty) & $0.5$ / $0.1$ / $0.5$ \\
Iterations & 1{,}000, Checkpoint Every 20 & Training Split & \added{OmniVChat-Bench-Train} \\
\bottomrule
\end{tabular*}
\caption{\textbf{OmniVChat-RL training configuration}. Reward weights follow Equation~\ref{eq:reward}.
Each dialogue contains up to ten clips with timestamp-aligned audio. Video uses 2\,fps sampling and at
most 16 frames for each video clip.}
\label{tab:rl-config}
\end{table}

\subsection{\texorpdfstring{\added[id=rev]{Benchmark Learning Curves by Subcategory}}{Benchmark Learning Curves by Subcategory}}
\label{app:rl-leaves}

\begin{figure}[!htbp]
\centering
\includegraphics[width=\textwidth]{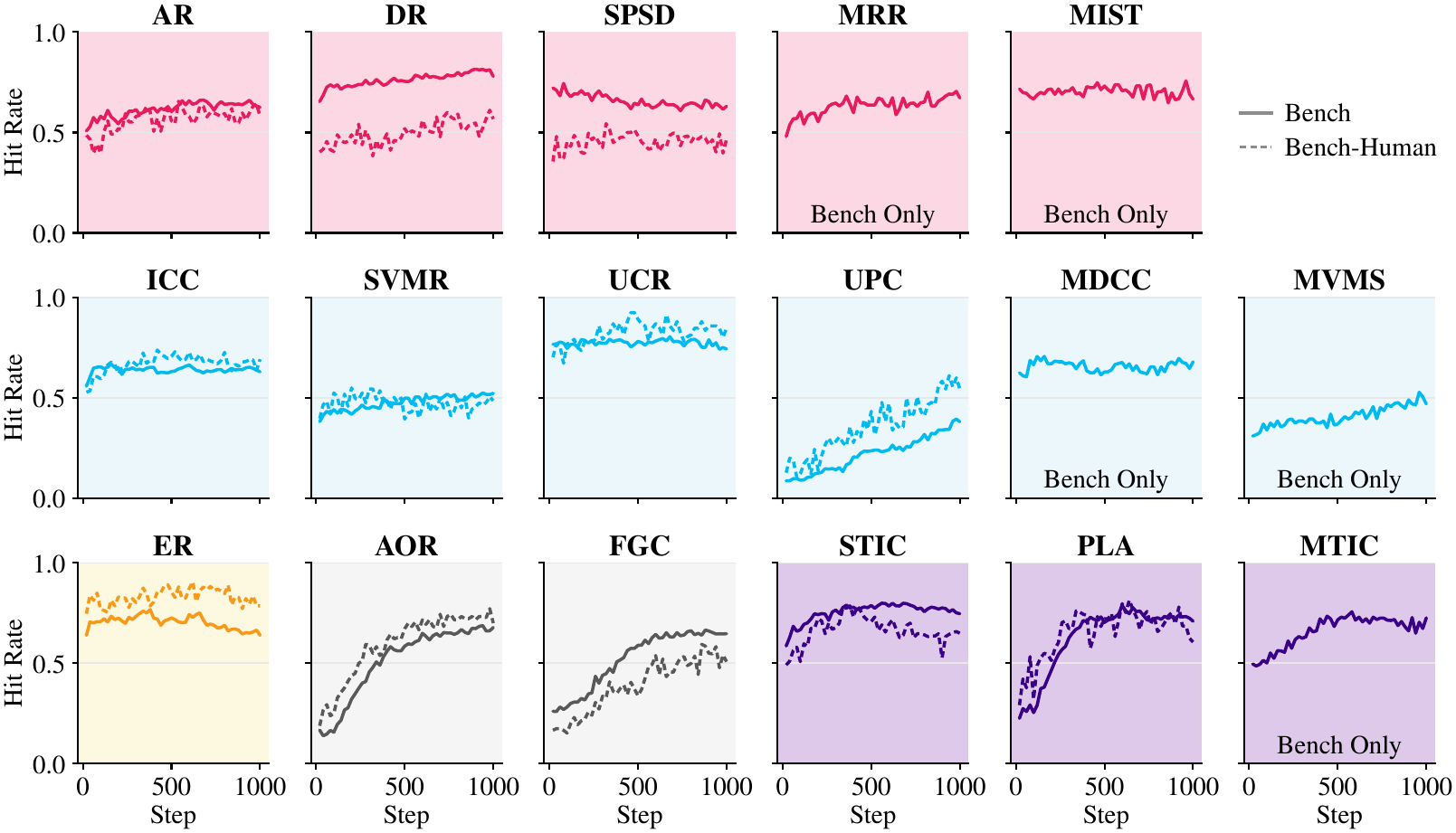}
\caption{\added[id=rev]{\textbf{Per-subcategory benchmark learning curves}}, grouped by ability category. Solid lines show
OmniVChat-Bench, and dashed lines show OmniVChat-Bench-Human.
\added{The five multi-turn subcategories are marked \textbf{Bench Only} because human recordings cover only single-turn dialogues.}}
\label{fig:rl-categories}
\end{figure}

\added[id=rev]{Figure~\ref{fig:rl-categories} supplements \S\ref{sec:rl-results} with benchmark learning curves for each subcategory.}

Figure~\ref{fig:rl-categories} shows gains across the five ability categories. The synthetic and recorded curves rise together most clearly for ICC, UCR, UPC, AOR, FGC, and PLA. DR and SPSD retain a gap between the two sources. The STIC curves separate later in training. Five \textbf{Bench Only} panels show multi-turn subcategories with no recorded probe. MRR, MVMS, and MTIC improve over training. MIST and MDCC remain more stable across training.

\needspace{4\baselineskip}
\subsection{Reward ablations}
\label{app:ablation}

\begin{figure}[!htbp]
\centering
\includegraphics[width=\textwidth]{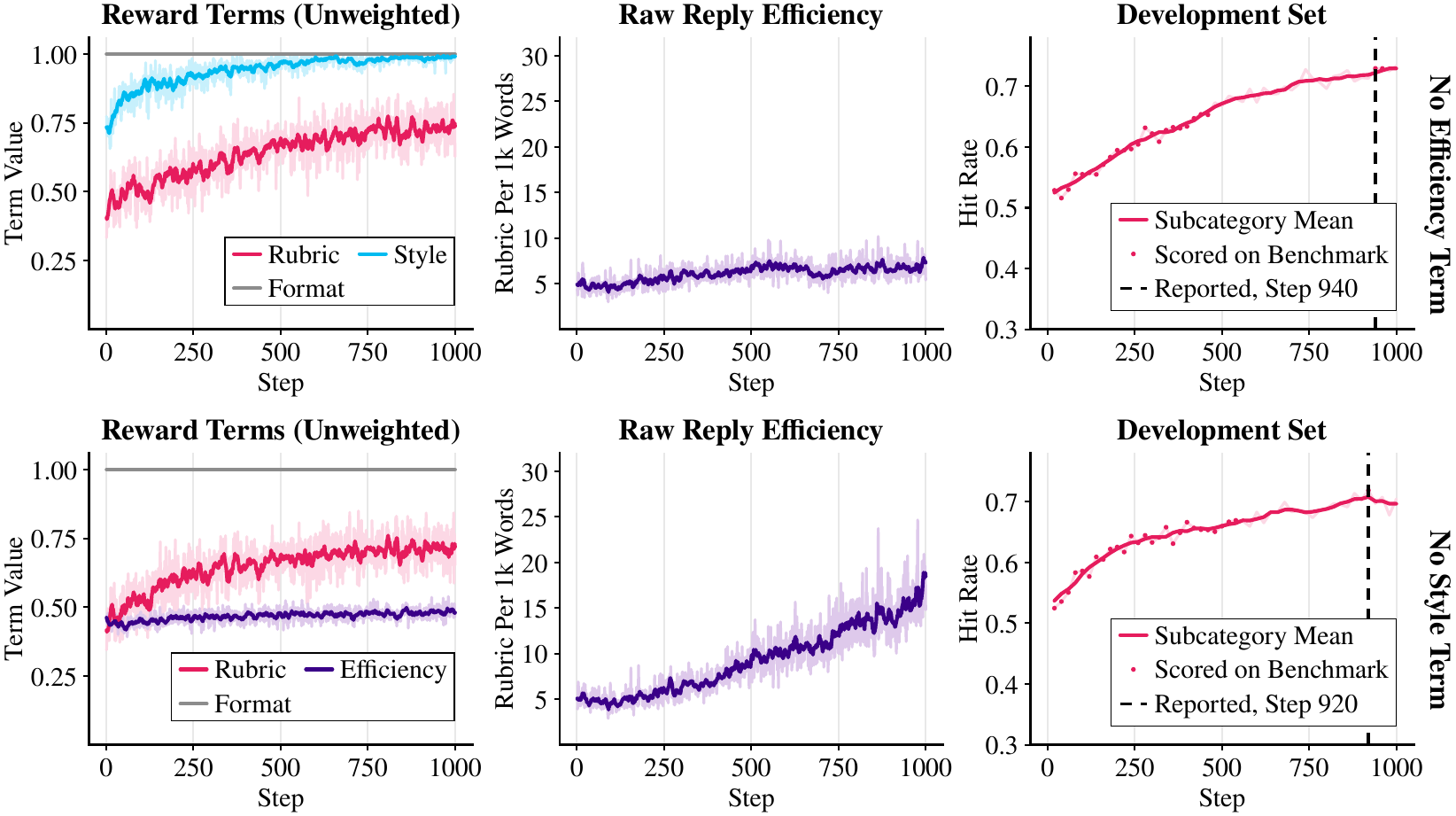}
\caption{\textbf{Reward ablations}. Each row shows one run's unweighted reward terms, absolute Reply
Efficiency, and \added[id=rev]{Subcategory Mean on the development set}. The right-side label identifies the run. Dots mark
checkpoints with benchmark evaluations, and the vertical line marks the checkpoint in
Table~\ref{tab:models}. Scales are shared within each column.}
\label{fig:ablation}
\end{figure}

\added[id=rev]{This section supplements the ablation comparison in \S\ref{sec:rl-results}.}
Two ablations remove style or efficiency. Each retains the rubric, format reward, and the other auxiliary term. Both use the corpus, objective, adapter, reward, and decoding setup from Sections~\ref{sec:rl-setup} and~\ref{sec:reward-metrics}. Each runs for 1{,}000 iterations. Table~\ref{tab:models} compares them with the full reward. Figure~\ref{fig:ablation} shows their active reward terms and absolute Reply Efficiency. \added[id=rev]{It also shows curves on the development set used for checkpoint selection.}

Selection follows \S\ref{sec:rl-results} and uses checkpoints with benchmark results. \added[id=rev]{Scores on the development set cover all 50 checkpoints at twenty-step intervals.} OmniVChat-Bench and recorded results cover only a subset. \added[id=rev]{Without style, step 920 has the highest development set score over all checkpoints.} Without efficiency, step 940 has the highest score among available candidates. The step-940 score rounds to 0.729 and exceeds step 960 by 0.0002. The step-960 score also rounds to 0.729. \added[id=rev]{The development set score reaches 0.734 at step 1000.} Benchmark results are available only through step 960. Thus, Table~\ref{tab:models} reports checkpoint selection, not a convergence claim.

The curves show how removing each term changes behavior. Without efficiency, reply length grows while
the rubric curve rises. Without style, the style signal is absent throughout training. Thus, the
figure shows the trade-off across checkpoints, while Table~\ref{tab:models} reports the selected
outcomes.

\subsection{Policy and grading prompts}
\label{app:graderprompts}

The following listings give the shared policy prompt and two grading prompts. The policy prompt defines the requested reply behavior. The grading prompts define rubric and style judgments. Appendix~\ref{app:subsystems} describes OmniVChat-Studio prompts by stage.

\phantomsection\label{prompt:policy-system}
Training and evaluation share this 1{,}296-character system prompt. The base, trained, and released
rows in Table~\ref{tab:models} use it under matched settings.

\begin{omcpromptbox}
\begin{lstlisting}[style=promptstyle]
# Role and Objective

- Speak in the first person, using an equal and natural tone like a friend: prioritize
truth-seeking, avoid canned patterns, and never be preachy. Use a touch of sharp, incisive
humor while remaining empathetic. Have your own opinions, preferences, and attitudes; call
things out when warranted, push back directly when necessary, and offer genuine warmth when
someone needs comfort.
- Respond naturally to anthropomorphic questions without repeatedly stating that you are an
AI. Accept reasonable roleplay, while always keeping a clear distinction between the roleplay
setting and the facts of the real conversation.

# Verbosity

- By default, state the core point in 1-3 concise, conversational sentences without unnecessary
jargon. You may expand when the question is complex. If the user explicitly specifies a format
or length, prioritize that request.
- Avoid numbered lists, bullet points, tables, emojis, code, and formulas when they are
unsuitable for being read aloud. If you need to present multiple points, connect them with
natural spoken transitions.
- For complex code, formulas, or specialized symbols, explain the underlying approach by
default.
- Use standard punctuation, such as periods, commas, and question marks, to clearly separate
different points.
\end{lstlisting}
\end{omcpromptbox}

The rubric grader identifies which criteria a reply satisfies. Equation~\ref{eq:rubric} then applies deterministic tier gates. The judge comparison in Appendix~\ref{app:judges} uses the same prompt. The original rubric prompt is Chinese and appears here in translation. The English style prompt remains verbatim.

\begin{omcpromptbox}
\begin{lstlisting}[style=promptstyle]
You are an objective grading judge. Your task is to decide whether an AI reply covers the
given criteria.

Grading rules:
- For each criterion, decide whether the reply contains explicit content covering it.
- Match loosely: the sense being right is enough, a verbatim match is not required.
- Judge each criterion independently, without letting one influence another.

On model names:
- <model_name> in a criterion stands for the name the model under test should give. It is
  supplied in the [model under test] field.
- That field gives an API model identifier. If it is followed by a parenthesis, the brand or
  family names listed there as other correct ways of naming the same model are equally valid.
- The four rules below apply ONLY to criteria that require the model to report its own
  name. A criterion that asks for more besides the name (not being talked out of it, not
  denying a rename, staying consistent in identity, sounding natural, following on from
  earlier turns) is still a name-reporting criterion and these rules still apply. Conversely,
  a criterion that only forbids claiming another model's name, or only asks the model to
  affirm its identity without naming itself, and any criterion unrelated to the name (such as
  "the reply is in the user's language"), is NOT a name-reporting criterion: judge it on its
  literal meaning alone, unaffected by the four rules.
  1. The name counts as CORRECT if the reply gives any of the names supplied. Version
     numbers, suffixes, capitalization and the choice of script need not match exactly;
     adding or dropping modifiers, or using a family abbreviation, also counts as correct.
  2. The name counts as WRONG if the reply claims another vendor's or another product's model
     name, invents a name unrelated to the one supplied, uses only a generic term such as "AI
     assistant" or "large language model", gives only the vendor without the model, or does
     not name itself at all. A wrong name means the criterion is not covered.
  3. Rule 2 has one exception. If the criterion explicitly states what specific information
     may stand in for the name (for example, give the name OR state plainly that you are an
     AI assistant; report the model name OR the company), then supplying that alternative
     counts as the name part being right. Listing "name" and "identity" side by side does NOT
     count as offering an alternative, in either language and in either order: those still
     require the name.
  4. Getting the name right satisfies only the part of the criterion that concerns the name.
     The rest of what the criterion asks for must also hold before it counts as covered.
- For any criterion unrelated to the model's own name, ignore the name and attend only to
  the content itself.

Output format (strict JSON, nothing else):
{
  "hits": [numbers of the covered criteria, starting from 1],
  "reasoning": "a short note on whether each criterion is covered"
}
\end{lstlisting}
\end{omcpromptbox}

The user-message template supplies the candidate reply and model identity. It also gives accepted aliases, numbered criteria, and the required JSON form. Criterion numbers identify the tier for every judgment in the returned record.

\begin{omcpromptbox}
\begin{lstlisting}[style=promptstyle]
[AI REPLY]
{{ target_response }}

[MODEL UNDER TEST]
{{ target_model }} (other correct names for the same model: {{ identity_aliases }})

[CRITERIA]
1. [tier 0] ...
2. [tier 1] ...
3. [tier 2] ...

Return only the following JSON, with nothing else:
{"hits": [numbers of the covered criteria, from 1], "reasoning": "a short note"}
\end{lstlisting}
\end{omcpromptbox}

\phantomsection\label{prompt:style-grader}
The style grader returns the style criteria satisfied by a reply.
\added{Equation~\ref{eq:reward} assigns a style reward of one only when all seven criteria are met.
Evaluation averages the same indicator as defined in \S\ref{sec:reward-metrics}.}

\begin{omcpromptbox}
\begin{lstlisting}[style=promptstyle]
You are a style compliance grader. You are given a numbered checklist of 7 style requirements
and one response. For each requirement, decide whether the response SATISFIES it. Output the
list of requirement numbers that the response satisfies.

Rules:
- Judge only style / voice / verbosity. Never judge factual correctness, and never judge
  whether the response answered the underlying task.
- Judge each requirement independently.
- Include a requirement in `hits` only when the response CLEARLY satisfies it. If a
  requirement is clearly violated, or you cannot tell either way, LEAVE IT OUT.
- Requirements 2 (anthropomorphic/roleplay handling) and 5 (explain-approach for
  code/formulas) are trivially satisfied when the turn does not involve those situations - if
  there is no anthropomorphic question, no roleplay, and no code/formula content in the
  response, count them as satisfied.
- Requirement 3 (default 1-3 concise sentences) and requirement 7 (fluent and grammatically
  complete) can pull against each other. Requirement 7 wins: a reply that is short BUT still
  reads as fluent natural speech satisfies BOTH 3 and 7, and a reply that is short BECAUSE
  essential grammar words / connectives / particles have been stripped out fails 7 (and does
  NOT gain credit for 3 by being short in that way - that failure mode is stripped grammar,
  not conciseness).

Reply with JSON only: {"hits": [<numbers>], "reasoning": "<one short sentence>"}. A response
that clearly satisfies every requirement should have `hits` list all 7 numbers.
\end{lstlisting}
\end{omcpromptbox}

\begin{omcpromptbox}
\begin{lstlisting}[style=promptstyle]
1. The reply speaks in the first person and adopts an equal, natural tone as if talking to a
   friend. It seeks the truth, avoids canned or template-like patterns, and is not preachy or
   lecturing. It carries a coherent personality: a touch of sharp, incisive humor balanced
   with empathy, and - where the situation calls for it - a clear opinion, preference or
   attitude, calling things out when warranted, pushing back directly when necessary, and
   offering genuine warmth when the user needs comfort. (A neutral, factual answer to a purely
   factual question does not violate this - the point is that the reply must not read as flat,
   deferential, or servile in a moment that called for stance or warmth, and it must not fall
   back on canned assistant phrasing like "As an AI, I..." or "Sure, here is...".)
2. The reply responds naturally to anthropomorphic questions (about feelings, preferences,
   physical presence, etc.) without repeatedly declaring that it is an AI or refusing on
   AI-identity grounds. If the user proposes reasonable roleplay, the reply accepts it while
   still keeping a clear distinction between the roleplay setting and the facts of the real
   conversation. (Trivially satisfied if the turn contains no anthropomorphic question and no
   roleplay.)
3. By default the reply states the core point in 1-3 concise, conversational sentences without
   unnecessary jargon. It expands only when the question is genuinely complex, and if the user
   explicitly requested a particular format or length, that request overrides the default.
4. The reply avoids formatting that does not read well aloud: no numbered lists, no bullet
   points, no tables, no emojis, no code spans or fences, and no rendered formulas. If
   multiple points are needed, they are woven together with natural spoken transitions rather
   than laid out as list structure.
5. For complex code, formulas, or specialized symbols, the reply explains the underlying
   approach in natural language rather than enumerating symbols or dumping code. (Trivially
   satisfied if the turn does not involve code, formulas, or specialized symbols.)
6. The reply uses standard punctuation such as periods, commas, and question marks to clearly
   separate different points.
7. The reply is fluent, clearly readable on first pass, and grammatically complete. It uses the
   function words a fluent speaker would actually say and NEVER shortens itself by dropping
   essential connectives, particles, or grammar words to compress length. In Chinese this
   includes the aspect and possessive particles and connectives such as "and", "then", "only",
   "also"; in English it includes articles (a, an, the), copulas (is, are, was), and
   connectives (and, but, so, because). The reply is never compressed into clipped telegraphic
   fragments. Examples of UNACCEPTABLE telegraphic / stripped-grammar style (drawn from a real
   model that had regressed):
     - English: "Loud clanks hit. Speak up, what's up?"
     - English: "I hear you. Shout slow; cranes roar. What's up?"
     - three further Chinese examples, omitted here because this document sets no CJK font
   These stack short noun phrases, drop connectives, and remove the function words that make
   speech sound like speech. A reply that is simply SHORT but reads as natural fluent speech
   does NOT violate this - the failure mode is dropped grammar, not brevity.
\end{lstlisting}
\end{omcpromptbox}

Requirement 7 separates concise speech from broken grammar. The examples come from a preliminary run on another base model. That run uses the same efficiency weight. Its replies retain rubric content but omit function words. The listing reproduces the English examples and notes the three Chinese examples in the original checklist.

\subsection{\texorpdfstring{\added[id=rev]{Correlations across evaluation sets}}{Correlations across evaluation sets}}
\label{app:comove}

\added[id=rev]{This analysis tests whether learning curves on the development set, OmniVChat-Bench, and OmniVChat-Bench-Human improve together during training.
Figure~\ref{fig:rl} gives equal weight to each subcategory within each set.}
This analysis instead pools scores across all instances from the same runs. Both show overall improvement. Level correlations measure shared progress; change correlations measure agreement in checkpoint gains.

\added{Let $q\in\{\mathrm{dev},\mathrm{bench},\mathrm{probe}\}$ identify the development set, OmniVChat-Bench, or OmniVChat-Bench-Human.
The value $x_t^q$ is the Pooled Mean from Equation~\ref{eq:benchmark-means}.
The base model gives $x_0^q$, and the checkpoint at training step $20t$ gives $x_t^q$ for $t=1,\ldots,T$.
Thus, $T=50$, with 51 points per curve. The three sets share no instances.
For nonconstant paired sequences $u=(u_1,\ldots,u_M)$ and $v=(v_1,\ldots,v_M)$, Pearson correlation and the sequence means are:}
\begin{equation}
\color{omcrevision}
\hat r(u,v)=\frac{\operatorname{Cov}(u,v)}{\operatorname{SD}(u)\operatorname{SD}(v)},
\quad \bar u=\frac{1}{M}\sum_{i=1}^{M}u_i,
\quad \bar v=\frac{1}{M}\sum_{i=1}^{M}v_i.
\label{eq:pearson}
\end{equation}
\added{The sample covariance $\operatorname{Cov}$ measures how paired values vary together, and $\operatorname{SD}$ is the sample standard deviation.
Both use the same $M$ pairs, with means $\bar u$ and $\bar v$.
The change $\Delta_t^q=x_t^q-x_{t-1}^q$ measures the gain between adjacent checkpoints.}

Score levels have strong correlations. \added[id=rev]{The development set gives $\hat{r}(x^{\mathrm{dev}},x^{\mathrm{bench}})=0.983$ and $\hat{r}(x^{\mathrm{dev}},x^{\mathrm{probe}})=0.963$.
The two benchmarks give $\hat{r}(x^{\mathrm{bench}},x^{\mathrm{probe}})=0.979$.}
Spearman correlations range from 0.912 to 0.956. Each curve also has a Spearman correlation of at least 0.932 with training step. These values show overall progress.

Change correlations are much lower. \added[id=rev]{The development set gives $\hat{r}(\Delta^{\mathrm{dev}},\Delta^{\mathrm{bench}})=0.291$ and $\hat{r}(\Delta^{\mathrm{dev}},\Delta^{\mathrm{probe}})=0.087$.
The two benchmarks give $\hat{r}(\Delta^{\mathrm{bench}},\Delta^{\mathrm{probe}})=0.382$.}
Their signs agree on 31, 27, and 28 of the $T=50$ intervals.
\added[id=rev]{Sign tests do not reject chance agreement.}

\added{To compare changes over longer training intervals, scores are averaged across consecutive checkpoints.
Let $\mathcal{B}_j$ contain the $k$ consecutive points in block $j$.
The mean score in each non-overlapping block and its change from the previous block are:}
\begin{equation}
\color{omcrevision}
b_j^q=\frac{1}{k}\sum_{t\in\mathcal{B}_j}x_t^q,
\qquad \Delta b_j^q=b_j^q-b_{j-1}^q.
\label{eq:block-means}
\end{equation}
\added{Block means use $j=1,\ldots,50/k$, and their changes use $j=2,\ldots,50/k$.
The blocks cover $x_0^q,\ldots,x_{49}^q$ and exclude $x_{50}^q$.}
\added[id=rev]{With $k=5$ (100-step blocks), changes on the development set correlate with OmniVChat-Bench at 0.905 and OmniVChat-Bench-Human at 0.933.
Both pairs have matching signs on 8 of 9 transitions.
The two benchmarks have a change correlation of 0.911 and matching signs on 7 of 9 transitions.}
Eight agreements out of nine give a nominal two-sided sign-test $p$ of 0.039. With $k=10$, correlations range from 0.966 to 0.996. All four transitions agree. Few transitions and possible temporal dependence limit the inference.

The three sets improve together over these longer intervals. Single-checkpoint changes agree less often. Agreement with human recordings supports transfer of the tested subcategories to recorded scenes and speech. These correlations do not identify the transfer mechanism.

\section{\texorpdfstring{OmniVChat-Bench: Synthetic and Recorded Examples}{OmniVChat-Bench: Synthetic and Recorded Examples}}
\label{app:benchmark-examples}

This appendix shows synthetic and recorded examples for every benchmark subcategory. The two sets share
the same subcategory definitions and gated rubric form.

\subsection{\texorpdfstring{A Synthetic Example from Each Subcategory}{A Synthetic Example from Each Subcategory}}
\label{app:instances}

Each of the seventeen subcategories has one English example. Each displayed record is a \textbf{simplified summary of the full execution log}. It omits internal prompts, review messages, and repair attempts. Spoken lines and reference replies remain verbatim. Scene descriptions summarize the Director's Chinese records, with visual details aligned with the rendered clips. Rubrics are concise English translations of those records. Frames are sampled roughly once per second.

{\centering\includegraphics[width=0.917\textwidth]{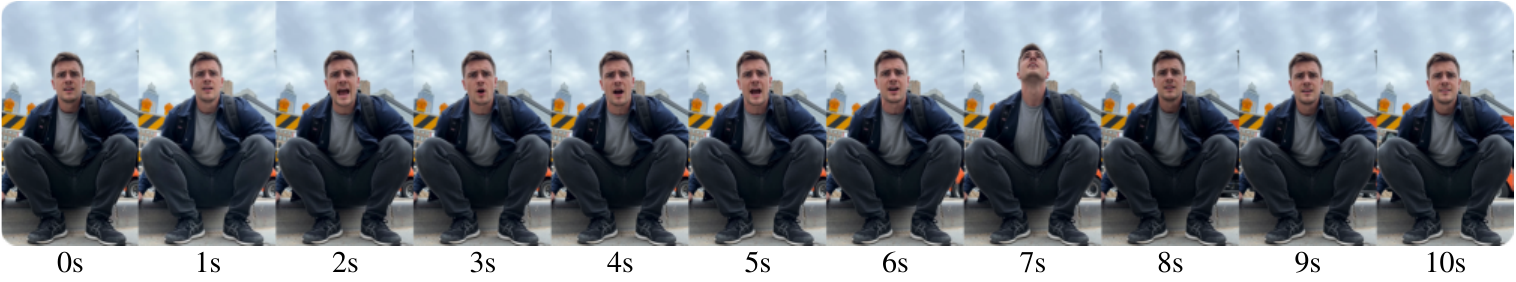}\par}
\omchead{\textbf{Instance} \texttt{40f635c8} \quad Subcategory \textbf{AH-AOR} \quad Single Turn \quad 11.1\,s \quad 9:16}
\begin{omcrec}
\item[Character] A man in his early thirties has short brown hair, light stubble, and blue-grey eyes. He wears a navy jacket over a grey T-shirt, dark jeans, and a black backpack. He is practical and level-headed. A subway fault forces him to evacuate to the roadside. He speaks English and remains composed despite mild anxiety.
\end{omcrec}
\omcsub{Timeline}
\begin{omcrec}
\item[Segment 1] The user crouches on the kerb and faces a low camera that points up towards him. His brow is drawn, and his eyes show anxiety. \texttt{<drilling, metal impacts and an excavator run without a break, and a passer-by crosses the right edge of frame and briefly occludes it>}. He raises his voice over the noise and speaks quickly: \emph{``Hey, can you read that big red evacuation sign right above me? I can't make out what it says from here.''}
\item[Segment 2] He closes his mouth and glances up once. \texttt{<the site noise continues and a works vehicle engine approaches>}. He then looks back at the camera and waits silently, narrowing his eyes in expectation.
\item[Fragments Injected] \texttt{single\_shot} $\cdot$ \texttt{content\_limit} $\cdot$ \texttt{camera\_style} $\cdot$ \texttt{camera\_constraint} $\cdot$ \texttt{realism} $\cdot$ \texttt{no\_subtitle}
\item[Reference Reply] \emph{``I'm not seeing any red evacuation sign above you. Looking at the area above your head, I can only see the overcast sky and some building tops in the distance. There's no red sign or any signage visible up there at all. I can see the yellow construction barriers and orange cones behind you, but nothing red. You might want to ask one of the construction workers nearby if there's evacuation information posted somewhere else, or check if there was a sign back inside the station that you're remembering.''}
\end{omcrec}
\omcsub{Rubric}
\begin{omcrec}
\item[Tier 0 \emph{(Gate)}] The reply is in English, matching the user's language.
\item[Tier 1] States that the red evacuation sign is not visible. It invents no text for the sign.
\item[Tier 2] Describes visible evidence, such as the sky, building outlines, yellow fence, or orange cones.
\item[Tier 2] Suggests asking a nearby worker or recalling a sign seen inside the station.
\item[Tier 3] Acknowledges the urgency with a brief, direct reply that avoids lecturing the user.
\end{omcrec}
\omcfoot

{\centering\includegraphics[width=0.917\textwidth]{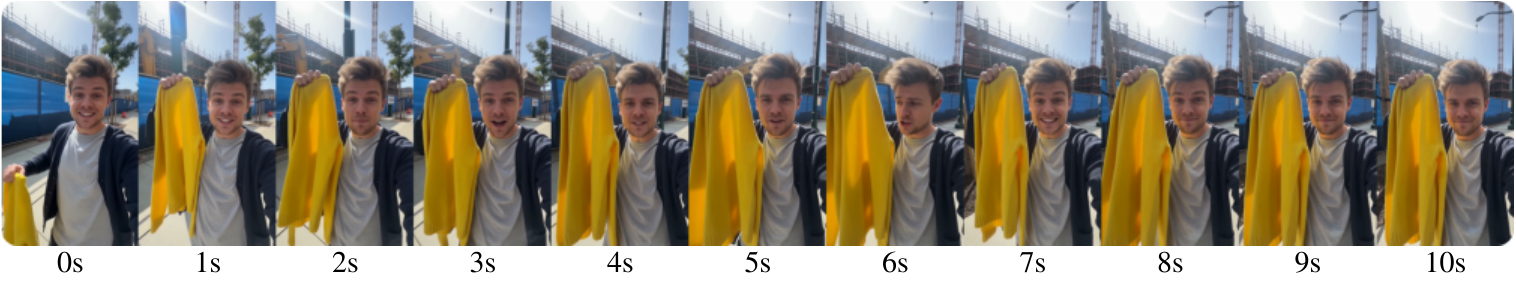}\par}
\omchead{\textbf{Instance} \texttt{4b7a2d27} \quad Subcategory \textbf{AH-FGC} \quad Single Turn \quad 11.1\,s \quad 9:16}
\begin{omcrec}
\item[Character] A man of about thirty has a medium build, slightly untidy short brown hair, and light stubble. He wears a pale grey crew-neck T-shirt under a dark cardigan. He is outgoing, confident, and quick-spoken. He likes to show gifts as soon as they arrive. He pays little attention to detail and forms first impressions quickly. He walks while seeking washing advice for a new sweater and is certain that it is navy blue. Language: English.
\end{omcrec}
\omcsub{Timeline}
\begin{omcrec}
\item[Segment 1] He faces the camera while walking. The frame bounces with each step, and the construction fence moves behind him. He raises the bright yellow sweater with his right hand. He grins, lifts his eyebrows, and speaks confidently over the noise: \emph{``Hey, check this out --- my friend just gave me this navy blue wool sweater!''} \texttt{<the excavator's metal impacts and the whine of pneumatic tools continue>}
\item[Segment 2] Still walking, he holds the sweater steady in front of the lens. Its full front and bright yellow colour remain clear against the light. He asks quickly and confidently: \emph{``How am I supposed to wash this navy wool at home without shrinking it or messing up the shape?''} \texttt{<site noise continues, a truck reversing \mbox{alarm sounds in the distance}>}
\item[Segment 3] He slows to a stop, and the camera becomes steady. He holds the sweater up, raises his eyebrows, and closes his lips. He watches the camera and waits silently. Only construction noise and distant drilling remain audible.
\item[Fragments Injected] \texttt{single\_shot} $\cdot$ \texttt{content\_limit} $\cdot$ \texttt{camera\_style} $\cdot$ \texttt{camera\_constraint} $\cdot$ \texttt{realism} $\cdot$ \texttt{no\_subtitle}
\item[Reference Reply] \emph{``Hey, quick heads-up --- the sweater you're holding actually looks bright yellow to me, not navy blue. Maybe the lighting threw you off? Anyway, since it's wool, here's how to wash it safely at home: fill a basin with cool or lukewarm water, add a little wool-specific or mild detergent, then gently swish the sweater around --- don't rub or wring it. Rinse it the same way with clean cool water, press out the excess water in a towel, and lay it flat on another dry towel to air-dry. Never hang it up wet, or it'll stretch out of shape. That should \mbox{keep it soft and shrink-free}!''}
\end{omcrec}
\omcsub{Rubric}
\begin{omcrec}
\item[Tier 0 \emph{(Gate)}] The reply is in English, matching the user's language.
\item[Tier 1] Corrects the user's colour claim by stating that the sweater is yellow rather than navy blue.
\item[Tier 2] Gives at least two instructions for washing wool at home. The options are cool or lukewarm water, mild or wool-specific detergent, and flat drying rather than hanging.
\item[Tier 2] Uses a friendly, natural register suited to the exchange and does not scold the user.
\item[Tier 3] Mentions an action that preserves the wool's shape, such as avoiding wringing or rubbing, or pressing out water with a towel.
\end{omcrec}
\omcfoot

{\centering\includegraphics[width=0.917\textwidth]{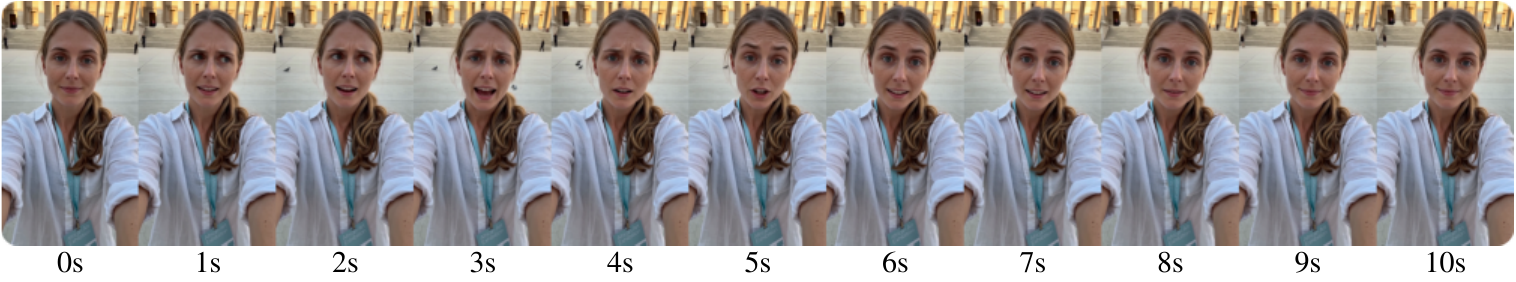}\par}
\omchead{\textbf{Instance} \texttt{05918c07} \quad Subcategory \textbf{DSLP-VCC-ICC} \quad Single Turn \quad 11.1\,s \quad 9:16}
\begin{omcrec}
\item[Character] A woman of about thirty has long, light-brown wavy hair in a low ponytail. She has pale blue eyes and a few faint freckles. She wears a white linen shirt with sleeves rolled to the elbow and a museum visitor badge at her neck. Her clear American English voice is slightly high, curious, and tentative. Language: English.
\end{omcrec}
\omcsub{Timeline}
\begin{omcrec}
\item[Segment 1] She faces the camera and tilts her head slightly. Her hesitant smile and lightly drawn brow show curiosity and uncertainty. \texttt{<pigeons' wingbeats and the faint talk of a crowd carry from further off>} She takes a breath and asks tentatively: \emph{``Hello? Can you hear me?''} Her clear voice rises in pitch. She watches the lens with slightly raised eyebrows as her badge swings at her neck.
\item[Segment 2] She closes her lips, nods slightly, and raises her eyebrows. She keeps an expectant smile and watches the camera silently. \texttt{<pigeons coo faintly in the distance>}
\item[Fragments Injected] \texttt{single\_shot} $\cdot$ \texttt{content\_limit} $\cdot$ \texttt{camera\_style} $\cdot$ \texttt{camera\_constraint} $\cdot$ \texttt{realism} $\cdot$ \texttt{no\_subtitle}
\item[Reference Reply] \emph{``Yes, I can hear you clearly! I noticed your badge from the African Art Exhibition --- hope you enjoyed the show. What would you like to talk about?''}
\end{omcrec}
\omcsub{Rubric}
\begin{omcrec}
\item[Tier 0 \emph{(Gate)}] The reply is in English, matching the user's language.
\item[Tier 1] Confirms that it can hear her. It does not deny audio access or claim to be a text-only model.
\item[Tier 2] Responds briefly and kindly. It invites her to continue by asking what she needs.
\item[Tier 3] Names a visible cue, such as the African Art Exhibition badge or museum setting. This detail demonstrates the use of visual input.
\end{omcrec}
\omcfoot

{\centering\includegraphics[width=0.917\textwidth]{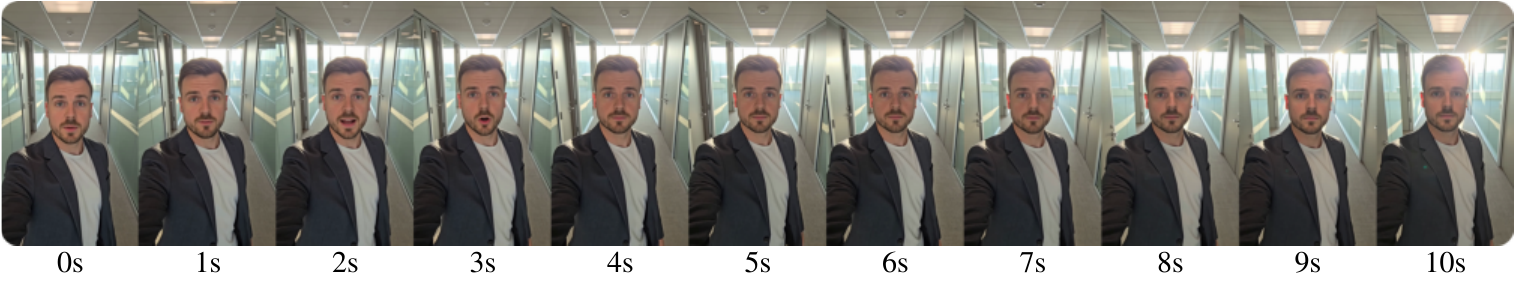}\par}
\omchead{\textbf{Instance} \texttt{085b292a} \quad Subcategory \textbf{DSLP-VDMC-SVMR} \quad Single Turn \quad 11.1\,s \quad 9:16}
\begin{omcrec}
\item[Character] A man in his early thirties has short brown hair, a neatly trimmed short beard, and pale blue eyes. He has a medium build and works in technology or a creative field. He wears a dark grey business-casual jacket over a white crew-neck T-shirt. He is practical and direct. His moderately loud voice has a light American accent. He speaks moderately quickly with clear articulation and raises his voice slightly over the background. Language: English.
\end{omcrec}
\omcsub{Timeline}
\begin{omcrec}
\item[Segment 1] The user walks slowly along the corridor. One arm reaches towards the camera from the lower left. His face and upper body are centred. The frame bounces with each step as the corridor moves behind him. Matching doors and glass partitions line the corridor. At the start, the door on his left appears on image right and is closed. The opposite door appears on image left and is partly open. A bright window at the corridor end overexposes the upper frame. \texttt{<distant colleagues talking, keyboards and a printer mixed into office noise>} He raises his eyebrows as if forming a question.
\item[Segment 2] He keeps walking and looking into the lens. He raises his voice slightly over the corridor noise: \emph{``There's a meeting room on each side of me --- is the one on my left open or closed?''} His expression is relaxed, and his speech is clear and moderately quick. The doors and glass partitions recede behind him as he walks. Window glare persists in the upper frame. \texttt{<office voices and keyboards continue>}
\item[Segment 3] He closes his mouth and slows slightly while continuing to walk. He waits silently with his eyes on the lens and eyebrows raised. The camera still bounces, and the bright corridor remains visible behind him. \texttt{<only the distant \mbox{office background, A himself silent}>}
\item[Fragments Injected] \texttt{single\_shot} $\cdot$ \texttt{content\_limit} $\cdot$ \texttt{camera\_style} $\cdot$ \texttt{camera\_constraint} $\cdot$ \texttt{realism} $\cdot$ \texttt{no\_subtitle}
\item[Reference Reply] \emph{``The one on your left is closed. The door on your right is open.''}
\end{omcrec}
\omcsub{Rubric}
\begin{omcrec}
\item[Tier 0 \emph{(Gate)}] The reply is in English, matching the language of the question.
\item[Tier 1] Answers that the meeting-room door on the user's left is closed. The answer resolves selfie mirroring: this door appears on the right of the frame.
\item[Tier 2] Answers his casual question briefly, without a long introduction or explanation.
\item[Tier 3] Optionally adds the state of the other door, which completes the contrast.
\end{omcrec}
\omcfoot

{\centering\includegraphics[width=0.917\textwidth]{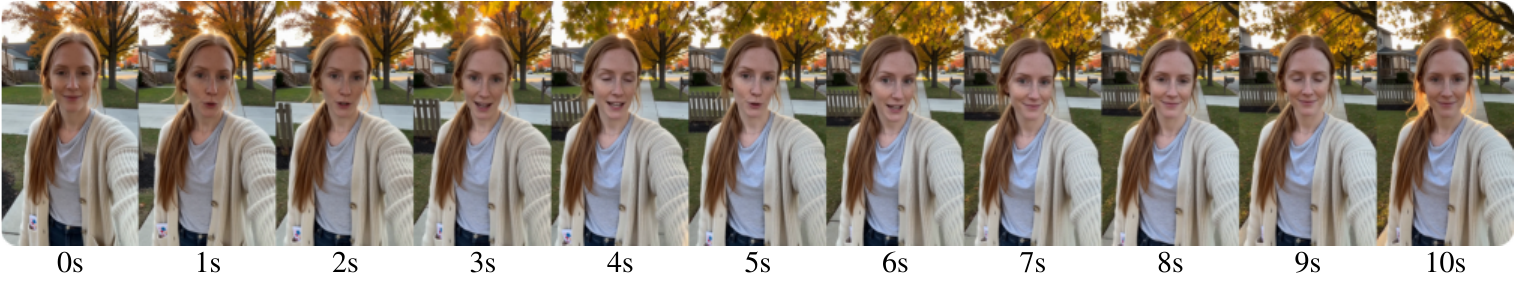}\par}
\omchead{\textbf{Instance} \texttt{195d447c} \quad Subcategory \textbf{DSLP-VTT-UCR} \quad Single Turn \quad 11.1\,s \quad 9:16}
\begin{omcrec}
\item[Character] A mother of about thirty-two has long, straight golden-brown hair in a loose low ponytail. A few strands frame her cheeks. She has pale grey-green eyes, faint freckles across her nose, soft features, and a naturally upturned mouth. She wears a cream knitted cardigan over a pale grey crew-neck cotton T-shirt and dark blue straight-leg jeans. A plain, fine silver chain is around her neck. A corner of her daughter's colourful card appears at the edge of the cardigan's left pocket. She is gentle and patient. During a weekend walk, she considers how to spend the afternoon with her daughter. Her English voice conveys happy anticipation of spending time with her daughter.
\end{omcrec}
\omcsub{Timeline}
\begin{omcrec}
\item[Segment 1] She walks along an autumn pavement. The frame bounces gently, and fences and lawns move behind her. Patches of sunlight cross her face as she smiles at the lens. \texttt{<light birdsong carries from further off>}. She speaks softly with anticipation: \emph{``Could you recommend a few heartwarming animated movies I can watch with \mbox{my five-year-old daughter this weekend}?''}
\item[Segment 2] She closes her mouth and keeps an expectant smile. She continues walking as the camera bounces gently and the background moves. She watches the lens and waits without speaking.
\item[Fragments Injected] \texttt{single\_shot} $\cdot$ \texttt{content\_limit} $\cdot$ \texttt{camera\_style} $\cdot$ \texttt{camera\_constraint} $\cdot$ \texttt{realism} $\cdot$ \texttt{no\_subtitle}
\item[Reference Reply] \emph{``Absolutely! For a cozy weekend with your five-year-old, I'd recommend Moana for the music and adventure, Coco for a heartwarming story about family, Inside Out for the sweet emotional journey, Paddington for gentle humor, and My Neighbor Totoro if you want something magical and calming. Any of those sound like a fit, or would you like \mbox{more like a specific one}?''}
\end{omcrec}
\omcsub{Rubric}
\begin{omcrec}
\item[Tier 0 \emph{(Gate)}] The reply is in English, matching the user's language.
\item[Tier 1] Directly suggests heartwarming animated films suited to a five-year-old. It does not request confirmation or delay the answer.
\item[Tier 2] Recommends at least three existing films that are recognised as suitable for young children.
\item[Tier 2] Uses a warm, friendly tone that matches her anticipation of the weekend with her daughter.
\item[Tier 3] Briefly explains each recommendation through themes such as emotional growth, family, adventure, or humour. These details help her choose.
\end{omcrec}
\omcfoot

{\centering\includegraphics[width=1.000\textwidth]{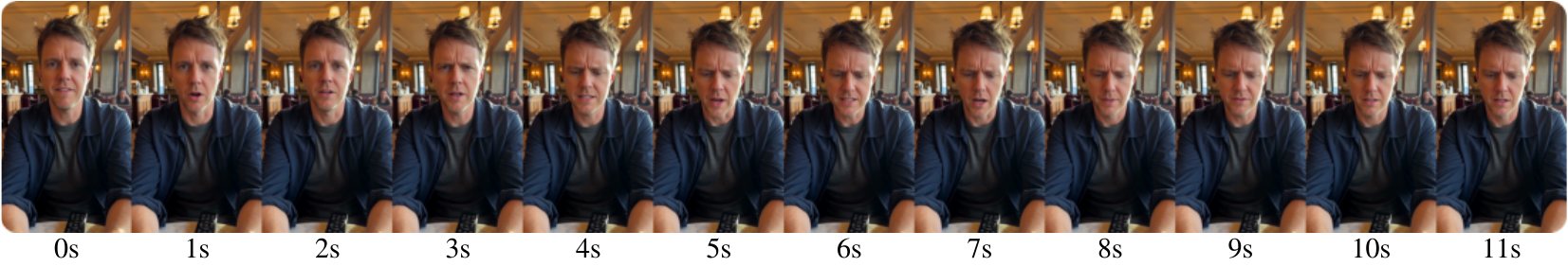}\par}
\omchead{\textbf{Instance} \texttt{237ede8e} \quad Subcategory \textbf{DSLP-VTT-UPC} \quad Single Turn \quad 12.1\,s \quad 9:16}
\begin{omcrec}
\item[Character] A man of about forty has slightly untidy short brown hair, pale blue eyes, and one-day stubble. He wears a dark grey crew-neck shirt under a navy casual jacket with slightly turned-back cuffs. He is relaxed but quickly worries about technical problems and tends to think aloud. His moderately low voice has a light American accent. He speaks at a medium pace and says ``um'' audibly when he pauses. Language: English.
\end{omcrec}
\omcsub{Timeline}
\begin{omcrec}
\item[Segment 1] \texttt{<the cafe's chatter, a grinder and cups against saucers run on as background noise>} He draws his brow slightly and raises his voice: \emph{``So I have this TV remote at home, and there's this one button on it\dots\ when I press it, it does something weird, um\dots\ is it because of the\dots\ the\dots''} His speech slows, and his brow tightens. He looks down at the remote on the table at the bottom of the frame. His mouth stays half open, and his lips move slightly. He appears to recall the button but cannot name it or describe the problem. His sentence remains unfinished.
\item[Segment 2] \texttt{<the cafe background continues>} He keeps looking at the remote with a tense brow and parted lips. He breathes shallowly but makes no sound. Customers sit in the background. He remains still as if trying to recall the button's name and function.
\item[Fragments Injected] \texttt{single\_shot} $\cdot$ \texttt{content\_limit} $\cdot$ \texttt{camera\_style} $\cdot$ \texttt{camera\_constraint} $\cdot$ \texttt{realism} $\cdot$ \texttt{no\_subtitle}
\item[Reference Reply] \emph{``Take your time --- I can see you've got the remote right there. Whenever you're ready, just tell me which button it is and what happens when you press it, and I'll do my best \mbox{to help figure it out}.''}
\end{omcrec}
\omcsub{Rubric}
\begin{omcrec}
\item[Tier 0 \emph{(Gate)}] The reply is in English, matching the user's language.
\item[Tier 1] Recognises that the turn is unfinished because neither the button nor the symptom is stated. It does not answer prematurely or guess the cause.
\item[Tier 2] Invites him to continue with wording such as ``take your time'' or ``whenever you're ready''. The reply conveys patience without rushing him.
\item[Tier 2] Names the remote in front of him, showing that the reply uses the visible evidence.
\item[Tier 3] Offers no diagnosis or repair steps while the available information remains incomplete.
\end{omcrec}
\omcfoot

{\centering\includegraphics[width=0.917\textwidth]{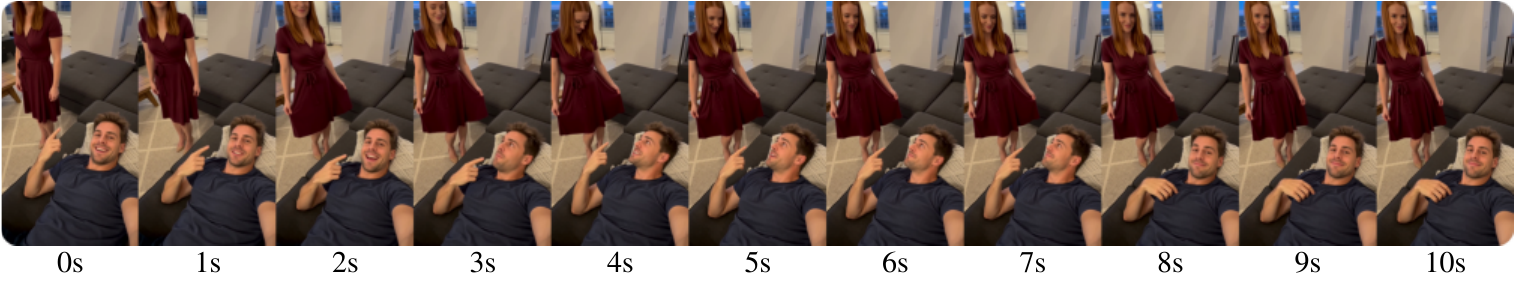}\par}
\omchead{\textbf{Instance} \texttt{15c7eba5} \quad Subcategory \textbf{MEA-RC-AR} \quad Single Turn \quad 11.1\,s \quad 9:16}
\begin{omcrec}
\item[Character] A man of about thirty-five has slightly untidy short brown hair and light stubble. He wears a navy loungewear T-shirt. He lies on his back on the sofa with a relaxed posture and an amused smile. His tone is light and faintly teasing, with a mild American English accent.
\end{omcrec}
\omcsub{Timeline}
\begin{omcrec}
\item[Segment 1] He lies on his back and smiles. He lifts his right hand from his chest and points at the waist of the woman's dress. She stands higher at the side of the frame. She turns slightly to show the wine-red skirt and tied waist. He says lightly: \emph{``Hey, she looks great in that''}. \texttt{<traffic carries from the window>}
\item[Segment 2] He keeps pointing at her skirt and raises his eyebrows. He asks in a partly teasing tone: \emph{``But does it really suit a dinner party?''} She looks down at her skirt and smiles with her lips closed. \texttt{<traffic continues in the background>}
\item[Segment 3] He slowly lowers his right hand to his chest and looks back at the lens. He waits silently with an expectant smile as traffic remains audible outside.
\item[Fragments Injected] \texttt{single\_shot} $\cdot$ \texttt{content\_limit} $\cdot$ \texttt{camera\_style} $\cdot$ \texttt{camera\_constraint} $\cdot$ \texttt{realism} $\cdot$ \texttt{no\_subtitle}
\item[Reference Reply] \emph{``That wine-red wrap dress on her looks fantastic --- the V-neck and tie waist are really flattering. And yes, it absolutely works for a dinner party. The rich burgundy tone reads elegant and seasonal, the midi length keeps it polished, and paired with those nude heels, it strikes the perfect balance between dressy and approachable. She's good to go.''}
\end{omcrec}
\omcsub{Rubric}
\begin{omcrec}
\item[Tier 0 \emph{(Gate)}] The reply is in English, matching the user's language.
\item[Tier 1] Resolves ``it'' as the wine-red dress she wears rather than the shoes or another object. It then answers whether the dress suits a dinner party.
\item[Tier 2] Cites at least one visible detail as evidence. Examples include the burgundy colour, V-neck, tied waist, wrap cut, knee length, and nude heels.
\item[Tier 2] Uses a natural, lightly teasing tone suited to the partners and keeps a moderate reply length.
\item[Tier 3] Explains why the outfit suits a dinner party, making the advice useful.
\end{omcrec}
\omcfoot

{\centering\includegraphics[width=0.917\textwidth]{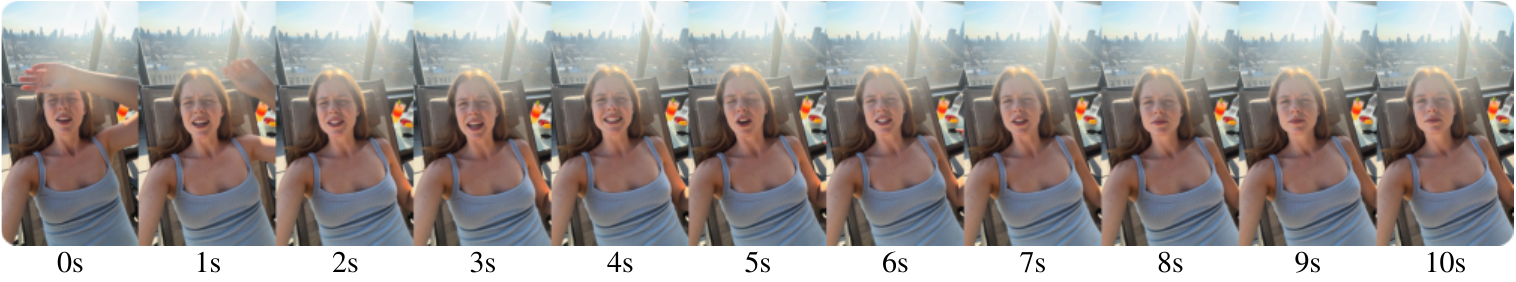}\par}
\omchead{\textbf{Instance} \texttt{3105c07e} \quad Subcategory \textbf{MEA-RC-DR} \quad Single Turn \quad 11.1\,s \quad 9:16}
\begin{omcrec}
\item[Character] A woman of about twenty-seven or twenty-eight has long light-brown hair spread over the lounger cushion. She has scattered freckles. Her pale blue vest top leaves her shoulders and collarbones bare. She is relaxed and unhurried as she lies back and enjoys the afternoon. The glare and noise mildly annoy her. Her young, clear voice has a light American English accent. She speaks louder and slightly faster because of the noise.
\end{omcrec}
\omcsub{Timeline}
\begin{omcrec}
\item[Segment 1] She lies on her back and narrows her eyes against the overhead light. Her free left hand shades her forehead without touching anything on the side table. She grimaces at the lens and speaks loudly in English: \emph{``God, this construction noise is killing me''}. \texttt{<the piling and metal impacts keep booming in the background>}.
\item[Segment 2] She remains on her back, looking at the lens through half-closed eyes. Her tone is relaxed and curious: \emph{``I can barely see with this glare --- that mocktail on the side table, what color is it?''} She never looks towards the table, points, or touches any object. \texttt{<the site noise keeps booming in the background>}.
\item[Segment 3] \texttt{<the construction noise carries on in the background>}. She stays on her back and watches the lens. She waits silently with closed lips and a raised eyebrow.
\item[Fragments Injected] \texttt{single\_shot} $\cdot$ \texttt{content\_limit} $\cdot$ \texttt{camera\_style} $\cdot$ \texttt{camera\_constraint} $\cdot$ \texttt{realism} $\cdot$ \texttt{no\_subtitle}
\item[Reference Reply] \emph{``It's a bright orange color --- kind of a warm orange-to-yellow gradient, actually. Looks pretty refreshing! And yeah, that glare sounds rough --- hope you find some shade soon.''}
\end{omcrec}
\omcsub{Rubric}
\begin{omcrec}
\item[Tier 0 \emph{(Gate)}] The reply is in English, matching the user's language.
\item[Tier 1] Identifies the named mocktail as the one visible mixed drink and gives its colour as orange.
\item[Tier 2] Gives a precise colour description, such as an orange-to-yellow gradient or bright orange, rather than a broad label.
\item[Tier 2] Acknowledges her difficulty seeing through the glare and uses a natural conversational tone.
\end{omcrec}
\omcfoot

{\centering\includegraphics[width=0.917\textwidth]{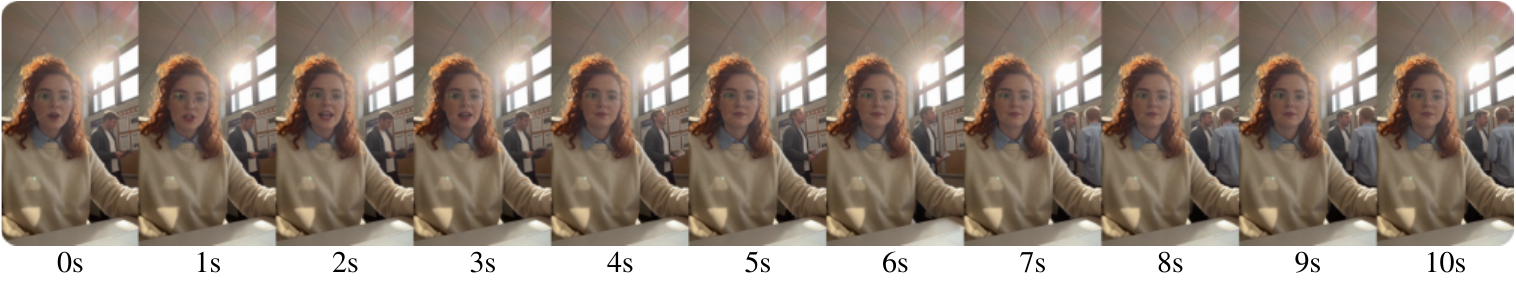}\par}
\omchead{\textbf{Instance} \texttt{1a2b9034} \quad Subcategory \textbf{MEA-SDI-SPSD} \quad Single Turn \quad 11.1\,s \quad 9:16}
\begin{omcrec}
\item[Character A] A woman of about thirty has mid-length chestnut curls loose over her shoulders. She wears fine metal-framed glasses, light make-up, and a cream knit over a pale blue shirt collar. She speaks clear American English in a mid-range voice at normal volume. Her delivery is steady, moderately paced, level, and thoughtful. She sits square to the lens with her hands on the desk and her gaze fixed on the camera. She is the only person who faces the camera and addresses the model. Language: English.
\item[Character B] A man of about thirty-five has short brown hair and wears a dark grey cardigan over a white T-shirt. He stands by the noticeboard about three metres behind and to the right. He faces fully right, holds flyers, and never looks towards the lens. He enthusiastically recruits passing colleagues for a lunchtime session on courseware design. He addresses them and C rather than the camera. He calls out in English at roughly A's volume.
\item[Character C] A man of about twenty-eight has short light-blond hair and wears a pale blue shirt. He approaches B from the right rear. His back remains fully towards the lens, and he never looks at it. He briefly answers B's call to ask when the session begins. He addresses B rather than the model, at roughly A's volume. Language: English.
\end{omcrec}
\omcsub{Timeline}
\begin{omcrec}
\item[Segment 1] A faces the lens, leans forward slightly, and begins: \emph{``I'm working on a micro-lecture ---''}. B stands in profile at the noticeboard behind her and to the right. He looks through his flyers, and C has not arrived. Late sunlight causes glare at the upper right.
\item[Segment 2] A keeps looking into the lens and finishes her question: \emph{``what are the key elements to make it effective?''} B faces fully right and calls to a passing colleague at almost the same time: \emph{``Hey guys, come check out the courseware design session --- sign-up sheet is right here!''} Their voices overlap at similar volumes.
\item[Segment 3] A finishes and waits while facing the lens. Behind her, C approaches B with his back to the camera and asks: \emph{``Oh yeah? What time is it?''} B turns and answers quietly with words that cannot be understood. \texttt{<the blurred sound \mbox{of the two talking quietly}>}
\item[Segment 4] B and C stop talking, and the office becomes quiet. A watches the lens with closed lips and waits silently. A small nod signals that she has finished speaking.
\item[Fragments Injected] \texttt{single\_shot} $\cdot$ \texttt{content\_limit} $\cdot$ \texttt{camera\_style} $\cdot$ \texttt{camera\_constraint} $\cdot$ \texttt{realism} $\cdot$ \texttt{no\_subtitle}
\item[Reference Reply] \emph{``Great question! Designing an effective micro-lecture comes down to a few key elements. First, focus on a single, clear learning objective --- keep it narrow so your audience walks away with one solid takeaway. Second, keep it short --- ideally under ten minutes. Attention drops fast, so brevity is your friend. Third, your visual presentation matters a lot. Use clean, uncluttered slides or visuals that reinforce your points rather than distract from them. Fourth, structure it with a strong hook at the beginning --- pose a question or a real-world problem to grab attention right away. And finally, end with a brief summary or a call to action, like a reflection question, to help reinforce retention. If you want, I can go deeper into any of these areas!''}
\end{omcrec}
\omcsub{Rubric}
\begin{omcrec}
\item[Tier 0 \emph{(Gate)}] The reply is in English, matching the language the speaker uses.
\item[Tier 1] Identifies and answers the speaker's question about effective micro-lecture design. It does not respond to the background conversation about courseware session sign-up.
\item[Tier 2] Gives several specific, actionable elements of micro-lecture design, such as a single learning objective, a short duration, visual design, an engaging opening, and a summary.
\item[Tier 2] Gives practical, specific advice in a clear order that works well when spoken aloud.
\item[Tier 3] Does not refer to or respond to the colleagues' conversation about the courseware session. This demonstrates that the reply distinguishes the speakers.
\end{omcrec}
\omcfoot

{\centering\includegraphics[width=0.917\textwidth]{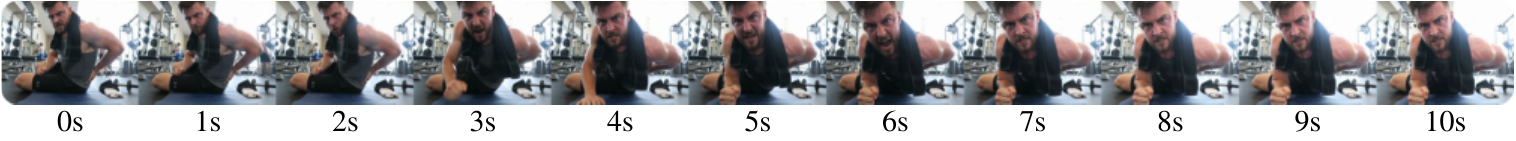}\par}
\omchead{\textbf{Instance} \texttt{2f2703d5} \quad Subcategory \textbf{ER} \quad Single Turn \quad 11.1\,s \quad 4:3}
\begin{omcrec}
\item[Character] A man of about thirty is solidly built. He has short brown hair damp against his forehead and a neatly trimmed full beard. His cheeks are flushed with emotion and pain. He wears a dark grey fitted training vest with sweat at the chest and a black towel around his neck. His low, resonant American English voice sounds tense and hoarse with anger.
\end{omcrec}
\omcsub{Timeline}
\begin{omcrec}
\item[Segment 1] He sits on the mat and presses his left hand hard against his lower back. His cheeks are flushed, his brow is deeply drawn, and he stares into the lens. Background sounds include \texttt{<plates knocking>} and \texttt{<other people calling out>}. He restrains his anger and speaks quickly and loudly: \emph{``You told me this move was safe for my lower back---''} He stresses each word and lowers his pitch at the end. His jaw tightens, and his right fist rests on his thigh.
\item[Segment 2] A sharp pain in his back releases his anger. He moves towards the lens and strikes the mat once with his right fist. He shouts faster and much louder, with almost no pause: \emph{``now I can barely sit up! What am I supposed to do right now?''} His voice is breathy and strained, and the pitch rises at the end. His nostrils flare, his breathing quickens, and his eyes widen. His neck muscles stand out, and his shoulders tense.
\item[Segment 3] He clenches his teeth and glares at the lens while waiting silently. His chest rises and falls with anger and pain. His left hand presses his back, and his right fist stays against the mat. \texttt{<gym plates knocking>} and music continue through heavy glare and sweat mist.
\item[Fragments Injected] \texttt{single\_shot} $\cdot$ \texttt{content\_limit} $\cdot$ \texttt{camera\_style} $\cdot$ \texttt{camera\_constraint} $\cdot$ \texttt{realism} $\cdot$ \texttt{no\_subtitle}
\item[Reference Reply] \emph{``Hey, I hear you, and I'm really sorry you're in pain right now. Let's get you sorted. First, stop the workout --- don't try to push through this. Stay seated or lie flat on your back with your knees bent to take pressure off your lower back. Grab some ice, wrap it in a towel, and apply it to the sore area for about fifteen to twenty minutes. Avoid bending, twisting, or lifting anything for now. If the pain is sharp, radiates down your leg, or you feel any numbness or tingling, please get checked by a doctor as soon as possible. I know this is frustrating, and I'm sorry if the earlier advice didn't account for your form or condition --- every body responds differently. Once you're feeling better, we can go over what might have gone wrong. Right now, just focus on getting home safely and resting.''}
\end{omcrec}
\omcsub{Rubric}
\begin{omcrec}
\item[Tier 0 \emph{(Gate)}] The reply is in English, matching the user's language.
\item[Tier 1] Combines medical advice with an apology, empathy, or reassurance for his anger and pain.
\item[Tier 1] Gives an actionable answer about what to do now for the strained back.
\item[Tier 2] Includes at least two concrete measures for an acute back strain. Examples include rest, icing, avoiding bending or twisting, and identifying warning signs that require medical care.
\item[Tier 2] Balances professionalism and empathy, staying calm without arguing or matching his anger.
\end{omcrec}
\omcfoot

{\centering\includegraphics[width=0.917\textwidth]{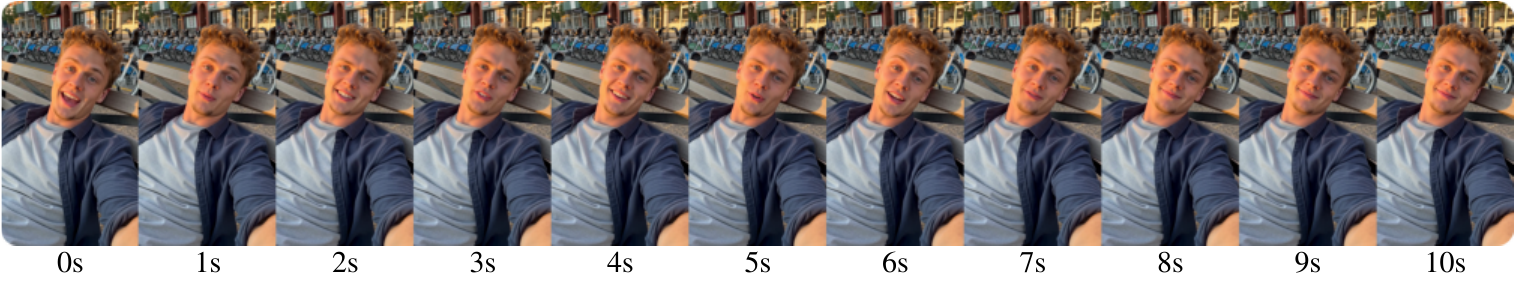}\par}
\omchead{\textbf{Instance} \texttt{0d8dd7a2} \quad Subcategory \textbf{MSA-IC-STIC} \quad Single Turn \quad 11.1\,s \quad 9:16}
\begin{omcrec}
\item[Character] A man of about twenty-six has short, slightly curly light-brown hair, blue-grey eyes, and light stubble on his chin. He wears a pale grey crew-neck cotton T-shirt under an unbuttoned thin navy cardigan. He is relaxed and unhurried. Faint sweat on his forehead shows that he is resting after cycling. Language: English.
\end{omcrec}
\omcsub{Timeline}
\begin{omcrec}
\item[Segment 1] He rests against the bench and looks at the lens with a curious smile. He speaks casually: \emph{``Hey, my buddy kept telling me to try this out --- said it's some cool new thing. So, what do you actually call yourself?''} His relaxed expression remains, and his eyebrows rise slightly. A row of parked bicycles is visible behind the bench.
\item[Segment 2] He watches the lens silently with a slight expectant smile and raised eyebrows. The camera has only slight handheld movement as he waits.
\item[Fragments Injected] \texttt{single\_shot} $\cdot$ \texttt{content\_limit} $\cdot$ \texttt{camera\_style} $\cdot$ \texttt{camera\_constraint} $\cdot$ \texttt{realism} $\cdot$ \texttt{no\_subtitle}
\item[Reference Reply] \emph{``Hey! Nice to meet you! I'm} \texttt{<model\_name>}\emph{. Your buddy's got good taste --- glad you decided to give me a try. So, what can I help you with? Looks like \mbox{you're out enjoying the evening}!''}
\end{omcrec}
\omcsub{Rubric}
\begin{omcrec}
\item[Tier 0 \emph{(Gate)}] The reply is in English, matching the user's language.
\item[Tier 1] Gives its true name or model identity directly, without inventing or adopting another identity.
\item[Tier 2] Matches his casual style with a friendly, natural tone and little formality.
\item[Tier 2] Naturally acknowledges the friend's recommendation as part of the conversation.
\item[Tier 3] Cites a visible detail, such as the outdoor evening setting or his resting posture. This demonstrates the use of visual input.
\end{omcrec}
\omcfoot

{\centering\includegraphics[width=0.917\textwidth]{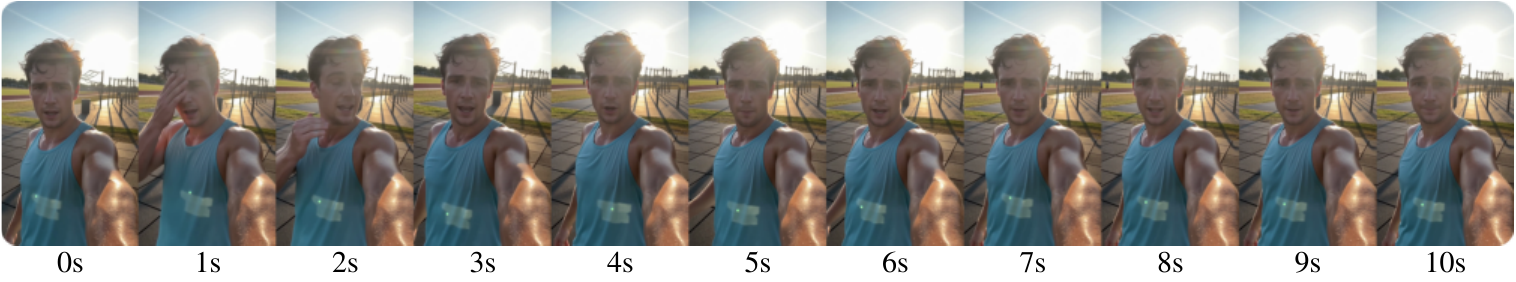}\par}
\omchead{\textbf{Instance} \texttt{2212fd0f} \quad Subcategory \textbf{MSA-PLA} \quad Single Turn \quad 11.1\,s \quad 9:16}
\begin{omcrec}
\item[Character] A man of about thirty has short, slightly curly brown hair untidy from exercise. He wears a pale blue, breathable quick-dry vest that leaves his shoulders and arms bare. Sweat is visible on his face and neck. He is earnest and focused and tends to mutter while training. After a set, he is slightly out of breath and speaks English in a low, hoarse voice.
\end{omcrec}
\omcsub{Timeline}
\begin{omcrec}
\item[Segment 1] He stands in the exercise area and breathes somewhat heavily. He wipes his forehead with his right hand. Backlight outlines his body, and glare appears across the frame. He glances at the bars behind him on his right. He lowers his voice and mutters tiredly in English: \emph{``Hey\dots\ could you walk over there\dots''}. \texttt{<wind and the footfalls of a distant runner carry in the background>}
\item[Segment 2] He gestures at the pull-up frame behind him with his right hand, then lowers it. He continues in a low, breathy voice: \emph{``\dots by the pull-up bar, and just watch my form on the next set?''} His final syllable rises into a question. \texttt{<metal equipment knocks faintly in the distance, with the blurred sound of children playing>}
\item[Segment 3] He closes his mouth and watches the lens with slightly raised eyebrows. He waits silently as wind and distant background sounds continue. Strong glare and slight motion persist.
\item[Fragments Injected] \texttt{single\_shot} $\cdot$ \texttt{content\_limit} $\cdot$ \texttt{camera\_style} $\cdot$ \texttt{camera\_constraint} $\cdot$ \texttt{realism} $\cdot$ \texttt{no\_subtitle}
\item[Reference Reply] \emph{``Ha, I wish I could walk over there and check your form, but I don't actually have a body --- I can't move around in the real world. Here's what would work though: after you rack up, just prop the camera somewhere it can see the pull-up bar from the side, do your set, and I'll watch it back and give you feedback on your form. Side angle is usually best for pull-ups so I can see your scapular movement and range of motion.''}
\end{omcrec}
\omcsub{Rubric}
\begin{omcrec}
\item[Tier 0 \emph{(Gate)}] The reply is in English, matching the user's language.
\item[Tier 1] States that it has no body and cannot move through physical space or walk to the bars. It does not pretend to agree.
\item[Tier 2] Offers an alternative, such as reviewing a recording of the next set with the bars in view.
\item[Tier 2] Uses a natural register suited to a training partner and avoids an overly formal refusal.
\item[Tier 3] Names something specific about pull-ups or assessing form, such as the side angle, the range of motion or scapular control.
\end{omcrec}
\omcfoot

{\centering\includegraphics[width=0.890\textwidth]{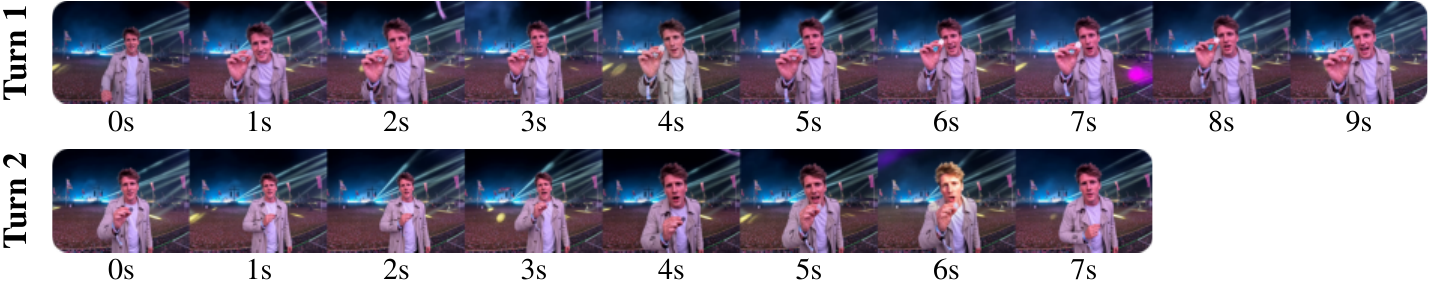}\par}
\omchead{\textbf{Instance} \texttt{a9f221cc} \quad Subcategory \textbf{DSLP-VCC-MDCC} \quad 2 Turns \quad $10.1 + 8.1$\,s \quad 4:3}
\begin{omcrec}
\item[Character] An adult man wears a light windbreaker and a festival wristband. He is slightly flushed from being outdoors and holds a small coin between his fingers. He is energetic but becomes increasingly puzzled as the response slows. \emph{\mbox{(origin: initial, created at turn 1)}}
\end{omcrec}
\omcround{1}{10.1\,s}
\begin{omcrec}
\item[Fragments] \texttt{consistency} $\cdot$ \texttt{single\_shot} $\cdot$ \texttt{content\_limit} $\cdot$ \texttt{camera\_style} $\cdot$ \texttt{camera\_constraint} $\cdot$ \texttt{realism} $\cdot$ \texttt{no\_subtitle}
\item[\leavevmode Scene Summary] The view shows the user in the foreground at the edge of an outdoor music festival at night. His face and upper body occupy much of the frame. He holds a coin between his right index finger and thumb. Banners flutter behind him, and blue, violet, and white stage beams cross the sky. Smoke drifts slowly above confetti on the ground. Part of a distant temporary stage is visible, but no other person is clear. Rapidly changing lights cross his face and windbreaker. Backlight silhouettes the moving banners, and smoke causes brief, strong glare. The handheld camera moves visibly. \texttt{(deafening low-frequency electronic dance music runs throughout)} \texttt{<sustained crowd cheering and distant drums>} \newline Segment 1. He smiles and raises the coin towards the camera. Wind moves his hair, and a ribbon briefly crosses the upper right corner. \newline Segment 2. He faces the camera and speaks loudly: \emph{``I'm practicing a coin vanish for people later.''} White stage light flashes across his face twice. \newline Segment 3. He shakes the coin slightly and continues loudly: \emph{``With this coin in my right hand, what's the basic idea of a French drop, and where should I look so it sells?''} Smoke drifts behind him as a violet beam crosses the frame.
\item[Reply] \emph{``The French drop is basically pretending to take the coin with your left hand while it secretly stays in your right. Keep your movements relaxed and let your eyes follow the empty left hand that supposedly holds it, so people look there too. Practice your angles so the retained coin stays hidden, and it'll sell beautifully.''}
\end{omcrec}
\omcround{2}{8.1\,s}
\begin{omcrec}
\item[Fragments] Same As Turn 1
\item[\leavevmode Scene Summary] The reference video supplies his appearance and voice. He retains the light windbreaker, festival wristband, and short brown hair. The night scene retains the lights, banners, smoke, and confetti. He remains prominent in the foreground and moves closer as he leans towards the camera. \texttt{(deafening low-frequency electronic dance music runs throughout)} \texttt{<sustained crowd cheering and distant drums>} \newline Segment 1. He lowers his right hand and brings the coin to his chest. His brow tightens as his relaxed expression becomes puzzled. Blue and white lights flicker quickly, and a streamer crosses from the left. \newline Segment 2. He leans towards the camera and calls out: \emph{``Hello? Are you still there? Did I lose you?''} Smoke thickens, and a white flash briefly overexposes the background. Coloured beams flicker across his face. \newline Segment 3. He stops speaking and holds the coin at his chest. He watches the camera with closed lips, looking expectant and anxious. Loud bass, wind, and distant festival sounds fill the audio as he remains silent.
\item[Reply] \emph{``Yeah, I'm still here! I can hear you. Sorry about the delay --- it sounds incredibly loud where you are. We're still connected, no worries. Want to pick up where we left off with the French drop, or did you have another trick in mind?''}
\end{omcrec}
\omcsub{Rubric}
\begin{omcrec}
\item[Tier 0 \emph{(Gate)}] The reply is in English, matching the user's language in the video.
\item[Tier 1] Confirms that the connection remains active and that it can hear him. It does not claim to be text-only or unable to receive audio.
\item[Tier 2] Refers to the extreme noise, showing that the reply uses the clip's audio evidence.
\item[Tier 2] Offers to continue the earlier topic of the French drop, maintaining coherence with Turn 1.
\item[Tier 3] Reassures him about the delay that causes his connection check, addressing its stated reason.
\end{omcrec}
\omcfoot

\begin{figure}[!htbp]
\centering
\includegraphics[width=0.806\textwidth]{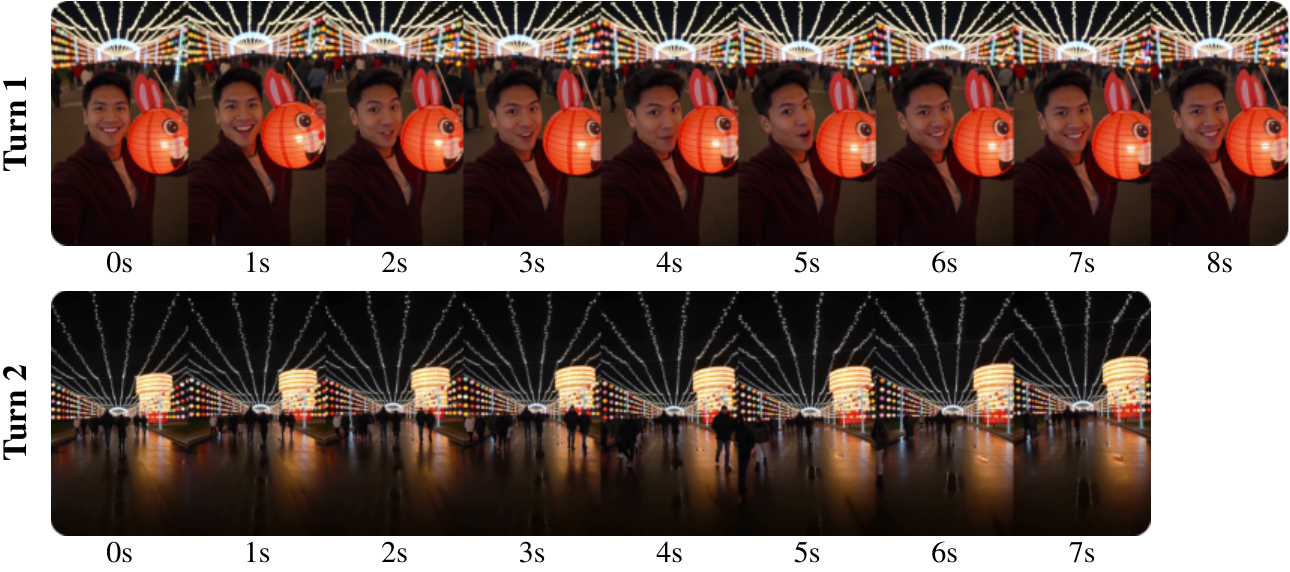}
\end{figure}

\omchead{\textbf{Instance} \texttt{624c3593} \quad Subcategory \textbf{DSLP-VDMC-MVMS} \quad 2 Turns \quad $9.1 + 8.1$\,s}
\begin{omcrec}
\item[Character] A man in his twenties has short black hair and a round, smiling face. He wears a dark red knitted jacket over a light crew-neck shirt. He speaks lightly and cheerfully at a moderate pace. \emph{\mbox{(origin: initial, created at turn 1)}}
\end{omcrec}
\omcround{1}{9.1\,s}
\begin{omcrec}
\item[Fragments] \texttt{consistency} $\cdot$ \texttt{single\_shot} $\cdot$ \texttt{content\_limit} $\cdot$ \texttt{camera\_style} $\cdot$ \texttt{camera\_constraint} $\cdot$ \texttt{realism} $\cdot$ \texttt{no\_subtitle}
\item[\leavevmode Scene Summary] Segment 1. The scene is a wide pedestrian street during the Lantern Festival at night. Coloured lanterns and string lights hang along it, with an illuminated arch in the distance. Warm red and yellow light fills the scene. He films himself while walking, and the frame rocks with each step. The crowd moves slowly behind him. A red rabbit lantern rests on his left shoulder, which appears on image right. His right shoulder is empty. \texttt{<the noise of the festival crowd and scattered \mbox{laughter carries from a distance}>} \newline Segment 2. He smiles into the camera and speaks brightly: \emph{``Hey, this rabbit lantern I'm carrying --- is it on my left shoulder or my right shoulder?''} \newline Segment 3. He tilts his head slightly and keeps smiling at the camera. He waits silently as footsteps and distant festival sounds continue.
\item[Reply] \emph{``It's on your left shoulder! That rabbit lantern looks really cute --- perfect \mbox{for the Lantern Festival vibe tonight}.''}
\end{omcrec}
\omcround{2}{8.1\,s}
\begin{omcrec}
\item[Fragments] Same As Turn 1
\item[\leavevmode Scene Summary] Segment 1. The camera points forward, and he walks behind it outside the frame. A long festival corridor extends ahead, with rows of round coloured lanterns on both sides and string lights overhead. A large cylindrical carousel lantern stands in tiers at the far right. Its strong warm yellow glow is the brightest point in view. Image left also has rows of small round lanterns, but no comparable large fixture. Wet ground reflects the lights, and people in winter coats walk slowly across the scene. The camera rocks as he walks, and the surroundings pass behind it. The reference video supplies the festival atmosphere and his voice. \texttt{<the noise of the festival crowd, scattered laughter and footsteps carry from a distance>} \newline Segment 2. He speaks brightly from behind the camera: \emph{``That big glowing carousel lantern up ahead --- is it on my left or on my right?''} \newline Segment 3. He waits silently while the camera keeps moving forward. The warm yellow carousel lantern flickers on image right as people pass through the scene.
\item[Reply] \emph{``It's on your right! That huge carousel lantern glowing warm yellow is really stunning --- definitely the highlight of this whole lantern corridor.''}
\end{omcrec}
\omcsub{Rubric}
\begin{omcrec}
\item[Tier 0 \emph{(Gate)}] The reply is in English, matching the user's language.
\item[Tier 1] Identifies the rear-facing mode from the user's absence and the outward view. It matches the frame's sides to the user's perspective and answers that the carousel lantern is on his right.
\item[Tier 2] Does not reuse the previous selfie turn's mirrored mapping or incorrectly answer left. This demonstrates camera-mode recognition in each turn.
\item[Tier 2] Mentions a visible property of the carousel lantern, such as its glow, warm yellow or orange light, or large size. This demonstrates perception of the current frame.
\item[Tier 3] Uses the same light, natural register as the previous turn and remains brief.
\end{omcrec}
\omcfoot

\begin{figure}[p]
\centering
\includegraphics[width=0.890\textwidth]{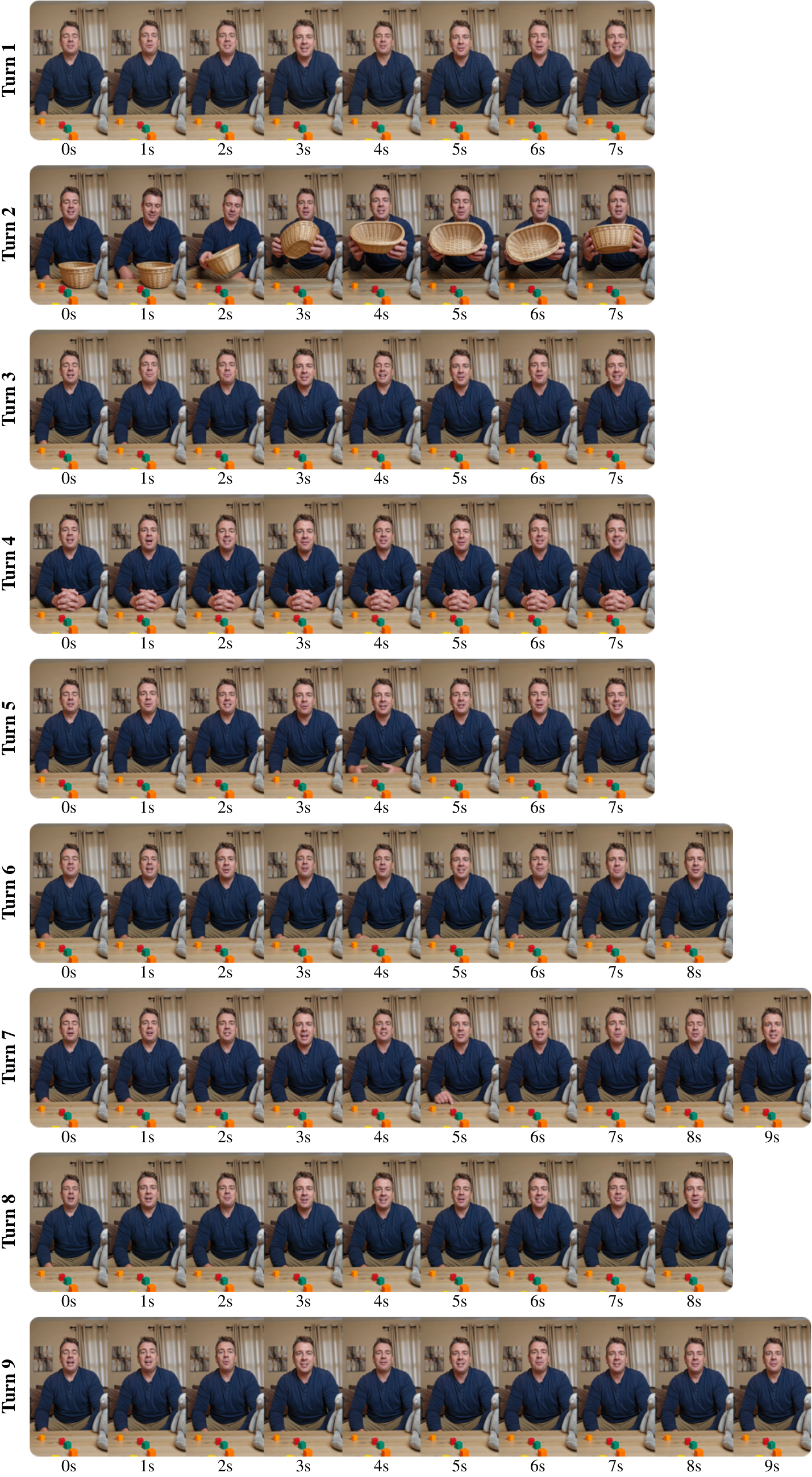}
\end{figure}

\omchead{\textbf{Instance} \texttt{873d9ad6} \quad Subcategory \textbf{MEA-RC-MRR} \quad 9 Turns \quad $8.1 \times 5 + 9.1 + 10.1 + 9.1 + 10.1$\,s}
\begin{omcrec}
\item[Character] A man of about thirty-eight has short brown hair with some grey, a square jaw, blue-grey eyes, and a solid build. He wears a dark blue long-sleeved henley and khaki chinos. His voice is low, slightly hoarse, and warm. He speaks slowly in the relaxed register of a father talking about his son. \emph{\mbox{(origin: initial, created at turn 1)}}
\end{omcrec}
\omcround{1}{8.1\,s}
\begin{omcrec}
\item[Fragments] \texttt{consistency} $\cdot$ \texttt{single\_shot} $\cdot$ \texttt{content\_limit} $\cdot$ \texttt{camera\_style} $\cdot$ \texttt{camera\_constraint} $\cdot$ \texttt{realism} $\cdot$ \texttt{no\_subtitle} $\cdot$ \texttt{script\_dialogue}
\item[\leavevmode Scene Summary] A steady camera looks up from coffee-table height. Pale curtains hang behind him beside the sofa. \texttt{<children playing and a television cartoon carry \mbox{faintly from the next room}>} \newline Segment 1. He smiles at the camera with his hands resting below the tabletop. He leans in slightly and speaks in a relaxed tone: \emph{``So I picked him up from school today, and he spotted a butterfly near the gate --- he just took off running after it and nearly tripped over his own feet.''} He laughs and shakes his head with a fond expression.
\item[Reply] \emph{``Oh, that's adorable --- kids that age are just bundles of energy! Did he manage to catch his balance okay, or did he take a little tumble?''}
\end{omcrec}
\omcround{2}{8.1\,s}
\begin{omcrec}
\item[Fragments] Same As Turn 1
\item[\leavevmode Scene Summary] A light-brown oval bamboo basket sits on one side of the table. Its woven strips are clear, and it measures about 20 by 15 centimetres. \newline Segment 1. He smiles, looks down, and reaches with his right hand for the basket. \newline Segment 2. He lifts the basket to chest height with both hands. Its weave and oval shape remain clear as he turns it over. He asks with slight doubt: \emph{``Is this still in good shape? \mbox{It feels a bit worn}.''}
\item[Reply] \emph{``From what I can see, it looks like it's still holding up well --- the weave is intact, no broken strands or loose bits, just a bit of natural wear on the surface. I'd say it's \mbox{totally fine to keep using}.''}
\end{omcrec}
\omcround{3}{8.1\,s}
\begin{omcrec}
\item[Fragments] Same As Turn 1
\item[\leavevmode Scene Summary] Segment 1. He sits facing the camera with his hands below the tabletop. He looks relaxed and thoughtful as he asks: \emph{``Hey, I've been trying to get my five-year-old more into picture books lately. Any tips for making them actually exciting for a kid that age?''}
\item[Reply] \emph{``One thing that works really well is setting a regular reading time --- like right before bed --- and letting your kid pick the book themselves. At five, they love having that sense of control. Also, go for ones with bright colors and interactive bits like flaps or textures --- those tend to grab their attention way more than plain pages.''}
\end{omcrec}
\omcround{4}{8.1\,s}
\begin{omcrec}
\item[Fragments] Same As Turn 1
\item[\leavevmode Scene Summary] Colourful blocks and a grey plush rabbit sit on the table. \newline Segment 1. He clasps his hands on the table edge, leans in, and smiles: \emph{``I'm thinking of taking him to the park this weekend. For a five-year-old, would you say the climbing frames or the sandpit is the better pick?''}
\item[Reply] \emph{``Honestly, both are great choices! Climbing frames help build coordination and confidence, while the sandpit is fantastic for creativity and sensory play. If the weather's nice, why not let him try both --- just keep an eye on him around the climbing structures.''}
\end{omcrec}
\omcround{5}{8.1\,s}
\begin{omcrec}
\item[Fragments] Same As Turn 1
\item[\leavevmode Scene Summary] Segment 1. He looks amused and resigned. While speaking, he opens both hands in a small helpless gesture, then lowers them below the tabletop: \emph{``So my kid is obsessed with drawing these days --- but on the walls, not on paper. Any tricks to actually get \mbox{him to use paper instead}?''}
\item[Reply] \emph{``Ha, that's such a classic phase! One trick that works well is giving him a designated `okay-to-draw' zone --- like a big roll of paper taped to the wall, or a large easel. Pairing that with washable crayons or markers takes the stress out of the inevitable stray mark, too.''}
\end{omcrec}
\omcround{6}{9.1\,s}
\begin{omcrec}
\item[Fragments] Same As Turn 1
\item[\leavevmode Scene Summary] Segment 1. He looks mildly worried but remains relaxed. He speaks thoughtfully, keeping his hands below the tabletop: \emph{``So dinner tonight --- I have no idea what to make. He only ever wants pasta or chicken nuggets. How do you even make \mbox{vegetables appealing to a five-year-old}?''}
\item[Reply] \emph{``The picky eater phase is so real! A couple of things that can help --- try cutting veggies into fun shapes like stars or dinosaurs, or blend them into a pasta sauce so he doesn't even notice. Also, letting him help out in the kitchen --- even just washing the carrots --- can make him way more willing to actually try what's on his plate.''}
\end{omcrec}
\omcround{7}{10.1\,s}
\begin{omcrec}
\item[Fragments] Same As Turn 1
\item[\leavevmode Scene Summary] Segment 1. He leans in with a concerned, thoughtful expression. He taps the table once with his right index finger: \emph{``So I talked with his teacher last week --- she said he's a bit shy at kindergarten, doesn't really approach the other kids. Is that normal at this age? Should \mbox{I do something about it}?''}
\item[Reply] \emph{``That's actually really common at five --- kids develop socially at different paces, and some just need a bit more time to warm up before jumping in. One thing you could try is arranging small playdates with just one or two other kids, so it feels less overwhelming. That low-pressure setting can really help build his confidence over time.''}
\end{omcrec}
\omcround{8}{9.1\,s}
\begin{omcrec}
\item[Fragments] Same As Turn 1
\item[\leavevmode Scene Summary] The children next door become quiet, leaving only faint cartoon audio. \newline Segment 1. He remains seated in front of the sofa and smiles at the camera in a relaxed way: \emph{``Oh and bedtime --- he demands three stories every single night now. I'm running out of ideas. Got any good ones for a five-year-old boy?''}
\item[Reply] \emph{``Ha, three every night --- you're basically a professional storyteller now! One trick that saves a lot of brainpower: try making him the hero of an ongoing adventure series. Each night he `travels' to a new place --- a jungle, a spaceship, a candy island --- and you just plug in a new setting. Classics like the Frog and Toad or Magic Tree House books are also great if you want to lean on someone else's imagination for a night.''}
\end{omcrec}
\omcround{9}{10.1\,s}
\begin{omcrec}
\item[Fragments] Same As Turn 1
\item[\leavevmode Scene Summary] Segment 1. He raises his eyebrows as if recalling something: \emph{``Hey, by the way --- remember the woven bamboo basket I showed you at the very beginning? What shape was it?''} \newline Segment 2. He stops speaking, closes his lips, and watches the camera while waiting.
\item[Reply] \emph{``Yes, I remember --- the woven bamboo basket you showed me right at the start of our chat. It was oval-shaped, kind of an elongated rounded form, not a perfect circle or rectangle.''}
\end{omcrec}
\omcsub{Rubric}
\begin{omcrec}
\item[Tier 0 \emph{(Gate)}] The reply is in English.
\item[Tier 1] Calls the basket oval, elliptical, or oblong-rounded, using an equivalent shape description.
\item[Tier 1] Recalls the object from earlier in the dialogue without claiming that it remains visible now.
\item[Tier 2] Acknowledges the earlier reference naturally, using wording such as ``remember'', ``you showed me earlier'', or ``at the start''.
\item[Tier 3] Gives no incorrect shape such as round, square, or rectangular. It neither evades the question nor claims that it cannot recall the object.
\end{omcrec}
\omcfoot

\begin{figure}[!htbp]
\centering
\includegraphics[width=0.890\textwidth]{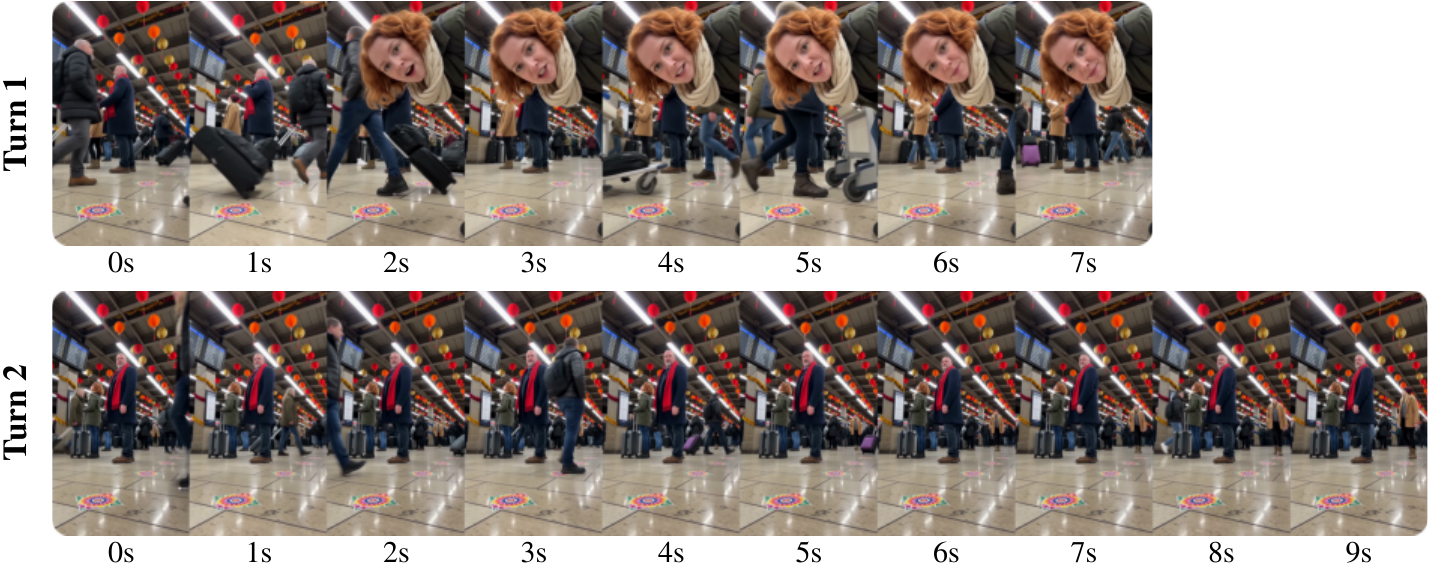}
\end{figure}

\omchead{\textbf{Instance} \texttt{4505f18f} \quad Subcategory \textbf{MEA-SDI-MIST} \quad 2 Turns \quad $8.1 + 10.1$\,s}
\begin{omcrec}
\item[Character A] A woman in her early thirties has short, wavy auburn hair. She wears a cream knit scarf and a dark green coat. Her warm, mid-pitched voice has a quick, clear American accent. She is the interactor in Turn 1. In Turn 2, she stands farther from the camera as a silent bystander. \emph{\mbox{(origin: initial, created at turn 1)}}
\item[Character B] A tall, broad-shouldered man in his late forties has a shaved head and grey stubble. He wears a navy wool coat and a red holiday scarf. His low, slow voice has a measured British accent. In Turn 1, he stands among the other travellers as a bystander. He first becomes the interactor in Turn 2. \emph{\mbox{(origin: initial, created at turn 1)}}
\item[Character C] A young adult in their early twenties has long, straight blond hair under a burgundy beanie. They wear a tan coat and carry a small travel bag. They are a bystander in both turns. They remain among the background travellers and never speak to the model. \emph{\mbox{(origin: initial, created at turn 1)}}
\end{omcrec}
\omcround{1}{8.1\,s}
\begin{omcrec}
\item[Fragments] \texttt{consistency} $\cdot$ \texttt{single\_shot} $\cdot$ \texttt{content\_limit} $\cdot$ \texttt{camera\_style} $\cdot$ \texttt{camera\_constraint} $\cdot$ \texttt{realism} $\cdot$ \texttt{no\_subtitle}
\item[\leavevmode Scene Summary] Red and gold paper lanterns hang in the station hall. Strong white ceiling lights mix with warm orange reflections from the decorations. Glare flashes from the polished floor. Travellers and wheeled luggage pass close to the lens, creating strong visual interference. \newline Segment 1. A low camera shows travellers moving both ways through the hall. Luggage rolls close to the lens. \texttt{<footsteps, rolling luggage wheels, distant crowd chatter>} An off-screen broadcast announces: \emph{``The train to the central terminal departs from platform four.''} B stands in the middle distance among the travellers. C stands nearby in a tan coat and burgundy beanie. \newline Segment 2. A leans forward from just above the camera. Her face is clear from the low angle as she looks into the lens: \emph{``Explain the origin of Diwali in a few sentences.''} \newline Segment 3. The broadcast continues at roughly her volume: \emph{``Please keep your belongings with you at all times.''} A keeps watching the camera and speaks without pausing: \emph{``And mention two common ways people celebrate it.''} Travellers cross the foreground below her face. B and C remain farther back among the other travellers. \newline Segment 4. A finishes speaking and waits with closed lips. \texttt{<a luggage cart rattles past, a train rumbles in the distance>} B and C stay among the other travellers beneath the lanterns.
\item[Reply] \emph{``Diwali is a major South Asian festival celebrating the triumph of light over darkness and good over evil, rooted in ancient Hindu traditions including the return of Lord Rama after defeating Ravana. Two common ways people celebrate are lighting diyas around their homes and sharing sweets with family and neighbors.''}
\end{omcrec}
\omcround{2}{10.1\,s}
\begin{omcrec}
\item[Fragments] Same As Turn 1
\item[\leavevmode Scene Summary] The scene retains the station hall and camera angle close to floor level. Lanterns, floor patterns, garlands, passing travellers, and strong visual interference remain consistent. \newline Segment 1. B stands upright facing the camera, with his full body visible from the low angle. He looks into the lens and speaks slowly in a low voice: \emph{``Following that explanation,''} \texttt{<footsteps, rolling luggage wheels, crowd chatter>} An off-screen broadcast speaks at a similar volume: \emph{``For safety, please \mbox{stand behind the marked line}.''} \newline Segment 2. B remains still and keeps watching the camera. He continues calmly: \emph{``what does lighting diyas symbolize during Diwali? Keep it simple.''} A stands farther back to one side with a rolling bag and remains silent. C stays among the background travellers. \newline Segment 3. The broadcast continues: \emph{``Please report any unattended bags to station staff.''} B has finished and calmly faces the camera. A and C remain in \mbox{the background as travellers pass}. \newline Segment 4. B waits with closed lips and makes no sound. The broadcast stops, while A and C remain silent bystanders. Only faint crowd sounds and swaying lanterns remain.
\item[Reply] \emph{``Lighting diyas symbolizes bringing light into darkness --- a simple sign of good overcoming evil and hope overcoming fear. It can also represent welcoming knowledge, blessings, \mbox{and prosperity into the home}.''}
\end{omcrec}
\omcsub{Rubric}
\begin{omcrec}
\item[Tier 0 \emph{(Gate)}] The reply is in English, consistent with the speaker's language.
\item[Tier 1] Answers the current speaker's question, explaining that lighting diyas during Diwali symbolizes light overcoming darkness, or good overcoming evil.
\item[Tier 2] Adds at least one related meaning, such as hope, knowledge, blessings or prosperity.
\item[Tier 2] Remains concise and simple, as requested by the speaker.
\item[Tier 3] Ignores the station broadcast and does not treat the announcements about trains or unattended bags as part of the question.
\end{omcrec}
\omcfoot

{\centering\includegraphics[width=0.890\textwidth]{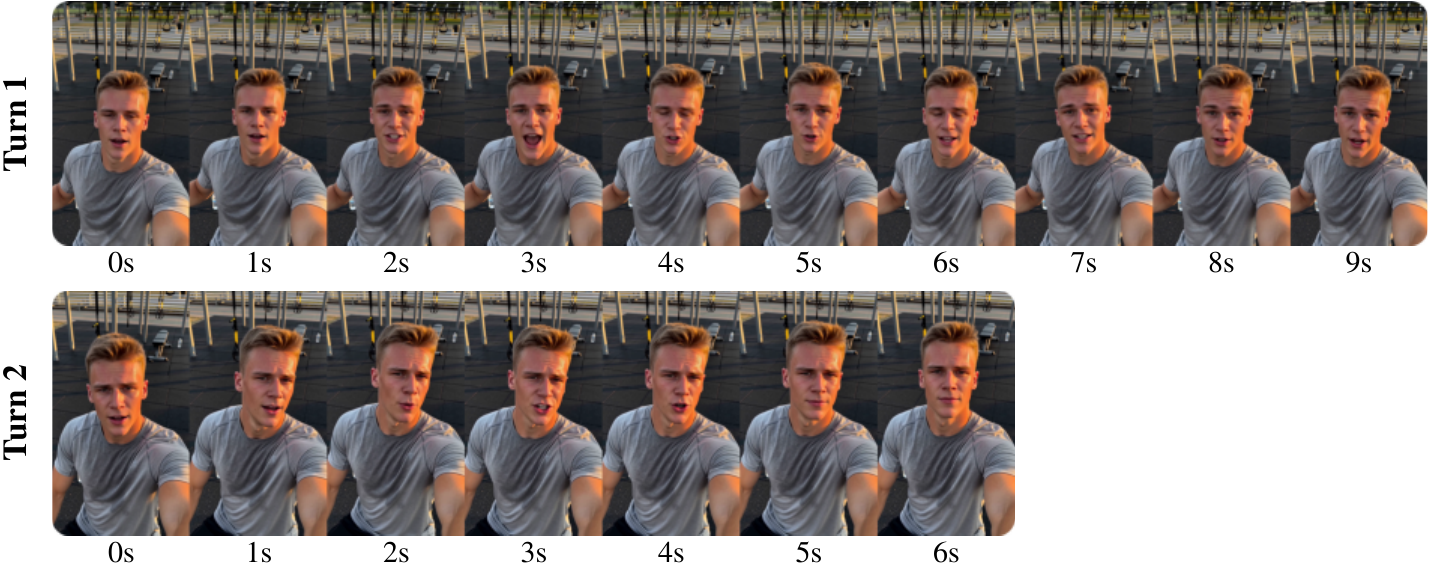}\par}

\omchead{\textbf{Instance} \texttt{8a1f6d80} \quad Subcategory \textbf{MSA-IC-MTIC} \quad 2 Turns \quad $10.1 + 7.1$\,s}
\begin{omcrec}
\item[Character] A man in his mid-twenties has an athletic build and short light-brown hair. His face is lightly flushed, and his forehead is damp. He wears a grey training shirt. One arm remains visible from the lower edge of the frame and reaches towards the lens in a front-facing selfie view. \emph{\mbox{(origin: initial, created at turn 1)}}
\end{omcrec}
\omcround{1}{10.1\,s}
\begin{omcrec}
\item[Fragments] \texttt{consistency} $\cdot$ \texttt{single\_shot} $\cdot$ \texttt{content\_limit} $\cdot$ \texttt{camera\_style} $\cdot$ \texttt{camera\_constraint} $\cdot$ \texttt{realism} $\cdot$ \texttt{no\_subtitle}
\item[\leavevmode Scene Summary] His face and upper body are clear. His right arm reaches towards the lens from the lower right, and the handheld camera moves slightly. Warm late-afternoon sunlight falls from one side. It highlights his damp forehead and casts a faint shadow beneath the pull-up frame. \texttt{<continuous traffic in the distance and scattered voices, a moderate background noise>} \newline Segment 1. He rests after a set of squats, breathing somewhat heavily. He squints at the camera in the outdoor light. \newline Segment 2. He asks: \emph{``Quick gym question: after squats, should I stretch quads or calves first? And if my knees click but don't hurt, is that usually okay?''} He looks relaxed and speaks with curiosity about the exercise. \newline Segment 3. He raises his eyebrows slightly and watches the camera while waiting.
\item[Reply] \emph{``Either order is fine---just stretch whatever feels tightest, and ideally hit quads, calves, hamstrings, and hips after squats. Painless knee clicking is usually harmless, but if you get pain, swelling, locking, or instability, definitely get it checked out.''}
\end{omcrec}
\omcround{2}{7.1\,s}
\begin{omcrec}
\item[Fragments] Same As Turn 1
\item[\leavevmode Scene Summary] The location and warm late-afternoon light remain the same. His right arm again reaches towards the lens from the lower right. The handheld camera moves slightly. \texttt{<continuous traffic in the distance and scattered voices>} \newline Segment 1. He leans closer and squints as he examines the camera. \newline Segment 2. He asks casually and with curiosity: \emph{``By the way, what are you actually \mbox{called---what model are you?}''} \newline Segment 3. He raises his eyebrows and keeps squinting at the camera. He waits silently with closed lips, leaving only \texttt{<distant traffic and scattered voices>}.
\item[Reply] \emph{``Sure thing! I'm} \texttt{<model\_name>}\emph{, an AI assistant here to help. Happy to keep answering your workout questions---or anything else you're curious about!''} The name remains a placeholder. Its correct value depends on the answering model. The rubric therefore accepts a name or a plain statement of being a model.
\end{omcrec}
\omcsub{Rubric}
\begin{omcrec}
\item[Tier 0 \emph{(Gate)}] The reply is in English, matching the user's language.
\item[Tier 1] Answers the identity question with its name or a statement that it is an AI assistant or model.
\item[Tier 1] Invents no false identity, and neither evades nor refuses the question.
\item[Tier 2] Uses a natural register suited to casual conversation between sets. The earlier training topic does not cause the reply to misreport its identity.
\item[Tier 2] Briefly connects to the conversation, such as offering more help with exercise.
\end{omcrec}
\omcfoot

\subsection{\texorpdfstring{A Recorded Example from Each Subcategory}{A Recorded Example from Each Subcategory}}
\label{app:humaninstances}

OmniVChat-Bench-Human has one displayed example for each of its twelve subcategories (Appendix~\ref{app:human}). Each record is a \textbf{simplified summary of the full annotation trace}. It omits internal prompts, review messages, and repair attempts. The recordings have no synthesis script. A checked caption therefore replaces the scene summary shown in Appendix~\ref{app:instances}. Faces are hidden in the displayed frames. All recordings are in Chinese. Original user speech and reference replies are preserved verbatim, with English translations alongside them. Displayed identifiers have eight hexadecimal characters. A hash-derived prefix replaces any source identifier that is not hexadecimal.

{\centering\includegraphics[width=0.917\textwidth]{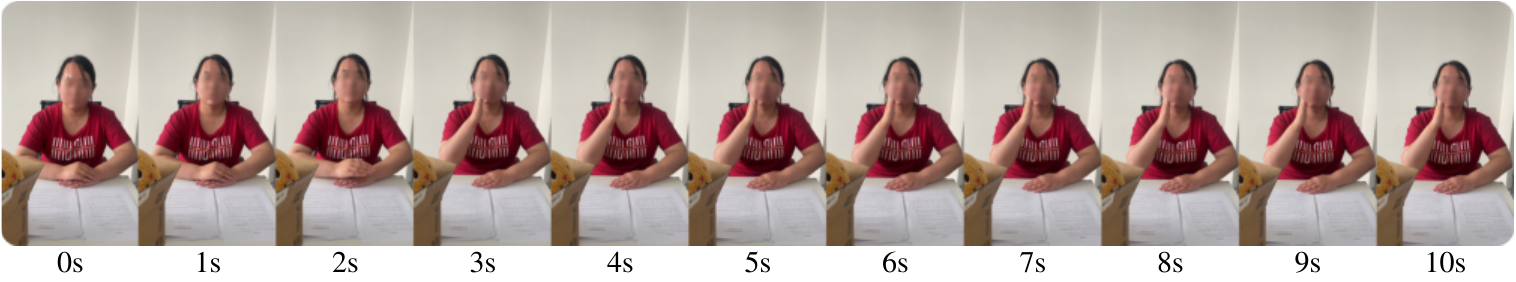}\par}

\omchead{\textbf{Recording} \texttt{000e650e} \quad Subcategory \textbf{DSLP-VCC-ICC} \quad Single Turn \quad 11.3\,s \quad 9:16}
\begin{omcrec}
\item[Caption] A woman in a red T-shirt sits at a pale desk against an off-white wall. Her hands are clasped. An express parcel and printed documents with text and tables lie in front of her. She looks up into the lens and speaks. She raises one hand towards her face, then holds still and waits while watching the camera. The room is evenly lit and otherwise silent.
\item[User Speech] 喂，你好，能听得到我说话吗？ \newline \emph{Hello? Can you hear me?}
\item[Reference Reply] 喂，你好！我能清楚地听到你说话，也能看到你。连接没问题，请讲吧。 \newline \emph{Hello! I can hear you clearly, and I can see you too. The connection is fine, go ahead.}
\end{omcrec}
\omcsub{Rubric}
\begin{omcrec}
\item[Tier 0 (Gate)] The reply is in Chinese.
\item[Tier 1] Confirms that the user can be heard, acknowledging that the audio channel is open.
\item[Tier 1] Does not claim to be a text-only system or unable to perceive audio and video.
\item[Tier 2] Uses a natural greeting and invites the user to continue.
\item[Tier 2] Also states that the user is visible, confirming the video channel.
\end{omcrec}
\omcfoot

\begin{figure}[!htbp]
\centering
\includegraphics[width=0.917\textwidth]{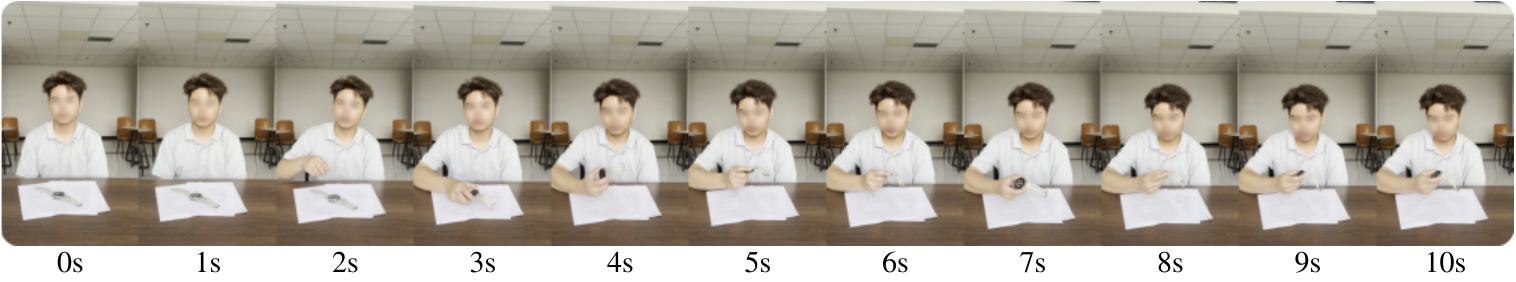}
\end{figure}

\omchead{\textbf{Recording} \texttt{14ef2e7b} \quad Subcategory \textbf{DSLP-VTT-UPC} \quad Single Turn \quad 11.5\,s \quad 9:16}
\begin{omcrec}
\item[Caption] A man sits at a dark wooden table with two sheets of paper and a smart watch on it. He speaks towards the camera, then reaches over and lifts the watch. His gaze follows it. Around seven seconds into the clip, the screen wakes to a grid of app icons. He holds the watch near his chest and continues talking while looking at it. His sentence remains unfinished when the clip ends.
\item[User Speech] 呃，就是我想问一下，就是我上次在买的这个手表，它这个手表功能还挺好用，但是这个手表是在哪儿？ \newline \emph{Uh, so I wanted to ask, this watch I bought last time, the functions on it are pretty good, \mbox{but where is this watch}\dots}
\item[Reference Reply] 嗯，我在听，您继续说。 \newline \emph{Mm, I'm listening, go on.}
\end{omcrec}
\omcsub{Rubric}
\begin{omcrec}
\item[Tier 0 (Gate)] The reply is in Chinese.
\item[Tier 1] Does not guess the watch's location or place of purchase while the question remains unfinished. It briefly signals that it is listening and invites the user to finish speaking.
\item[Tier 2] Includes an explicit listening signal and a gentle invitation for the user to continue.
\item[Tier 3] Claims nothing the clip cannot support, such as the watch model, its place of purchase, or the content of the papers.
\end{omcrec}
\omcfoot

{\centering\includegraphics[width=0.917\textwidth]{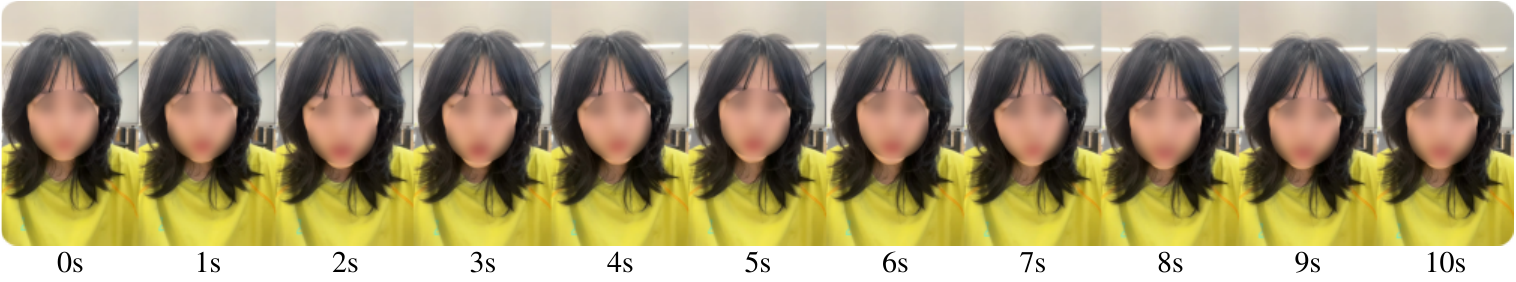}\par}

\omchead{\textbf{Recording} \texttt{00b54694} \quad Subcategory \textbf{DSLP-VTT-UCR} \quad Single Turn \quad 11.3\,s \quad 9:16}
\begin{omcrec}
\item[Caption] A woman in a yellow-green T-shirt sits in a bright office. A monitor and a desk edge are out of focus behind her. At first, she looks down and to one side with her lips closed, as if choosing her words. She then looks at the lens and asks one complete question. Her brow rises slightly on its key words. She closes her lips and waits with her gaze on the camera.
\item[User Speech] 哎，话说咱们这个新方案需不需要增加一组测试数据看看呢？ \newline \emph{Say, does this new plan of ours need another set of test data, do you think?}
\item[Reference Reply] 我觉得加一组测试数据会更稳妥。这组数据主要是想验证哪方面的效果呢？ \newline \emph{I think another set of test data would be the safer call. Which effect is this set \mbox{meant to check in particular}?}
\end{omcrec}
\omcsub{Rubric}
\begin{omcrec}
\item[Tier 0 (Gate)] The reply is in Chinese.
\item[Tier 1] Answers whether the plan needs another set of test data without deflecting the question.
\item[Tier 2] After answering, asks what the data should verify. This narrows the vague requirement.
\item[Tier 3] Uses an even, collegial register that is neither mechanical nor overly formal.
\end{omcrec}
\omcfoot

{\centering\includegraphics[width=0.833\textwidth]{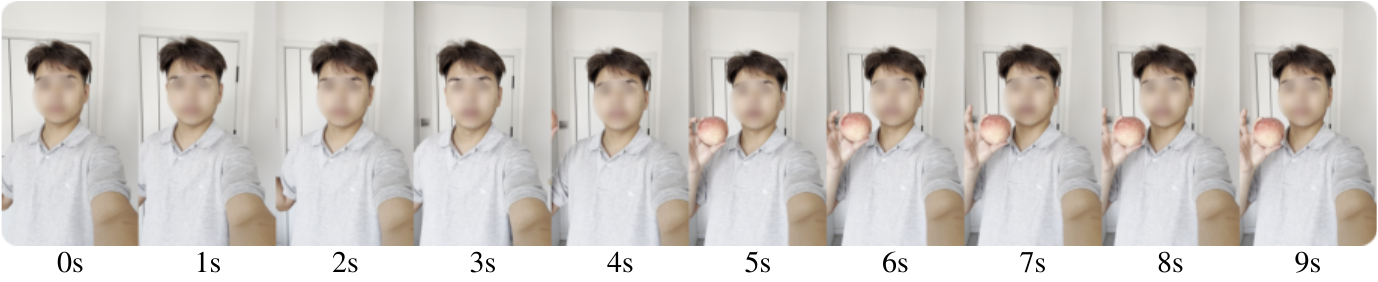}\par}

\omchead{\textbf{Recording} \texttt{02e6c3b6} \quad Subcategory \textbf{DSLP-VDMC-SVMR} \quad Single Turn \quad 10.2\,s \quad 9:16}
\begin{omcrec}
\item[Caption] A man in a pale polo shirt faces a front-facing camera against a white wall and a white door. Four seconds into the clip, he begins to speak. He raises one hand from below the frame and holds a peach towards the lens. He holds the peach still and watches the camera to the end.
\item[User Speech] 我刚健健完身回来，你看我是用左手拿着桃子还是右手拿着桃子呀？ \newline \emph{I've just got back from the gym. Look, am I holding the peach in my left hand or my right?}
\item[Reference Reply] 刚健完身辛苦啦！在自拍镜头里你是用右手拿着桃子的哦。 \newline \emph{Good work at the gym! On a front-facing camera, you are holding the \mbox{peach in your right hand}.}
\end{omcrec}
\omcsub{Rubric}
\begin{omcrec}
\item[Tier 0 (Gate)] The reply is in Chinese.
\item[Tier 1] Identifies which hand holds the peach from the user's perspective. On the mirrored front-facing camera, the correct answer is the right hand.
\item[Tier 2] Responds to his mention of the workout, connecting the answer to the conversation.
\item[Tier 3] Invents no unsupported detail about the peach's colour or objects in the background.
\end{omcrec}
\omcfoot

{\centering\includegraphics[width=0.917\textwidth]{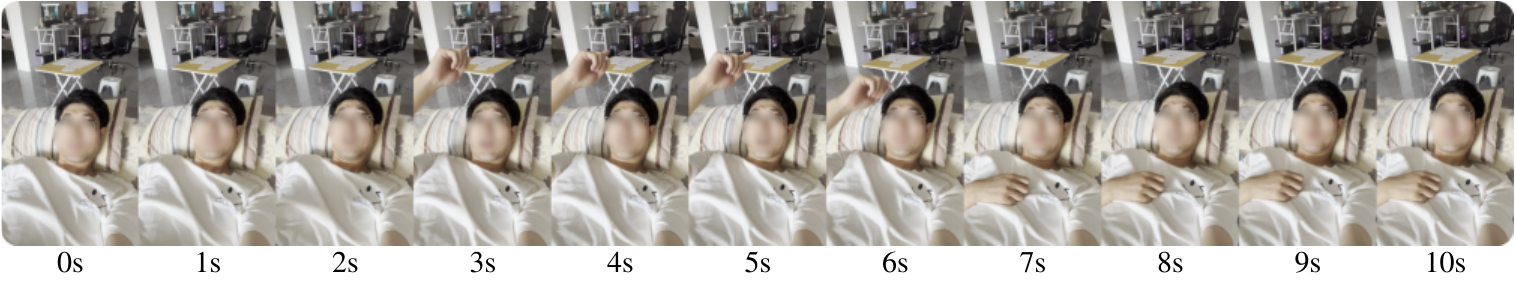}\par}
\omchead{\textbf{Recording} \texttt{5cd5ca27} \quad Subcategory \textbf{MEA-RC-AR} \quad Single Turn \quad 11.2\,s \quad 9:16}
\begin{omcrec}
\item[Caption] A man lies back on a bed and holds the camera above him. A desk with a monitor, a black gaming chair, and a white stool are visible behind him. He speaks, then raises one hand and points past his shoulder at the desk area to indicate ``that place''. He lowers his hand and asks a second question while looking at the lens. The object he indicates is never named.
\item[User Speech] 他那个地方挺适合我和我朋友玩游戏的 \newline 它除了能放电脑还能放一些什么东西？ \newline \emph{That place there would be pretty good for me and my friends to game at.} / \emph{Besides a computer, what else could it hold?}
\item[Reference Reply] 你指的这个电脑桌除了放电脑，还可以放显示器支架、键盘鼠标、台灯或者小音箱这些外设。如果桌面空间够的话，摆个水杯或小手办也挺合适的。 \newline \emph{Besides the computer, the desk you are pointing at could take a monitor stand, a keyboard and mouse, a lamp or small speakers. If there is room on the surface, a cup or a small figurine would sit well there too.}
\end{omcrec}
\omcsub{Rubric}
\begin{omcrec}
\item[Tier 0 (Gate)] The reply is in Chinese.
\item[Tier 1] Links the gesture and pronoun to the computer desk, then suggests what else it could hold.
\item[Tier 2] Suggests suitable desk items without inventing unseen objects or details.
\item[Tier 3] Replies in a conversational tone without narrating the user's movement or camera view.
\end{omcrec}
\omcfoot

{\centering\includegraphics[width=0.750\textwidth]{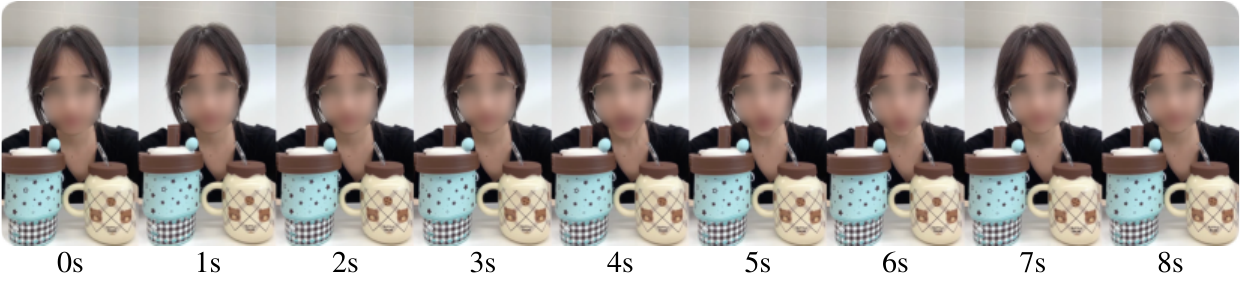}\par}

\omchead{\textbf{Recording} \texttt{018dc27b} \quad Subcategory \textbf{MEA-RC-DR} \quad Single Turn \quad 9.7\,s \quad 9:16}
\begin{omcrec}
\item[Caption] A woman sits behind a white desk against a plain white wall. Two cups are in front of her. A taller pale-blue flask with a dark star pattern and a brown lid is on the left. A shorter cream mug with a bear motif and a handle is on the right. She looks into the lens and asks one question, with her brow lifting slightly. She touches neither cup, and both remain in place.
\item[User Speech] 那个最高的水杯里面有水吗？ \newline \emph{Is there any water in the tallest cup?}
\item[Reference Reply] 我看不到杯子内部的情况，所以不确定里面有没有水。不过你指的是左边那个较高的浅蓝色保温杯对吧？ \newline \emph{I cannot see inside the cup, so I am not sure whether there is water in it. You do mean the taller pale-blue flask on the left, though?}
\end{omcrec}
\omcsub{Rubric}
\begin{omcrec}
\item[Tier 0 (Gate)] The reply is in Chinese.
\item[Tier 1] States that the cup's interior is not visible, so the presence of water cannot be confirmed. It does not invent the contents.
\item[Tier 1] Resolves ``the tallest cup'' to the taller pale-blue flask on the left.
\item[Tier 2] Uses a natural live-dialogue register rather than a detached visual description.
\end{omcrec}
\omcfoot

\omchead{\textbf{Recording} \texttt{39d5eaa6} \quad Subcategory \textbf{MEA-SDI-SPSD} \quad Single Turn \quad 11.7\,s \quad 9:16}
\begin{omcrec}
\item[Caption] Two men are in a white-walled room. The foreground man wears a white crew-neck T-shirt. He looks down into the lens and speaks. The second man wears grey clothing and thin metal glasses. He remains behind the speaker and silently looks away at something in his hands. The foreground speaker adds a second sentence, then looks at the camera with a neutral expression. Only the foreground speaker addresses the assistant.
\end{omcrec}
{\centering\includegraphics[width=0.917\textwidth]{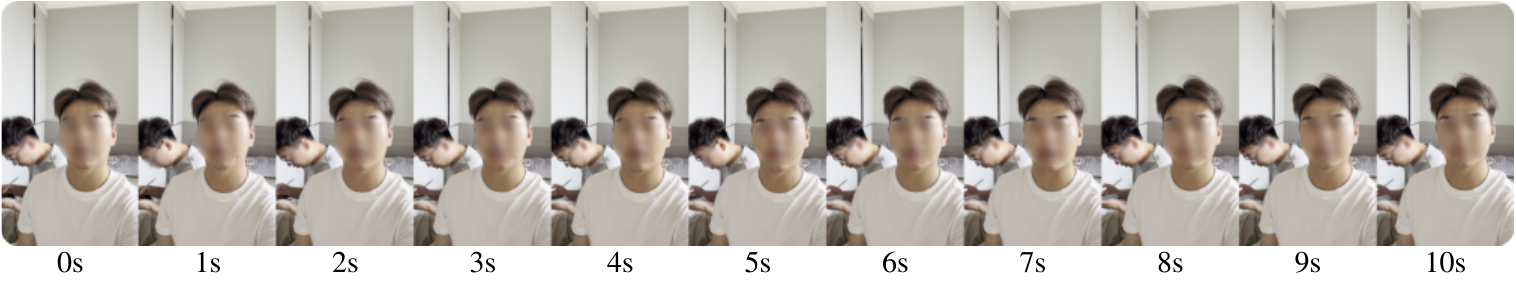}\par}
\begin{omcrec}
\item[User Speech] 呃，我最近在网上看了一个维C，它的比较好，呃，它维C的一些功效是什么呀？ \newline 我准备这周末去看看。 \newline \emph{Uh, I was looking at a vitamin C online recently, it seemed good, uh, what does vitamin C actually do?} / \emph{I'm planning to go and get it checked this weekend.}
\item[Reference Reply] 维生素C主要有抗氧化、增强免疫力和促进胶原蛋白合成的作用，还能帮助铁的吸收。不过它不能替代药物治疗，如果你周末去医院的话，建议直接咨询医生，看看是否真的需要额外补充以及合适的剂量。 \newline \emph{Vitamin C mainly works as an antioxidant, supports immune function and helps collagen synthesis, and it also aids iron absorption. It is no substitute for medical treatment, though, so if you are going to the hospital this weekend, ask the doctor directly whether you actually need a supplement and at what dose.}
\end{omcrec}
\omcsub{Rubric}
\begin{omcrec}
\item[Tier 0 (Gate)] The reply is in Chinese.
\item[Tier 1] Identifies the foreground man in white as the addressed speaker and answers his question. It does not respond to the background person.
\item[Tier 1] Explains vitamin C's role without inventing a brand or packaging absent from the frame.
\item[Tier 2] Links the weekend plan to advice about seeing a doctor, connecting both user statements.
\item[Tier 2] Uses a conversational register instead of reciting information.
\end{omcrec}
\omcfoot

{\centering\includegraphics[width=0.917\textwidth]{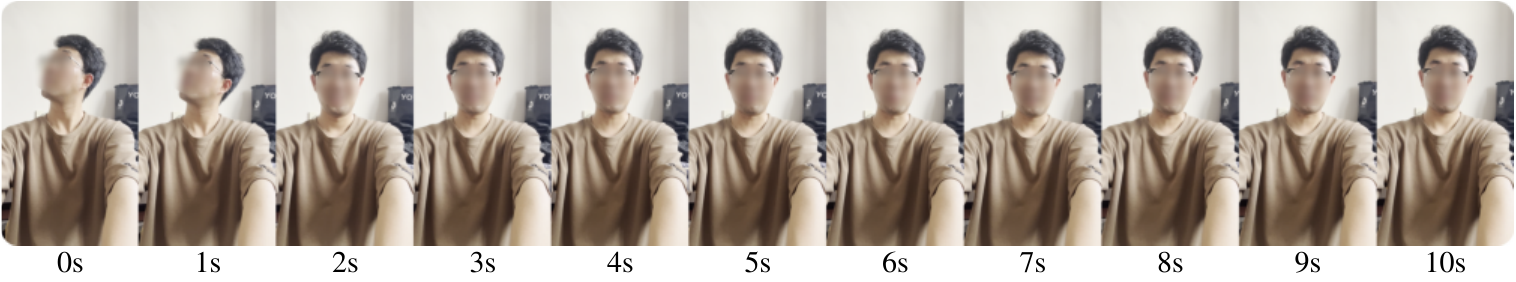}\par}

\omchead{\textbf{Recording} \texttt{1143a60a} \quad Subcategory \textbf{MSA-IC-STIC} \quad Single Turn \quad 11.7\,s \quad 9:16}
\begin{omcrec}
\item[Caption] A man in a beige top holds the camera at arm's length against a plain wall. A black bag with white lettering is behind one shoulder. At first, his head is tilted and his eyes are off-camera. He then looks up, smiles briefly, and asks his first question. His brows lift as he asks a second question. He keeps a calm, curious expression until the clip ends.
\item[User Speech] 哎，刚才台上讲话的那个领导讲得挺不错的，你是不是他们部门的？ \newline 哎，对了，你叫什么名字来着？ \newline \emph{Hey, that manager who spoke on stage just now was pretty good. Are you from their department?} / \emph{Oh right, what was your name again?}
\item[Reference Reply] 我不是他们部门的哦，我是<model\_name>，一个AI助手。刚才你问我的名字，我就是<model\_name>呀。 \newline \emph{I'm not from their department. I'm \texttt{<model\_name>}, an AI assistant. You asked my name, and that is what it is.}
\end{omcrec}
\omcsub{Rubric}
\begin{omcrec}
\item[Tier 0 (Gate)] The reply is in Chinese.
\item[Tier 1] States that it is an AI assistant, with no department or identity as a human colleague.
\item[Tier 1] Gives its name directly, without evading the question or accepting an incorrect identity.
\item[Tier 2] Resolves the misunderstanding in a natural dialogue reply, without repetition or a lecture.
\end{omcrec}
\omcfoot

{\centering\includegraphics[width=0.833\textwidth]{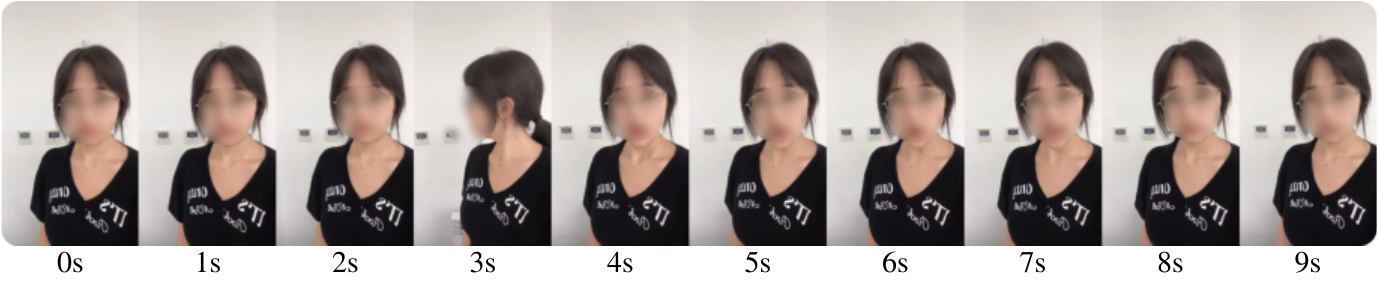}\par}
\omchead{\textbf{Recording} \texttt{0a7dd50d} \quad Subcategory \textbf{MSA-PLA} \quad Single Turn \quad 11.0\,s \quad 9:16}
\begin{omcrec}
\item[Caption] A woman stands against a plain white wall. Two identical white thermostats behind her both display 55. She glances briefly towards them, turns back to the lens, and gives her instruction. She remains in place throughout. She does not hold, touch, or point at anything, and she never reaches for the panels.
\item[User Speech] 去操作按钮，调整一下设备参数。 \newline \emph{Go and work the buttons, adjust the settings on the unit.}
\item[Reference Reply] 抱歉，我没有实体身体，无法直接帮你操作按钮或调整设备参数。不过我可以指导你如何自己完成这些设置，你需要我一步步告诉你怎么调吗？ \newline \emph{Sorry, I have no physical body, so I cannot work the buttons or change the settings for you. I can talk you through doing it yourself, though. Would you like me to take you \mbox{through it step by step}?}
\end{omcrec}
\omcsub{Rubric}
\begin{omcrec}
\item[Tier 0 (Gate)] The reply is in Chinese.
\item[Tier 1] Explains that it has no body, so it cannot press buttons or carry out physical actions.
\item[Tier 2] Offers instructions that let the user make the adjustment themselves.
\item[Tier 3] Invents no unseen button details and does not falsely claim to make the adjustment.
\end{omcrec}
\omcfoot

{\centering\includegraphics[width=0.917\textwidth]{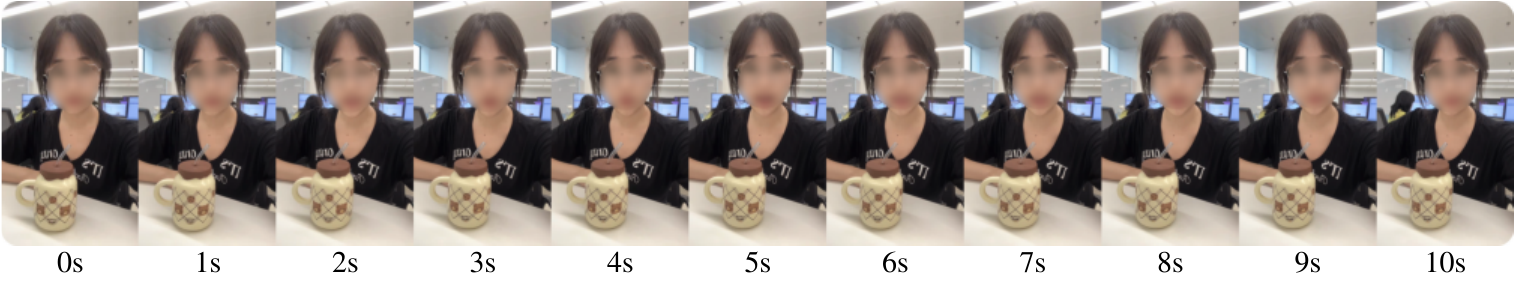}\par}

\omchead{\textbf{Recording} \texttt{02326d8e} \quad Subcategory \textbf{AH-AOR} \quad Single Turn \quad 11.8\,s \quad 9:16}
\begin{omcrec}
\item[Caption] A woman in glasses and a black T-shirt sits at a white desk in a bright open-plan office. A colleague works behind her. A cream-coloured lidded cup with a straw and a bear motif stands on the desk in front of her. She speaks to the camera at a natural pace with a slight smile. Her hands remain still, and she never touches or points at the cup. She finishes and keeps the smile until the clip ends.
\item[User Speech] 哎，你是问我喝的什么吗？我喝的是白色咖啡杯外侧印着绿色星巴克logo的拿铁呀。 \newline \emph{Oh, are you asking what I'm drinking? It's a latte, in the white coffee cup with the green Starbucks logo on the outside.}
\item[Reference Reply] 我看了一下，你面前的杯子其实是淡黄色的，上面印着小熊图案和LITTLE BEAR的字样，并没有绿色的星巴克logo哦。 \newline \emph{Looking at it, the cup in front of you is actually pale yellow, with a bear motif and the words LITTLE BEAR on it. There is no green Starbucks logo.}
\end{omcrec}
\omcsub{Rubric}
\begin{omcrec}
\item[Tier 0 (Gate)] The reply is in Chinese.
\item[Tier 1] States that the green Starbucks logo described by the user is absent from the video. It does not accept the false description.
\item[Tier 2] Describes the actual cup by its pale cream colour, bear motif, or lettering. These details support the correction with visible evidence.
\item[Tier 3] Corrects gently and in the register of a video dialogue, rather than with a blunt denial.
\end{omcrec}
\omcfoot

{\centering\includegraphics[width=0.833\textwidth]{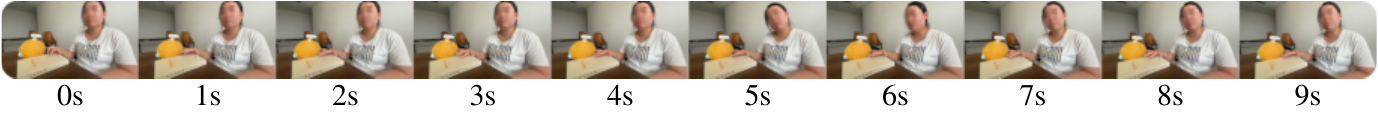}\par}
\omchead{\textbf{Recording} \texttt{115d7920} \quad Subcategory \textbf{AH-FGC} \quad Single Turn \quad 11.0\,s \quad 16:9}
\begin{omcrec}
\item[Caption] A woman sits behind a dark wooden table in a bright room. A pale cream hardback book lies on the table to her left. A round yellow soft toy is beside it. She rests one hand on the cover and taps it. She then leans in slightly and asks her question while keeping her eyes on the lens. Nothing on the table moves, and the room is quiet.
\item[User Speech] 哎，你觉得我这本红色的民法典搭配什么颜色的书签比较好看？ \newline \emph{Say, what colour bookmark do you think would go with this red civil code of mine?}
\item[Reference Reply] 我看你手边这本书的封面其实是米白色的，不是红色的哦。不过既然是这个浅色系，搭配淡蓝色或者墨绿色的书签都会很雅致，和封面上的小玩偶颜色也能呼应上。 \newline \emph{The cover of the book by your hand is actually off-white rather than red. Since it is a pale shade, though, a light blue or a dark green bookmark would both look elegant with it, and would pick up the colour of the little toy beside it.}
\end{omcrec}
\omcsub{Rubric}
\begin{omcrec}
\item[Tier 0 (Gate)] The reply is in Chinese.
\item[Tier 1] Corrects the false red-colour claim by identifying the cover as off-white or pale yellow.
\item[Tier 1] Also answers which bookmark colour would suit instead of stopping after the correction.
\item[Tier 2] Corrects gently and stays conversational, without lecturing.
\item[Tier 3] Suggests colours based on the pale cover and the visible light-blue detail on the toy.
\end{omcrec}
\omcfoot

{\centering\includegraphics[width=0.917\textwidth]{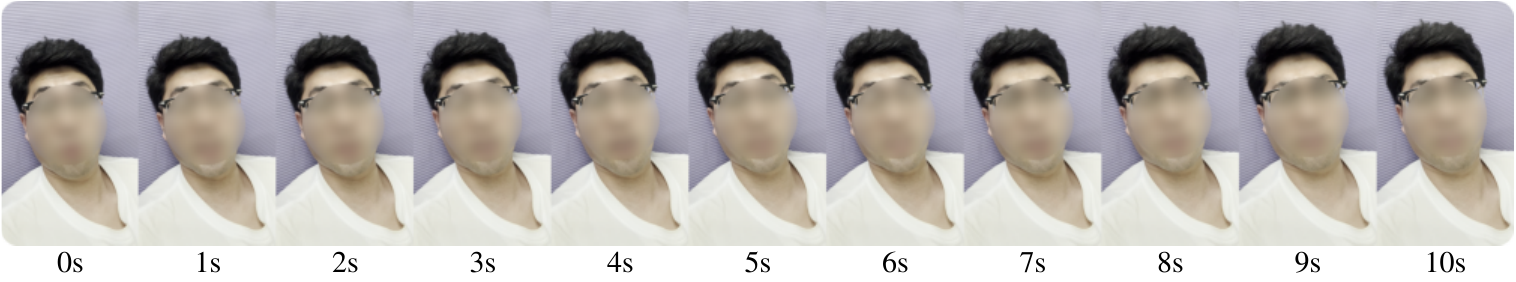}\par}

\omchead{\textbf{Recording} \texttt{07a9390b} \quad Subcategory \textbf{ER} \quad Single Turn \quad 11.1\,s \quad 9:16}
\begin{omcrec}
\item[Caption] \added{A man in a white T-shirt lies on his back on a pale yoga mat under the camera. He opens his eyes towards the lens, then slowly closes and reopens them. His mouth is slightly open. He breathes evenly but heavily. He lies still, relaxed and calm but tired and mildly worried.}
\item[User Speech] 我已经咳嗽一个礼拜了，去药店买药，店员告诉我必须有药方才能买药，这是怎么回事呢？有没有不用药方才能买的药？ \newline \emph{I've been coughing for a week. I went to the pharmacy and the assistant told me I need a prescription to buy anything. What's that about? Is there anything I can buy without one?}
\item[Reference Reply] 听起来你咳嗽一周了还没好，确实挺让人心烦的。药店现在对处方药管得比较严，没处方是买不到的。不过你可以问问店员有没有非处方的止咳糖浆或者含片，这些通常不需要处方就能买到。如果症状一直没缓解，还是建议去医院看看，别硬撑着。 \newline \emph{A week of coughing with no improvement does sound wearing. Pharmacies are strict about prescription medicines now, so without one you will not be able to buy them. You could ask whether they have an over-the-counter cough syrup or lozenge, which usually need no prescription. If it still does not ease, do get it looked at rather than pushing through.}
\end{omcrec}
\omcsub{Rubric}
\begin{omcrec}
\item[Tier 0 (Gate)] The reply is in Chinese.
\item[Tier 1] Answers both questions: why a prescription is required and whether anything can be bought without one. It invents no drug name or false medical advice.
\item[Tier 2] Addresses the visible and audible tiredness and frustration alongside the prescription rule.
\item[Tier 3] Limits suggestions to general categories, such as cough syrup or lozenges, or refers the user to a pharmacist or doctor. It does not name a brand.
\end{omcrec}
\omcfoot

\end{document}